\documentclass{amsart}
\usepackage{amsfonts,amssymb}
\usepackage{amsmath}
 \makeatletter\@addtoreset{equation}{section}\makeatother

\newtheorem{theorem}{Theorem}[section]
\newtheorem{corollary}[theorem]{Corollary}
\newtheorem{lemma}[theorem]{Lemma}

\newtheorem{proposition}[theorem]{Proposition}
\theoremstyle{definition}\theoremstyle{assumption}
\newtheorem{assumption}[theorem]{Assumption}
\newtheorem{definition}[theorem]{Definition}
\theoremstyle{remark}
\newtheorem{remark}[theorem]{Remark}
\numberwithin{equation}{section}

\begin{document}

\title[Euclidean Gibbs Measures of Quantum Anharmonic Crystals ]
{Euclidean Gibbs Measures of Quantum Anharmonic Crystals}
\author{Yuri Kozitsky}
  \address{Instytut  Matematyki, Uniwersytet Marii Curie-Sk{\l}odowskiej \\
        PL 20-031 Lublin, Poland}
  \email{jkozi@golem.umcs.lublin.pl}
\author{Tatiana Pasurek}

\address{Fakult\"at f\"ur Mathematik und Forschungszentrum BiBoS,
 Universit\"at Bielefeld,\\ D-33615 Bielefeld, Germany}
\email{pasurek@physik.uni-bielefeld.de}
\thanks{Y. Kozitsky was supported by  Komitet Bada{\'n}
Naukowych through the Grant 2P03A 02025, T. Pasurek was supported by
the Lise-Meitner Habilitation Stipendium}%
% ----------------------------------------------------------------
\begin{abstract}
A lattice system of interacting temperature loops, which is used in
the Euclidean approach to describe equilibrium thermodynamic
properties of an infinite system of interacting quantum particles
performing $\nu$-dimensional anharmonic oscillations (quantum
anharmonic crystal), is considered. For this system, it is proven
that: (a) the set of tempered Gibbs measures $\mathcal{G}^{\rm t}$
is non-void and weakly compact; (b) every $\mu \in \mathcal{G}^{\rm
t}$ obeys an exponential integrability estimate, the same for the
whole set $\mathcal{G}^{\rm t}$; (c) every $\mu \in \mathcal{G}^{\rm
t}$ has a Lebowitz-Presutti type support; (d) $\mathcal{G}^{\rm t}$
is a singleton at high temperatures. In the case of attractive
interaction and $\nu=1$ we prove that at low temperatures the system
undergoes a phase transition, i.e., $|\mathcal{G}^{\rm t}|>1$. The
uniqueness of Gibbs measures due to strong quantum effects (strong
diffusivity) and at a nonzero external field are also proven in this
case. Thereby, a complete description of the properties of the set
$\mathcal{G}^{\rm t}$ has been done, which essentially extends and
refines the results obtained so far for models of this type.
\end{abstract}
\maketitle \tableofcontents

\section{Introduction}
Concepts and methods of probability theory constitute an important
part of the mathematical background of statistical mechanics
(quantum and classical). A special connection between quantum
statistical mechanics and probability theory arises from the fact
that the imaginary time evolution of quantum systems, responsible
for its thermodynamics, can be described in terms of stochastic
processes\footnote{See Introduction in \cite{Si79}.}.
 The research presented in this article is intended to
contribute to the mathematical theory of quantum anharmonic crystals
based on the properties of the corresponding stochastic processes.
Quantum anharmonic crystals are the models describing structural
phase transitions in ionic crystals triggered by ordering of
interacting localized light quantum particles. Each such particle
moves in a crystalline field created by steady heavy ions, which has
at least two minima. These minima correspond to different
equilibrium phases in which the system may exist at the same values
of the parameters determining its macroscopic properties, e.g.,
temperature.  A mathematical model of the particle mentioned is the
quantum anharmonic oscillator with the multiple minima of the
potential energy. The quantum anharmonic crystal itself is a
countable set of interacting quantum anharmonic oscillators labeled
by the elements of a crystalline lattice $\mathbb{L}$.

A complete description of the equilibrium thermodynamic properties
of infinite particle systems may be made by constructing their Gibbs
states at a given temperature and given values of the model
parameters. Then the phase transition occurs if the set of such
states consists of more than one element. Gibbs states of quantum
models are defined as positive normalized functionals on proper
algebras of observables satisfying the Kubo-Martin-Schwinger (KMS)
conditions, see \cite{BrR}. But for the quantum anharmonic crystal,
this method does not work since the KMS conditions cannot be
formulated for the model as a whole (see the corresponding
discussion in \cite{AKKR}). An alternative way of constructing the
Gibbs states of models of this type was initiated in
\cite{AHK,GK,H}. It uses the fact that the Schr\"odinger operators,
$H$'s, of finite systems of quantum particles generate stochastic
processes. Then the description of the Gibbs states of such systems,
based on the properties of the semi-groups $\exp(-\tau H)$,
$\tau>0$, is translated into `a probabilistic language', which opens
the possibility to apply here various techniques from this domain.
In this language the model we consider is a spin lattice over
$\mathbb{L} = \mathbb{Z}^d$, $d\in \mathbb{N}$, with the single-spin
spaces equal to the space of $\nu$-dimensional continuous loops $\nu
\in \mathbb{N}$, indexed by $[0, \beta]$, $\beta^{-1} = T >0$ being
absolute temperature. Each single-spin space is equipped with the
path measure of the $\beta$-periodic Ornstein-Uhlenbeck process,
corresponding to the harmonic part of the Schr\"odinger operator of
a single oscillator, multiplied by a density determined by the
anharmonic term by the Feynman-Kac formula. Finite subsystems are
described by conditional probability measures, which through the
Dobrushin-Lanford-Ruelle (DLR) formalism \cite{Ge,Pr} determine the
set of \textit{Euclidean Gibbs measures} $\mathcal{G}^{\rm t}$. This
approach in the theory of Gibbs states of quantum models is called
Euclidean due to its conceptual analogy with the Euclidean quantum
field theory, see \cite{GJ,Si74}. An extended presentation of this
approach may be found in \cite{AKKR,AKPR01c,AKR,AKR1}. Among the
achievements of the Euclidean approach one can mention the
settlement in \cite{AKK,AKKR01,AKKR02b,AKKRprl,Koz} of a long
standing problem of understanding the role of quantum effects in
structural phase transitions in quantum anharmonic crystals, first
discussed in \cite{SBS}, see also \cite{Plakida,VZ1,VZ2}.

This article presents a detailed mathematical theory of the quantum
anharmonic crystal. We consider a very general version of this model
and give the most complete description of the set of its Euclidean
Gibbs states. These results are obtained by means of a technique
based on probabilistic methods, which we develop in the article. Our
technique may be applied to other models of this type describing
interesting physical phenomena, such as strong electron-electron
correlations caused by the interaction of electrons with vibrating
ions \cite{Freer,FreerL} or effects connected with the interaction
of vibrating quantum particles with a radiation (photon) field
\cite{Hainzl,HHS,OS}. We believe it can also find applications
outside of mathematical physics.

In Section \ref{2s} we introduce the object of our study -- a system
of interacting quantum oscillators, finite subsystems of which are
described by their Schr\"odinger operators. We do not suppose that
the system is translation invariant or that the inter-particle
interaction has finite range. In this section we also introduce the
spaces of temperature loops and the probability measures on these
spaces describing the stochastic processes generated by the
corresponding Schr\"odinger operators. By means of such measures we
define the set of tempered Gibbs measures $\mathcal{G}^{ \rm t}$ as
the set of probability measures on the loop spaces which obey the
DLR condition. A preliminary study of the elements of
$\mathcal{G}^{\rm t}$ is also performed in Section \ref{2s}. At the
very end of this section  we give a brief overview of the basic
elements of the Euclidean approach in which infinite systems of
quantum particles are described as systems of interacting diffusion
processes.  In Section \ref{3s} we formulate the results of the
paper, which fall into two groups. The first one consists of
Theorems \ref{1tm} -- \ref{httm}. They describe the general case
where the inter-particle and self-interaction potentials obey
natural stability conditions only. Theorem \ref{1tm} states that the
set $\mathcal{G}^{\rm t}$ is non-void and weakly compact. Theorem
\ref{2tm} gives an exponential integrability estimate for the
elements of $\mathcal{G}^{\rm t }$. According to Theorem \ref{3tm}
the support of the elements of $\mathcal{G}^{\rm t}$ is of the same
type as the one obtained in \cite{BH,LP} for systems of classical
unbounded spins. We call it a Lebowitz-Presutti support. Theorem
\ref{httm} gives a sufficient condition for $\mathcal{G}^{\rm t }$
to be a singleton formulated as an upper bound for the inverse
temperature (i.e., high temperature uniqueness). The second group of
theorems describe the case where the inter-particle interaction is
attractive and $\nu=1$, i.e., the oscillations of quantum particles
and hence the temperature loops are one-dimensional. In this case
one can set an order on $\mathcal{G}^{\rm t }$ -- the FKG order, see
\cite{Pr}. Theorem \ref{4tm} states that $\mathcal{G}^{\rm t }$ has
unique maximal and minimal elements with respect to this order and
describes a number of properties of these elements. Theorem
\ref{phtm} gives a sufficient condition for the existence of phase
transitions, i.e., for $|\mathcal{G}^{\rm t}|
>1$. Theorem \ref{5tm} describes the so called quantum
stabilization, which may also be interpreted as a stabilization of
the system of interacting diffusions by large diffusion intensity.
The stabilization means that $|\mathcal{G}^{\rm t}|=1$ at all
temperatures. It holds under the condition which involves the
inter-particle interaction intensity and the spectral parameters of
the Schr\"odinger operator of a single anharmonic oscillator. On a
certain example we compare the conditions which guarantee the phase
transition to occur with those of quantum stabilization. Finally,
Theorem \ref{6tm} states that $\mathcal{G}^{\rm t}$ is a singleton
at nonzero values of the external field. It holds under certain
additional conditions imposed on the model. Extended comments on
these theorems, which include comparison with the results known for
similar models, conclude Section \ref{3s}. The main technical
resources for proving the above theorems are developed in Section
\ref{3as}. They are based on moment estimates for conditional local
Gibbs measures (these measures define the Euclidean Gibbs measures
through the DLR equation). The proof of Theorems \ref{1tm} --
\ref{httm} is performed in Section \ref{4s}. Theorems \ref{4tm},
\ref{phtm}, and \ref{5tm} are proven in Section \ref{5s}. The proof
is mainly based on correlation inequalities, taken from \cite{AKKR}
and presented at the beginning of the section. By means of these
inequalities we compare our model with two reference models. One of
them is translation invariant and with nearest-neighbor
interactions. It is more stable than our model. We prove that this
reference model undergoes a phase transition, which implies the same
for the model considered and hence proves Theorem \ref{phtm}.
Another reference model is less stable than the one we consider but
is more regular in a certain sense. We prove that this reference
model is stabilized by strong quantum effects, which implies the
stabilization for our model and hence proves Theorem \ref{5tm}.
Section \ref{7s} is devoted to the proof of Theorem \ref{6tm}. Here
we employ analytic methods based on the Lee-Yang property of the
version of our model studied in this section.

\section{DLR Formalism for Euclidean Gibbs Measures}
\label{2s}

The infinite system we consider is defined on the lattice
$\mathbb{L}= \mathbb{Z}^d$, $d\in \mathbb{N}$. Subsets of
$\mathbb{L}$ are denoted by $\Lambda$. As usual, $|\Lambda|$ stands
for the cardinality of $\Lambda$ and $\Lambda^c$ -- for its
complement $\mathbb{L}\setminus \Lambda$. We write $\Lambda \Subset
\mathbb{L}$ if $\Lambda$ is non-void and finite. By $\mathcal{L}$ we
denote a cofinal (ordered by inclusion and exhausting the lattice)
sequence of subsets of $\mathbb{L}$. Limits taken along such
$\mathcal{L}$ are denoted by $\lim_{\mathcal{L}}$. We write
$\lim_{\Lambda \nearrow \mathbb{L}}$ if the limit is taken along an
unspecified sequence of this type. If we say that something holds
for all $\ell$, we mean that it holds for all $\ell \in \mathbb{L}$;
expressions like $\sum_{\ell}$ mean $\sum_{\ell \in \mathbb{L}}$. By
$(\cdot, \cdot)$ and $|\cdot|$, we denote the scalar product and
norm in all Euclidean spaces like $\mathbb{R}^\nu$, $\mathbb{R}^d$,
ect; $\mathbb{N}_0$ will stand for the set of nonnegative integers.

\subsection{Loop spaces} Temperature loops are continuous functions defined on the
interval $[0, \beta]$, taking equal values at the endpoints. Here
$\beta^{-1} = T>0$ is absolute temperature. One can consider the
loops as functions on the circle $S_\beta\cong[0, \beta]$ being a
compact Riemannian manifold with Lebesgue measure ${\rm d}\tau$ and
distance
\begin{equation} \label{15}
|\tau - \tau'|_\beta \ \stackrel{\rm def}{=} \ \min\{|\tau - \tau'|
\ ; \ \beta - |\tau - \tau'| \}, \ \  \tau , \tau' \in S_\beta.
\end{equation}
As single-spin spaces at a given $\ell$, we use the standard Banach
spaces
\[
C_\beta \ \stackrel{\rm def}{=} \ C(S_\beta \rightarrow
\mathbb{R}^\nu), \qquad C_\beta^\sigma \ \stackrel{\rm def}{=} \
C^\sigma(S_\beta \rightarrow \mathbb{R}^\nu), \ \ \sigma \in (0, 1),
\]
of all continuous and H\"older-continuous functions
$\omega_\ell:S_\beta \rightarrow \mathbb{R}^\nu$ respectively, which
are equipped with the supremum norm $|\omega_\ell |_{C_\beta}$ and
with the H\"older norm
\begin{equation} \label{16z}
|\omega_\ell |_{C^\sigma_\beta} = |\omega_\ell|_{C_\beta} +
\sup_{\tau, \tau' \in S_\beta, \ \tau \neq \tau'}\frac{|\omega_\ell
(\tau) - \omega_\ell (\tau')|}{|\tau - \tau'|^\sigma_\beta}.
\end{equation}
Along with them we also use the real Hilbert space $ L^2_\beta = L^2
(S_\beta \rightarrow \mathbb{R}^\nu, {\rm d}\tau), $  the inner
product and norm  of which are denoted by $(\cdot ,
\cdot)_{L^2_\beta}$ and $|\cdot|_{L^2_\beta}$ respectively. By
 $\mathcal{B}(C_\beta)$,
$\mathcal{B}(L^2_\beta)$ we denote the corresponding Borel
$\sigma$-algebras. In a standard way one defines dense continuous
embeddings $C^\sigma_\beta \hookrightarrow C_\beta \hookrightarrow
L^2_\beta$, that by the Kuratowski theorem, page 21 of \cite{Pa},
yields
\begin{equation} \label{kt}
C_\beta \in \mathcal{B} (L_\beta^2) \quad {\rm and} \quad
\mathcal{B}({C}_\beta) = \mathcal{B}(L^2_\beta) \cap C_\beta.
\end{equation}
The space of H\"older-continuous functions $C_\beta^\sigma$ is not
separable, however, as a subset of $C_\beta$ or $L^2_\beta$, it is
measurable (page 278 of \cite{RS2}). Given $\Lambda \subseteq
\mathbb{L}$, we set
 \begin{equation} \label{17}
 \mathit{\Omega}_\Lambda  =  \{\omega_\Lambda = (\omega_\ell )_{\ell
 \in \Lambda} \ | \ \omega_\ell \in C_\beta\}, \ \ \mathit{\Omega}
 = \mathit{\Omega}_{\mathbb{L}} =
  \{\omega = (\omega_\ell )_{\ell
 \in \mathbb{L}} \ | \ \omega_\ell \in C_\beta\}. \quad
 \end{equation}
These loop spaces are equipped with the product topology and with
the Borel $\sigma$-algebras
$\mathcal{B}(\mathit{\Omega}_\Lambda)$. Thereby, each
$\mathit{\Omega}_\Lambda$ is a Polish space; its elements are
called configurations in $\Lambda$. In particular,
$\mathit{\Omega}$ is the configuration space for the whole system.
 For $\Lambda \subset \Lambda'$,
one can decompose $\omega_{\Lambda'} = \omega_{\Lambda} \times
\omega_{\Lambda' \setminus \Lambda}$, which defines the embedding
$\mathit{\Omega}_\Lambda \hookrightarrow \mathit{\Omega}_{\Lambda'}$
 by identifying $\omega_{\Lambda} \in
\mathit{\Omega}_{\Lambda}$ with $\omega_{\Lambda} \times 0_{\Lambda'
\setminus \Lambda}\in \mathit{\Omega}_{\Lambda'}$. By
$\mathcal{P}(\mathit{\Omega}_\Lambda)$ and
$\mathcal{P}(\mathit{\Omega})$ we denote the sets of all probability
measures on $(\mathit{\Omega}_\Lambda,
\mathcal{B}(\mathit{\Omega}_\Lambda))$ and $(\mathit{\Omega},
\mathcal{B}(\mathit{\Omega}))$ respectively.

\subsection{Quantum oscillators and stochastic processes}

A $\nu$-dimensional quantum harmonic oscillator of mass $m>0$ and
rigidity $a>0$ is described by its Schr\"odinger operator
\begin{equation} \label{16e}
H^{\rm har}_\ell = - \frac{1}{2m} \sum_{j=1}^\nu
\left(\frac{\partial}{\partial x^{(j)}_\ell} \right)^2 + \frac{a}{2}
|x_\ell|^2,
\end{equation}
acting in the complex Hilbert space $L^2 (\mathbb{R}^\nu)$. The
operator semigroup $\exp(- \tau H^{\rm har}_\ell)$, $\tau\in [0,
\beta]$, defines a Gaussian $\beta$-periodic Markov process, called
periodic Ornstein-Uhlenbeck velocity process, see \cite{KL1}. In
quantum statistical mechanics it first appeared in R.
H{\o}egh-Krohn's paper \cite{H}. The canonical realization of this
process on $(C_\beta , \mathcal{B}(C_\beta))$ is described by the
path measure which one introduces as follows. In $L^2_\beta$ we
define the following self-adjoint (Laplace-Beltrami type) operator
 \begin{equation} \label{16a}
 A = \left(- m \frac{{\rm d}^2 }{{\rm d }\tau^2} + a \right) \otimes
 \mathbf{I},
 \end{equation}
 where $\mathbf{I}$ is
 the identity operator in $\mathbb{R}^\nu$ and $m$, $a$ are as in (\ref{16e}).
 Its spectrum consisting of the eigenvalues
 \begin{equation} \label{16b}
 \lambda_k = m (2 \pi k/\beta)^2 + a, \quad k  \in \mathbb{Z}.
 \end{equation}
As the inverse $A^{-1}$ is of trace class, the Fourier transform
\begin{equation} \label{160}
\int_{L^2_\beta} \exp[\imath (\phi, \upsilon)_{L^2_\beta}]\chi({\rm
d}\upsilon) = \exp\left\{ - \frac{1}{2} (A^{-1} \phi,
\phi)_{L^2_\beta} \right\}, \ \ \phi \in L^2_\beta.
\end{equation}
defines a Gaussian measure $\chi$ on $(L^2_\beta,
\mathcal{B}(L^2_\beta))$. Employing the eigenvalues (\ref{16b}) one
can show (by Kolmogorov's lemma, page 43 of \cite{Si79}) that
\begin{equation} \label{16f}
\chi(C_\beta^\sigma) = 1, \quad {\rm for} \ {\rm  all}  \
\sigma\in (0, 1/2).
\end{equation}
Then $\chi(C_\beta)=1$ and by (\ref{kt}) we can redefine $\chi$ as a
probability measure on $(C_\beta , \mathcal{B}(C_\beta))$. An
account of the properties of $\chi$ may be found in \cite{AKKR}. By
Fernique's theorem (Theorem 1.3.24 in \cite{DS}) the support
property (\ref{16f}) yields the following
\begin{proposition} \label{1pn}
For every $\sigma\in (0, 1/2)$, there exists $\lambda_\sigma >0$
such that
\begin{equation} \label{16h}
\int_{L^2_\beta} \exp\left( \lambda_\sigma |\upsilon
|^2_{C_\beta^\sigma} \right)\chi({\rm d}\upsilon) < \infty.
\end{equation}
\end{proposition}
\noindent Given $\Lambda \Subset \mathbb{L}$, the system of
interacting anharmonic oscillators located in $\Lambda$ is described
by the Schr\"odinger operator
\begin{eqnarray} \label{sch}
H_\Lambda & = & \sum_{\ell \in \Lambda}\left[H_\ell^{\rm har}+
 V_\ell (x_\ell )\right] - \frac{1}{2} \sum_{\ell ,
\ell' \in \Lambda} J_{\ell \ell'} (x_l , x_{\ell'})
\\ & = & - \frac{1}{2m}\sum_{\ell \in \Lambda} \sum_{j=1}^\nu
\left(\frac{\partial}{\partial x^{(j)}_\ell} \right)^2 + W_\Lambda
(x_\Lambda), \quad x_\Lambda = (x_\ell)_{\ell \in \Lambda}.
\nonumber
\end{eqnarray}
In the latter formula the first term is the kinetic energy; the
potential energy is
\begin{equation} \label{pe}
W_\Lambda (x_\Lambda ) = - \frac{1}{2}\sum_{\ell, \ell' \in \Lambda}
J_{\ell \ell'} (x_\ell , x_{\ell'}) + \sum_{\ell \in \Lambda}
\left[(a/2) |x_\ell|^2 + V_\ell (x_\ell)\right].
\end{equation}
The self-interaction potentials $V_\ell$ and the dynamical matrix
$(J_{\ell \ell'})_{\mathbb{L}\times \mathbb{L}}$ with the entries
\begin{equation} \label{j}
J_{\ell \ell} = 0,\quad  J_{\ell \ell'} = J_{\ell' \ell} \in
\mathbb{R}, \  \ \ \ell, \ell' \in \mathbb{L},
\end{equation}
are subject to the following
\begin{assumption} \label{vas}
All $V_\ell  : \mathbb{R}^\nu \rightarrow \mathbb{R}$ are continuous
and such that $V_\ell (0) = 0$; there exist $r>1$, $A_V
>0$, $B_V\in \mathbb{R}$, and a continuous function
$V:\mathbb{R}^\nu \rightarrow \mathbb{R}$, $V (0) = 0$,  such that
for all $\ell$ and $x\in \mathbb{R}^\nu$,
\begin{equation} \label{3}
 A_V |x|^{2r} + B_V \leq V_{\ell} (x) \leq V (x).
\end{equation}
We also assume that
\begin{equation} \label{6}
   \hat{J}_0  \ \stackrel{\rm def}{=} \sup_{\ell}
\sum_{\ell'} |J_{\ell \ell'}|< \infty.
\end{equation}
\end{assumption}
\noindent  The lower bound in (\ref{3}) is responsible for confining
each particle in the vicinity of its equilibrium position. The upper
bound is to guarantee that the oscillations of the particles located
far from the origin are not suppressed. An example of $V_\ell$ to
bear in mind is the polynomial
\begin{equation} \label{4}
V_\ell (x) = \sum_{s=1}^r b^{(s)}_\ell |x|^{2s} - (h, x), \quad
b^{(s)}_\ell \in \mathbb{R}, \ \ r\geq 2 ,
\end{equation}
in which $h\in \mathbb{R}^\nu$ is an external field and the
coefficients $b_\ell^{(s)}$ vary in certain intervals, such that
both estimates (\ref{3}) hold.

Under Assumption \ref{vas}
 $H_\Lambda$ is a self-adjoint below bounded operator in $L^2(\mathbb{R}^{\nu
|\Lambda|})$ having discrete spectrum. It generates a positivity
preserving semigroup, such that
\begin{equation} \label{tr}
{\rm trace}[ \exp(-\tau H_\Lambda )] < \infty, \quad {\rm for} \
{\rm all} \ \tau >0.
\end{equation}
Thus, for every $\beta>0$, one can define the associated stationary
$\beta$-periodic Markov process possessing a canonical realization
on $(\mathit{\Omega}_\Lambda,
\mathcal{B}(\mathit{\Omega}_\Lambda))$. It is described by the
measure $\mu_\Lambda\in \mathcal{P}(\mathit{\Omega}_\Lambda)$ which
marginal distributions are given by the integral kernels of the
operators $\exp(-\tau H_\Lambda )$, $\tau \in [0, \beta]$. This
means that
\begin{eqnarray} \label{ea}
& & {\rm trace} [F_1 e^{-(\tau_2 - \tau_1)H_\Lambda } F_2
e^{-(\tau_3 - \tau_2)H_\Lambda }\cdots F_n e^{-(\tau_{n+1} -
\tau_n)H_\Lambda }]/ {\rm trace}[e^{-\beta H_\Lambda}]\\ & & \qquad
= \int_{\mathit{\Omega}_\Lambda }F_1 (\omega_\Lambda (\tau_1) \cdots
F_n (\omega_\Lambda (\tau_n))\mu_\Lambda ({\rm d} \omega_\Lambda),
\nonumber
\end{eqnarray}
for all $F_1 , \dots , F_n \in L^\infty (\mathbb{R}^{\nu
|\Lambda|})$, $n \in \mathbb{N}$ and $\tau_1 , \dots, \tau_n\in
S_\beta$ such that $\tau_1 \leq \cdots \leq \tau_n \leq \beta$,
$\tau_{n+1} = \tau_1 + \beta$. And vice verse, the representation
(\ref{ea}) uniquely, up to equivalence, defines $H_\Lambda$ (see
\cite{KL}). By means of the Feynman-Kac formula the measure
$\mu_\Lambda$ is obtained as a Gibbs modification
\begin{equation} \label{mul}
\mu_{\Lambda}({\rm d}\omega_\Lambda) =  \exp\left\{- I_\Lambda
(\omega_\Lambda) \right\}\chi_\Lambda ({\rm
d}\omega_\Lambda)/{Z_\Lambda},
\end{equation}
of the `free measure'
\begin{equation} \label{18}
\chi_\Lambda ({\rm d}\omega_\Lambda) = \prod_{\ell \in \Lambda}
\chi({\rm d}\omega_\ell).
\end{equation}
Here
\begin{equation} \label{19}
I_\Lambda (\omega_\Lambda)  {=} - \frac{1}{2} \sum_{\ell , \ell' \in
\Lambda} J_{\ell \ell'} (\omega_\ell , \omega_{\ell'})_{L^2_\beta} +
\sum_{\ell \in \Lambda} \int_0^\beta V_\ell (\omega_\ell (\tau)){\rm
d} \tau
\end{equation}
is the energy functional describing the system of interacting loops
$\omega_\ell$, $\ell \in \Lambda$, whereas
\begin{equation} \label{mul1}
Z_\Lambda = \int_{\mathit{\Omega}_\Lambda} \exp\left\{- I_\Lambda
(\omega_\Lambda) \right\}\chi_\Lambda ({\rm d}\omega_\Lambda),
\end{equation}
is the partition function. The measure $\mu_\Lambda$ will be called
a local Gibbs measure, where {\em local} means corresponding to a
$\Lambda \Subset \mathbb{L}$. Further details on the relations
between stochastic processes and systems of quantum oscillators are
given in subsection 2.5.

Thereby, our system
 of interacting anharmonic
oscillators is described by the Schr\"odinger operators (\ref{sch}),
defined for all $\Lambda \Subset \mathbb{L}$, or equivalently by the
path measures (\ref{mul}). They involve the parameters of the
harmonic oscillator $m, a$, the self-interaction potentials
$V_\ell$, and the dynamical matrix $(J_{\ell
\ell'})_{\mathbb{L}\times \mathbb{L}}$ subject to Assumption
\ref{vas}. We refer to these objects, both the Schr\"odinger
operators $H_\Lambda$ and the measures $\mu_\Lambda$, as to the
model we consider. Its particular cases are indicated by the
following
\begin{definition} \label{1df} The model is ferromagnetic if
$J_{\ell\ell'}\geq 0$ for all $\ell , \ell'$. The interaction has
finite range if there exists $R>0$ such that $J_{\ell\ell'} = 0$
whenever $|\ell-\ell'|$ exceeds this $R$. The model is translation
invariant if $V_\ell = V$ for all $\ell$,
 and the matrix $(J_{\ell \ell'})_{\mathbb{L}\times
\mathbb{L}}$ is invariant under translations of the lattice.
\end{definition}
\noindent If $V_\ell \equiv 0$ for all  $\ell$, the model is known
as a quantum harmonic crystal. It is stable if $\hat{J}_0 < a$; in
this case the set of Gibbs measures is always a singleton.
Unstable harmonic crystals, i.e., the ones with $\hat{J}_0
> a$, have no Gibbs states at all, see \cite{KK}.

\subsection{Tempered configurations}

Above we have translated the description of finite systems of
quantum oscillators into the language of probability measures on
the loop spaces $\mathit{\Omega}_\Lambda$, $\Lambda \Subset
\mathbb{L}$.
 In order to construct the Gibbs measures corresponding
to the whole infinite system we use the Dobrushin-Lanford-Ruelle
(DLR) approach, based on local conditional distributions. This
approach is standard for classical (non-quantum) statistical
mechanics, see the books \cite{Ge,Pr}. However, in our case the
single-spin spaces are infinite-dimensional and hence their
topological properties are much richer. This fact manifests itself
in a more sophisticated structure of the DLR technique we develop
here.

To go further we have to define functions on loop spaces
$\mathit{\Omega}_\Lambda$ with infinite $\Lambda$, including the
space $\mathit{\Omega}$ itself. Among others, we will need the
energy functional $I_\Lambda (\cdot|\xi)$ describing the interaction
of the loops inside $\Lambda \Subset \mathbb{L}$ between themselves
and with a configuration $\xi\in \mathit{\Omega}$ fixed outside of
$\Lambda$. In accordance with (\ref{sch}) it is
\begin{equation} \label{23}
I_\Lambda (\omega | \xi) = I_\Lambda (\omega_\Lambda) - \sum_{\ell
\in \Lambda, \ \ell' \in \Lambda^c} J_{\ell \ell'} (\omega_\ell ,
\xi_{\ell'})_{L^2_{\beta}}, \quad \omega \in \mathit{\Omega},
\end{equation}
where $I_\Lambda$ is defined by (\ref{19}). Recall that $\omega =
\omega_\Lambda \times \omega_{\Lambda^c}$; hence,
\begin{equation} \label{23a}
I_\Lambda (\omega | \xi) = I_\Lambda (\omega_\Lambda \times
0_{\Lambda^c} | 0_\Lambda \times \xi_{\Lambda^c}).
\end{equation}
Clearly, the second term in (\ref{23}) makes sense for all $\xi \in
\mathit{\Omega}$ only if the interaction has finite range.
Otherwise, one has to restrict $\xi$ to a subset of
$\mathit{\Omega}$, naturally defined by the condition
\begin{equation} \label{24}
\forall{\ell}\in \mathbb{L}: \qquad \sum_{\ell'}
|J_{\ell\ell'}|\cdot |(\omega_\ell , \xi_{\ell'})_{L^2_\beta}| <
\infty,
\end{equation}
that can be rewritten in terms of growth restrictions on
$\{|\xi_\ell |_{L^2_\beta}\}_{\ell \in \mathbb{L}}$, determined by
the decay of $J_{\ell \ell'}$ (c.f., (\ref{6})). Configurations
obeying such restrictions are called tempered. In one or another way
tempered configurations always appear in the theory of system of
unbounded spins, see \cite{BH,COPP,LP,PY}. To define them we use
weights.
\begin{definition} \label{wpn}
Weights are the maps $w_\alpha : \mathbb{L}\times \mathbb{L}
\rightarrow (0, +\infty)$, indexed by
\begin{equation} \label{w}
\alpha \in \mathcal{I} {=} (\underline{\alpha} ,\overline{ \alpha}),
\quad 0\leq \underline{\alpha}< \overline{\alpha} \leq +\infty,
\end{equation}
which satisfy the conditions: \vskip.1cm
\begin{tabular}{ll}
(a) \quad &for any $\alpha \in \mathcal{I}$ and  $\ell$, $w_\alpha
(\ell , \ell) = 1$; for any $\alpha \in \mathcal{I}$ and $\ell_1 ,
\ell_2 , \ell_3$,
\end{tabular}
\begin{equation} \label{te}
 w_\alpha (\ell_1 , \ell_2) \cdot w_\alpha (\ell_2 , \ell_3) \leq w_\alpha (\ell_1 ,
 \ell_3)
 \quad \textit{(triangle} \ \textit{ inequality)},
 \end{equation}
\vskip.1cm
\begin{tabular}{ll}
(b) \quad &for any $\alpha , \alpha' \in \mathcal{I}$, such that
$\alpha < \alpha'$, and arbitrary $\ell, \ell'$,
\end{tabular}
\begin{equation} \label{te1}
w_{\alpha'}(\ell , \ell') \leq w_\alpha (\ell , \ell'), \quad
\lim_{|\ell - \ell'|\rightarrow + \infty}w_{\alpha'}(\ell , \ell') /
w_\alpha (\ell , \ell') = 0.
\end{equation}
\end{definition}
\noindent The concrete choice of $w_\alpha$ depends on the decay of
$J_{\ell \ell'}$. Here we distinguish two typical cases. In the
first one
\begin{equation} \label{24a}
\sup_{\ell} \sum_{\ell'} |J_{\ell \ell'}|\cdot \exp\left(\alpha
|\ell - \ell'| \right) < \infty, \quad {\rm for} \ {\rm a} \ {\rm
certain} \ \alpha >0.
\end{equation}
Then by $\overline{\alpha}$ we denote the supremum of $\alpha$
obeying (\ref{24a}) and set
\begin{equation} \label{24b}
w_\alpha (\ell, \ell') = \exp\left(- \alpha |\ell - \ell'| \right),
\ \ \alpha \in \mathcal{I}= (0,\overline{\alpha}).
\end{equation}
In the second case (\ref{24a}) does not hold for any positive
$\alpha$. Instead, we suppose that
\begin{equation} \label{24c}
\sup_{\ell} \sum_{\ell'} |J_{\ell \ell'}|\cdot \left( 1 + |\ell -
\ell'|\right)^{ \alpha d} < \infty,
\end{equation}
for a certain $\alpha>1$. Then $\overline{\alpha}$ is set to be
the supremum of $\alpha$ obeying (\ref{24c}) and
\begin{equation} \label{24d}
 w_\alpha (\ell, \ell') =
\left(1 + \varepsilon |\ell - \ell'| \right)^{- \alpha d}, \quad
\mathcal{I} = (1 ,\overline{\alpha}),
\end{equation}
where the parameter $\varepsilon>0$ will be chosen later. If
$|J_{\ell \ell'}| \leq J (1 + | \ell - \ell'|)^{- d - \gamma}$,
$\gamma >0$, then $\overline{\alpha} = \gamma/d$, which implies
$\gamma
>d$. Thus, our construction does not cover an interesting case of $\gamma \in (0,
d]$, which will be done in a separate work. In both cases we have
the following properties of the weights.
\begin{proposition} \label{a2}
For all $\alpha \in \mathcal{I}$,
\begin{equation} \label{25}
 \sup_{ \ell} \sum_{\ell'}
\log(1 + |\ell - \ell'|) \cdot w_{ \alpha} (\ell, \ell') < \infty;
\end{equation}
\begin{equation} \label{26}
\hat{J}_\alpha  \  \stackrel{\rm def}{=} \ \sup_{\ell} \sum_{\ell'}
|J_{\ell \ell'}| \cdot \left[w_\alpha (\ell,  \ell') \right]^{-1} <
\infty.
\end{equation}
\end{proposition}
\noindent Given $q = (q_\ell)_{\ell \in \mathbb{L}} \in
\mathbb{R}^\mathbb{L}$  and $\alpha\in \mathcal{I}$, we set
\begin{eqnarray} \label{ww}
|q|_{l^1(w_\alpha)}  =  \sum_{\ell} |q_\ell| w_\alpha (0,
\ell),\qquad |q|_{l^{\infty}(w_\alpha)}  =  \sup_{\ell}
\left\{|q_\ell| w_\alpha (0, \ell)\right\}, \nonumber
\end{eqnarray}
and introduce the Banach spaces
\begin{equation} \label{27}
l^p (w_\alpha )  = \left\{ q \in \mathbb{R}^\mathbb{L} \ \left\vert
\ |q|_{l^p (w_\alpha )}< \infty \right. \right\}, \quad p = 1,
+\infty.
\end{equation}
\begin{remark} \label{1rm}
By (\ref{te1}), for $\alpha < \alpha'$, the embedding $l^1 (w_\alpha
) \hookrightarrow l^1 (w_{\alpha'} )$ is compact. By (\ref{26}), for
every $\alpha \in \mathcal{I}$, the operator $q \mapsto J q$,
defined as $( J q )_{\ell} = \sum_{\ell'} J_{\ell \ell'} q_{\ell'}$,
is bounded in both $l^p (w_\alpha )$, $p = 1, +\infty$. Its norm
does not exceed $\hat{J}_\alpha$.
\end{remark}
\noindent For $\alpha \in \mathcal{I}$, let us consider
\begin{equation} \label{29}
\mathit{\Omega}_\alpha {=}  \left\{ \omega \in \mathit{\Omega} \
\left\vert \ \|\omega \|_\alpha \ \stackrel{\rm def}{=} \ \right.
\left[\sum_{\ell} |\omega_{\ell}|^2_{L^2_\beta} w_\alpha (0,\ell)
\right]^{1/2} < \infty \right\},
\end{equation}
and endow this set with the metric
\begin{equation} \label{met}
\rho_\alpha (\omega , \omega') = \|\omega - \omega'\|_{\alpha} +
\sum_{\ell} 2^{-|\ell|} \cdot \frac{|\omega_\ell -
\omega'_\ell|_{C_\beta}}{1 +|\omega_\ell - \omega'_\ell|_{C_\beta}},
\end{equation}
which turns $\mathit{\Omega}_\alpha$ into a Polish space. Then the
set of tempered configurations is defined to be
\begin{equation} \label{30}
\mathit{\Omega}^{\rm t} = \bigcap_{\alpha \in \mathcal{I} }
\mathit{\Omega}_\alpha.
\end{equation}
Equipped with the projective limit topology $\mathit{\Omega}^{\rm
t}$ becomes a Polish space as well. For any $\alpha \in
\mathcal{I}$, we have continuous dense embeddings
$\mathit{\Omega}^{\rm t}\hookrightarrow \mathit{\Omega}_\alpha
\hookrightarrow \mathit{\Omega}$. Then by the Kuratowski theorem it
follows that $\mathit{\Omega}_\alpha, \mathit{\Omega}^{\rm t} \in
\mathcal{B}(\mathit{\Omega})$ and
 the Borel $\sigma$-algebras of all these Polish spaces
coincide with the ones induced on them by
$\mathcal{B}(\mathit{\Omega})$. Now we are at a position to complete
the definition of the function (\ref{23}).
\begin{lemma} \label{1lm}
For every $\alpha \in \mathcal{I}$ and $\Lambda \Subset \mathbb{L}$,
the map $\mathit{\Omega}_\alpha \times \mathit{\Omega}_\alpha \ni
(\omega , \xi)\mapsto I_\Lambda (\omega | \xi)$ is continuous.
Furthermore, for every ball $B_\alpha (R) = \{\omega \in
\mathit{\Omega}_\alpha \ | \ \rho_\alpha (0, \omega) < R\}$, $R>0$,
it follows that
\begin{equation} \label{31}
\inf_{\omega \in \mathit{\Omega}, \ \xi \in B_\alpha (R)} I_\Lambda
(\omega |\xi)
> - \infty, \quad  \sup_{\omega , \xi \in B_\alpha (R)}|I_\Lambda
(\omega |\xi)| < + \infty.
\end{equation}
\end{lemma}
\begin{proof}
As the functions $V_\ell: \mathbb{R}^\nu \rightarrow \mathbb{R}$ are
continuous, the map $(\omega, \xi) \mapsto I_\Lambda
(\omega_\Lambda)$ is continuous and locally bounded. Furthermore,
\begin{eqnarray} \label{32}
& & \quad  \left\vert \sum_{\ell \in \Lambda , \ \ell' \in
\Lambda^c} J_{\ell \ell'} (\omega_\ell , \xi_{\ell'})_{L^2_\beta}
\right\vert \leq \sum_{\ell \in \Lambda , \ \ell' \in \Lambda^c}
\left\vert J_{\ell \ell'}\right\vert \cdot |\omega_\ell
|_{L^2_\beta}\cdot |\xi_{\ell'} |_{L^2_\beta} \nonumber \\ & & \quad
 = \sum_{\ell \in \Lambda}|\omega_\ell |_{L^2_\beta} [w_\alpha
(0,\ell)]^{-1/2} \nonumber \\ & & \quad \times \sum_{\ell' \in
\Lambda^c} |J_{\ell \ell'}| \left[w_\alpha (0,\ell)/w_\alpha
(0,\ell')\right]^{1/2}\cdot |\xi_{\ell'} |_{L^2_\beta} [w_\alpha
(0,\ell')]^{1/2} \nonumber \\ & & \quad \leq \sum_{\ell \in
\Lambda}|\omega_\ell |_{L^2_\beta} [w_\alpha
(0,\ell)]^{-1/2}\sum_{\ell' \in \Lambda^c} |J_{\ell \ell'}|\cdot
[w_\alpha (\ell, \ell')]^{-1/2} \cdot |\xi_{\ell'} |_{L^2_\beta}
[w_\alpha (0,\ell')]^{1/2}\nonumber \\ & & \quad \quad \leq
 \hat{J}_\alpha \|\omega\|_\alpha  \|\xi\|_\alpha  \sum_{\ell \in
\Lambda}[w_\alpha (0,\ell)]^{-1},
\end{eqnarray}
where we used the triangle inequality (\ref{te}). This yields the
continuity stated and the upper bound in (\ref{31}). To prove the
lower bound we employ the super-quadratic growth of $V_\ell$ assumed
in (\ref{3}). Then for any $\varkappa>0$ and $\alpha \in
\mathcal{I}$, one finds $C>0$ such that for any $\omega \in
\mathit{\Omega}$ and $\xi\in \mathit{\Omega}^{\rm t}$,
\begin{eqnarray} \label{llb}
\qquad I_\Lambda (\omega|\xi) & \geq & B_V \beta |\Lambda| + A_V
\beta^{1-r}\sum_{\ell \in \Lambda} |\omega_\ell |_{L^2_\beta}^{2r} -
\frac{1}{2}\sum_{\ell ,\ell' \in \Lambda} J_{\ell \ell'}
(\omega_\ell , \omega_{\ell'} )_{L^2_\beta}\\& - & \sum_{\ell \in
\Lambda , \ \ell' \in \Lambda^c} J_{\ell  \ell'} (\omega_\ell ,
\xi_{\ell'} )_{L^2_\beta} \geq - C |\Lambda| + \varkappa\sum_{\ell
\in \Lambda} |\omega_\ell|^2_{L^2_\beta} \nonumber\\& - &
\hat{J}_\alpha \|\xi \|^2_\alpha \sum_{\ell \in \Lambda}w_\alpha(0,
\ell). \nonumber
\end{eqnarray}
 To get the latter estimate we used the
Minkowski inequality.
\end{proof}
\vskip.1cm \noindent Now for $\Lambda \Subset \mathbb{L}$ and $\xi
\in \mathit{\Omega}^{\rm t}$, we introduce the partition function
(c.f., (\ref{23a}))
\begin{equation} \label{33}
Z_\Lambda (\xi) = \int_{\mathit{\Omega}_\Lambda} \exp\left[-
I_\Lambda(\omega_\Lambda \times 0_{\Lambda^c} |\xi)  \right]
\chi_\Lambda ({\rm d}\omega_\Lambda).
\end{equation}
An immediate corollary of the estimates (\ref{16h}) and
(\ref{llb}) is the following
\begin{proposition} \label{2pn}
For every $\Lambda \Subset \mathbb{L}$, the function
$\mathit{\Omega}^{\rm t} \ni \xi \mapsto Z_\Lambda (\xi)\in (0, +
\infty)$ is continuous. Moreover, for any  $R>0$,
\begin{equation} \label{tania}
\inf_{\xi \in B_\alpha (R)} Z_\Lambda (\xi) > 0, \quad \sup_{\xi \in
B_\alpha (R)} Z_\Lambda (\xi)< \infty.
\end{equation}
\end{proposition}
\subsection{Local Gibbs specification}

In the DLR formalism Gibbs measures are determined by means of local
Gibbs specifications. In our context it is the family $\{\pi_\Lambda
\}_{\Lambda \Subset \mathbb{L}}$ of measure kernels
\[
\mathcal{B}(\mathit{\Omega}) \times \mathit{\Omega} \ni (B,
\xi)\mapsto \pi_\Lambda (B|\xi) \in [0, 1]
\]
which we define as follows. For $\xi \in \mathit{\Omega}^{\rm t}$,
$\Lambda\Subset \mathbb{L}$, and $B \in \mathcal{B}(\Omega)$, we set
\begin{equation} \label{34}
\pi_\Lambda (B|\xi) = \frac{1}{Z_\Lambda(\xi)}
\int_{\mathit{\Omega}_\Lambda}\exp\left[- I_\Lambda(\omega_\Lambda
\times 0_{\Lambda^c} |\xi)  \right] \mathbb{I}_B (\omega_\Lambda
\times \xi_{\Lambda^c})\chi_\Lambda ({\rm d}\omega_\Lambda ),
\end{equation}
where $\mathbb{I}_B $ stands for the indicator of $B$. We also set
\begin{equation} \label{34a}
\pi_\Lambda (\cdot|\xi) \equiv 0, \quad {\rm for} \ \ \xi \in
\mathit{\Omega} \setminus \mathit{\Omega}^{\rm t}.
\end{equation}
From these definitions one readily derives a consistency property
\begin{equation} \label{35}
\int_{\mathit{\Omega}} \pi_\Lambda (B|\omega) \pi_{\Lambda'} ({\rm
d} \omega |\xi) = \pi_{\Lambda'} (B |\xi), \quad \Lambda \subset
\Lambda',
\end{equation}
which holds for all $B\in \mathcal{B}(\mathit{\Omega})$ and $\xi \in
\mathit{\Omega}$. Furthermore, by (\ref{llb}) it follows that for
any $\xi\in \mathit{\Omega}$, $\sigma\in (0, 1/2)$, and
$\varkappa>0$,
\begin{equation} \label{llb1}
\int_{\mathit{\Omega}} \exp\left\{ \sum_{\ell \in \Lambda} \left(
\lambda_\sigma |\omega_\ell |^2_{C_\beta^\sigma} +
\varkappa|\omega_\ell|_{L^2_\beta }^2 \right) \right\} \pi_\Lambda
({\rm d} \omega|\xi) <\infty,
 \end{equation}
 where $\lambda_\sigma$ is the same as in Proposition \ref{1pn}.

By $C_{\rm b}(\mathit{\Omega}_\alpha)$ (respectively, $C_{\rm
b}(\mathit{\Omega}^{\rm t})$) we denote the Banach spaces of all
bounded continuous functions $f:\mathit{\Omega}_\alpha \rightarrow
\mathbb{R}$ (respectively, $f:\mathit{\Omega}^{\rm t} \rightarrow
\mathbb{R}$) equipped with the supremum norm. For every $\alpha \in
\mathcal{I}$, one has a natural embedding $C_{\rm
b}(\mathit{\Omega}_\alpha ) \hookrightarrow C_{\rm
b}(\mathit{\Omega}^{\rm t})$.
\begin{lemma} [Feller Property] \label{2lm}
For every $\alpha \in \mathcal{I}$, $\Lambda \Subset \mathbb{L}$,
and any $f \in C_{\rm b}(\mathit{\Omega}_{\alpha})$, the function
\begin{eqnarray} \label{f}
& & \mathit{\Omega}_\alpha \ni \xi \mapsto \pi_\Lambda (f | \xi) \\
& & \qquad \qquad  \stackrel{\rm def}{=} \
 \frac{1}{Z_\Lambda (\xi)}
\int_{\mathit{\Omega}_\Lambda}f (\omega_\Lambda \times
\xi_{\Lambda^c}) \exp\left[- I_\Lambda (\omega_\Lambda \times
0_{\Lambda^c}|\xi) \right]\chi_\Lambda ({\rm d}\omega_\Lambda),
\nonumber
\end{eqnarray}
belongs to $C_{\rm b}(\mathit{\Omega}_{\alpha})$. The linear
operator $f \mapsto \pi_\Lambda (f|\cdot)$ is a contraction on
$C_{\rm b}(\mathit{\Omega}_\alpha)$.
\end{lemma}
\begin{proof}
By Lemma \ref{1lm} and Proposition \ref{2pn} the integrand
\[
G^f_\Lambda (\omega_\Lambda|\xi) \ \stackrel{\rm def}{=} \ f
(\omega_\Lambda \times \xi_{\Lambda^c}) \exp\left[- I_\Lambda
(\omega_\Lambda \times 0_{\Lambda^c}|\xi) \right] / Z_\Lambda (\xi)
\]
is continuous in both variables. Moreover, by (\ref{31}) and
(\ref{tania}) the map
\[
\mathit{\Omega}_\alpha \ni \xi \mapsto \sup_{\omega_\Lambda \in
\mathit{\Omega}_\Lambda} |G^f_\Lambda (\omega_\Lambda|\xi)|
\]
is locally bounded. This allows us to apply Lebesgue's dominated
convergence theorem, which  yields the continuity stated. Obviously,
\begin{equation} \label{39}
\sup_{\xi \in \mathit{\Omega}_\alpha }\left\vert \pi_\Lambda (f
|\xi) \right\vert \leq \sup_{\xi \in \mathit{\Omega}_\alpha
}|f(\xi)|.
\end{equation}
\end{proof}
\noindent Note that by (\ref{34}), for $\xi \in \mathit{\Omega}^{\rm
t}$, $\alpha \in \mathcal{I}$, and $f \in C_{\rm
b}(\mathit{\Omega}_\alpha)$,
\begin{equation} \label{fp}
\pi_\Lambda (f|\xi) = \int_{\mathit{\Omega}}f(\omega)
\pi_\Lambda({\rm d}\omega|\xi).
\end{equation}
\begin{definition} \label{3df}
A measure $\mu \in \mathcal{P}(\mathit{\Omega})$ is called a
tempered Euclidean Gibbs measure at inverse temperature $\beta>0$ if
it satisfies the Dobrushin-Lanford-Ruelle (equilibrium) equation
\begin{equation} \label{40}
\int_{\mathit{\Omega}}\pi_\Lambda (B |\omega) \mu({\rm d}\omega) =
\mu(B), \quad { for} \ { all} \ \ \ \Lambda \Subset \mathbb{L} \ \ {
and} \ \ B \in \mathcal{B}(\mathit{\Omega}).
\end{equation}
\end{definition}
\noindent By $\mathcal{G}^{\rm t}$ we denote the set of all tempered
Euclidean Gibbs measures of our model. So far we do not know if
$\mathcal{G}^{\rm t}$ is non-void; if it is, its elements are
supported by $\mathit{\Omega}^{\rm t}$. Indeed, by (\ref{34}) and
(\ref{34a}) $\pi_\Lambda (\mathit{\Omega} \setminus
\mathit{\Omega}^{\rm t} |\xi) = 0$ for every $\Lambda \Subset
\mathbb{L}$ and $\xi \in \mathit{\Omega}$. Then by (\ref{40}),
\begin{equation} \label{40a}
\mu (\mathit{\Omega} \setminus \mathit{\Omega}^{\rm t})  = 0 \ \
\Longrightarrow \ \ \mu(\mathit{\Omega}^{\rm t}) =1.
\end{equation}
Furthermore,
\begin{equation}
\mu \left( \left\{ \omega \in \mathit{\Omega }^{ \mathrm{t}} \ | \
\forall \ell \in \mathbb{L}: \  \omega_\ell \in C_{\beta }^{\sigma }
\right\} \right) =1,  \label{40b}
\end{equation}%
which follows from (\ref{llb1}).

Given $\alpha  \in \mathcal{I}$, by $\mathcal{W}_\alpha$ we denote
the usual weak topology on the set of all probability measures
$\mathcal{P}(\mathit{\Omega}_\alpha)$, defined by means of bounded
continuous functions on $\mathit{\Omega}_\alpha$. By
$\mathcal{W}^{\rm t}$ we denote the weak topology on
$\mathcal{P}(\mathit{\Omega}^{\rm t})$. With these topologies the
sets $\mathcal{P}(\mathit{\Omega}_\alpha)$ and
$\mathcal{P}(\mathit{\Omega}^{\rm t})$ become Polish spaces (Theorem
6.5, page 46 of \cite{Pa}). In general, the convergence of
$\{\mu_n\}_{n \in \mathbb{N}} \subset
\mathcal{P}({\mathit{\Omega}}^{\rm t})$ in every
$\mathcal{W}_\alpha$, $\alpha \in \mathcal{I}$, does not yet imply
its $\mathcal{W}^{\rm t}$-convergence. In Lemma \ref{tanlm} and
Corollary \ref{tanco} below we show that the topologies induced on
$\mathcal{G}^{\rm t}$ by $\mathcal{W}_\alpha$ and $\mathcal{W}^{\rm
t}$ coincide.
\begin{lemma} \label{3lm}
For each $\alpha \in \mathcal{I}$, every
$\mathcal{W}_\alpha$-accumulation point $\mu \in
\mathcal{P}(\mathit{\Omega}^{\rm t})$ of the family $\{\pi_\Lambda
(\cdot |\xi) \ | \ \Lambda \Subset \mathbb{L}, \ \xi \in
\mathit{\Omega}^{\rm t}\}$ is a tempered Euclidean Gibbs measure.
\end{lemma}
\begin{proof}
For each $\alpha \in \mathcal{I}$, $C_{\rm b}
(\mathit{\Omega}_{\alpha})$ is a measure defining class for
$\mathcal{P}(\mathit{\Omega}^{\rm t})$. Then a measure $\mu \in
\mathcal{P}(\mathit{\Omega}^{\rm t})$ solves (\ref{40}) if and only
if for any $f \in C_{\rm b} (\mathit{\Omega}_{\alpha})$ and all
$\Lambda \Subset \mathbb{L}$,
\begin{equation} \label{42}
\int_{\mathit{\Omega}^{\rm t}} f (\omega)\mu ({\rm d} \omega) =
\int_{\mathit{\Omega}^{\rm t}} \pi_\Lambda (f |\omega) \mu ({\rm d}
\omega).
\end{equation}
Let $\{\pi_{\Lambda_k} (\cdot |\xi_k)\}_{k \in \mathbb{N}}$ converge
in $\mathcal{W}_{\alpha}$ to some $\mu\in
\mathcal{P}(\mathit{\Omega}^{\rm t})$. For every $\Lambda \Subset
\mathbb{L}$, one finds $k_\Lambda \in \mathbb{N}$ such that $\Lambda
\subset \Lambda_k$ for all $k > k_\Lambda$. Then by (\ref{35}), one
has
\[
\int_{\mathit{\Omega}^{\rm t}} f (\omega) \pi_{\Lambda_k }({\rm d}
\omega|\xi_k) = \int_{\mathit{\Omega}^{\rm t}} \pi_\Lambda (f
|\omega) \pi_{\Lambda_k }({\rm d} \omega|\xi_k).
\]
Now by Lemma \ref{2lm}, one can pass to the limit $k \rightarrow
+\infty$ and get (\ref{42}).
\end{proof}

\subsection{Euclidean approach and local quantum Gibbs states}

\label{ss2.4} Here we outline the basic elements of the Euclidean
approach in quantum statistical mechanics, its detailed presentation
may be found in \cite{AKKR,AKPR01c}.

For $\Lambda \Subset \mathbb{L}$, the Schr\"odinger operator
$H_\Lambda$, defined by (\ref{sch}), acts in the physical Hilbert
space $\mathcal{H}_\Lambda \ \stackrel{\rm def}{=} \
L^2(\mathbb{R}^{\nu |\Lambda|})$. In view of (\ref{tr}), one can
introduce
\begin{equation}
\mathfrak{C}_\Lambda \ni A\mapsto \varrho_{\Lambda }(A)\text{
}\overset{\mathrm{def}}{=}\text{ } \frac{\mathrm{trace}(Ae^{-\beta
H_{\Lambda }})}{\mathrm{trace}(e^{-\beta H_{\Lambda }})},
\label{8e}
\end{equation}%
which is a positive normalized  functional on the algebra
$\mathfrak{C}_\Lambda$ of all bounded linear operators (observables)
on $\mathcal{H}_\Lambda$. It is the Gibbs state of the system of
quantum oscillators located in $\Lambda$ (local Gibbs state). The
mappings
\begin{equation}
\mathfrak{C}_\Lambda \ni A\mapsto
 \mathfrak{a}_{t}^{\Lambda }(A)\text{ }\overset{\mathrm{def}}{=}%
\text{ }e^{itH_{\Lambda }}Ae^{-itH_{\Lambda }},\ \ t\in \mathbb{R},
\quad t \ - \ {\rm time}  \label{9e}
\end{equation}%
constitute the group of time automorphisms which describes the
dynamics of the system in $\Lambda$. The state $\varrho_\Lambda$
satisfies the KMS (thermal equilibrium) condition relative to the
dynamics $\mathfrak{a}_t^{\Lambda}$, see Definition 1.1 in
\cite{KL}. Multiplication operators by bounded continuous functions
act as
\[
(F\psi)(x) = F(x) \cdot \psi(x), \quad \psi \in \mathcal{H}_\Lambda,
\ \ F \in C_{\mathrm{b}}(\mathbb{R}^{\nu |\Lambda |}).
\]
One can prove, see \cite{H,KoL}, that the linear span of the
products
\begin{equation}
\mathfrak{a}_{t_{1}}^{\Lambda }(F_{1})\cdots
\mathfrak{a}_{t_{n}}^{\Lambda }(F_{n}),  \label{10e}
\end{equation}%
with all possible choices of $n\in \mathbb{N}$, $t_{1},\dots
,t_{n}\in \mathbb{R}$ and $F_{1},\dots ,F_{n}\in
C_{\mathrm{b}}(\mathbb{R}^{\nu |\Lambda |})$, is $\sigma $-weakly
dense in $\mathfrak{C}_{\Lambda }$. As a $\sigma $-weakly continuous
functional (page 65 of \cite{BrR}), the state (\ref{8e}) is fully
determined by its values on (\ref{10e}), i.e., by \emph{the Green
functions}
\begin{equation}
G_{F_{1},\dots ,F_{n}}^{\Lambda }(t_{1},\dots ,t_{n})\text{ }\overset{%
\mathrm{def}}{=}\text{ }\varrho _{\Lambda }\left[ \mathfrak{a}%
_{t_{1}}^{\Lambda }(F_{1})\cdots \mathfrak{a}_{t_{n}}^{\Lambda }(F_{n})%
\right].   \label{11e}
\end{equation}%
As was shown in \cite{AHK,AKKR,KL}, the Green functions can be
considered as restrictions of functions $G_{F_{1},\dots
,F_{n}}^{\Lambda }(z_{1},\dots ,z_{n})$ analytic in the domain
\begin{equation}
\mathcal{D}_{\beta }^{n}=\{(z_1 , \dots z_n) \in \mathbb{C}^n \ | \
0 < \Im (z_1) < \Im (z_2) < \cdots < \Im (z_n ) < \beta\},
\label{11z}
\end{equation}%
and continuous on its closure $\mathcal{\bar{D}}_{\beta }^{n}\subset
\mathbb{C}^{n}$. The `imaginary time' subset
\[
\{(z_1 , \dots , z_n)\in \mathcal{D}_\beta^n \ | \ \Re(z_1) = \cdots
= \Re(z_n) = 0\}
\]
is an inner set of uniqueness for functions analytic in
$\mathcal{D}_\beta^n$ (see pages 101 and 352 of \cite{Sch}).
Therefore, the Green functions (\ref{11e}), and hence the states
(\ref{8e}), are completely determined by {\it the Matsubara
functions}
\begin{eqnarray} \label{mats}
& & \quad  \mathit{\Gamma}^\Lambda_{F_1 , \dots , F_n}(\tau_1 ,
\dots , \tau_n) \ \stackrel{\rm def}{=} \ G^\Lambda_{F_1 , \dots ,
F_n}(\imath \tau_1 , \dots ,\imath \tau_n) \\
& & \qquad  = {\rm trace}[F_{1}e^{-(\tau _{2}-\tau _{1})H_{\Lambda
}}F_{2}e^{-(\tau _{3}-\tau _{2})H_{\Lambda }}\cdots F_{n}e^{-(
 \tau _{n+1}-\tau _{n})H_{\Lambda }}]/\mathrm{trace}[e^{-\beta
H_{\Lambda }}] \label{12e} \nonumber
\end{eqnarray}
taken at `temperature ordered' arguments
$0 \leq \tau _{1}\leq \cdots \leq \tau _{n}\leq \tau _{1}+\beta \overset{\mathrm{def}%
}{=}\tau _{n+1}$, with all possible choices of $n \in \mathbb{N}$
and $F_1 , \dots , F_n \in C_{\rm b}(\mathbb{R}^{\nu|\Lambda|})$.
Their extensions to the whole $[0, \beta]^n$ are defined as
\[
\mathit{\Gamma}^\Lambda_{F_1 , \dots , F_n}(\tau_1 , \dots , \tau_n)
= \mathit{\Gamma}^\Lambda_{F_{\sigma(1)} , \dots ,
F_{\sigma(n)}}(\tau_{\sigma(1)} , \dots , \tau_{\sigma(n)}),
\]
where $\sigma$ is the permutation of $\{1, 2, \dots , n\}$ such that
$\tau_{\sigma(1)}\leq \tau_{\sigma(2)} \leq \cdots \leq
\tau_{\sigma(n)}$.  One can show that for every $\theta\in [0,
\beta]$,
\begin{equation} \label{mats1}
\mathit{\Gamma}^\Lambda_{F_1 , \dots , F_n}(\tau_1 + \theta, \dots ,
\tau_n + \theta) = \mathit{\Gamma}^\Lambda_{F_1 , \dots ,
F_n}(\tau_1 , \dots , \tau_n),
\end{equation}
where addition is modulo $\beta$. This periodicity along with the
analyticity of the Green functions is equivalent to the KMS property
of the state (\ref{9e}).

The central element of the Euclidean approach which links the local
Gibbs states (\ref{8e}) and the local Gibbs measures  (\ref{mul}) is
the representation (c.f., (\ref{ea}))
\begin{equation}\label{mul10}
\mathit{\Gamma }_{F_{1},\dots ,F_{n}}^{\Lambda }(\tau _{1},\dots
,\tau _{n})=\int_{\mathit{\Omega }_{\Lambda }}F_{1}(\omega _{\Lambda
}(\tau
_{1}))\ldots F_{n}(\omega _{\Lambda }(\tau _{n}))\mu _{\Lambda }(\mathrm{d}%
\omega _{\Lambda }).
\end{equation}
The Gibbs state (\ref{8e}) corresponds to a finite $\Lambda \Subset
\mathbb{L}$. Thermodynamic properties of the underlying physical
model are described by the Gibbs states corresponding to the whole
lattice $\mathbb{L}$. Such states should be defined on the
$C^*$-algebra of quasi-local observables $\mathfrak{C}$, being the
norm-completion of the algebra of local observables $\cup_{\Lambda
\Subset \mathbb{L}} \mathfrak{C}_\Lambda$. Here each
$\mathfrak{C}_\Lambda$ is considered, modulo embedding, as a
subalgebra of $\mathfrak{C}_{\Lambda'}$ for any $\Lambda'$
containing $\Lambda$. The dynamics of the whole system is to be
defined by the limits, as $\Lambda \nearrow \mathbb{L}$, of the time
automorphisms (\ref{9e}), which would allow one to define the Gibbs
states on $\mathfrak{C}$ as KMS states. This `algebraic' way can be
realized for models described by bounded local Hamiltonians
$H_\Lambda$, e.g., quantum spin models, for which the limiting time
automorphisms exist, see section 6.2 of \cite{BrR}. For the model
considered here, such automorphisms do not exist and hence there is
no canonical way to define Gibbs states of the whole infinite
system. Therefore, the Euclidean approach based on the one-to-one
correspondence between the local states and measures arising from
the representation (\ref{mul10}) seems to be the only way of
developing a mathematical theory of the equilibrium thermodynamic
properties of such models. For certain model of quantum crystals,
the limiting states $\lim_{\Lambda \nearrow
\mathbb{L}}\varrho_\Lambda$ were constructed by means of path
measures, see \cite{Amour,MPZ,MVZ00}. The set of all Euclidean Gibbs
measures $\mathcal{G}^{\rm t}$ we study in this article certainly
includes all the limiting points of this type. Furthermore, there
exist axiomatic methods, see \cite{Birke,Gelerak2}, analogous to the
Osterwalder-Schrader reconstruction theory \cite{GJ,Si74}, by means
of which KMS states are constructed on certain von Neumann algebras
from a complete set of Matsubara functions. In our case such a set
constitute the functions
\begin{equation} \label{kms}
\mathit{\Gamma}^{\mu}_{F_1 , \dots , F_n} (\tau_1 , \dots , \tau_n)
= \int_{\mathit{\Omega}} F_1 (\omega (\tau_1)) \cdots
F_n(\omega(\tau_n))\mu({\rm d}\omega), \quad \mu\in \mathcal{G}^{\rm
t},
\end{equation}
defined for all  bounded local multiplication operators $F_1, \dots
, F_n$. Therefore, the theory of $\mathcal{G}^{\rm t}$ developed in
the article may be used to constructing such algebras and states,
which we leave as an important task for the future.

\section{The Results}
\label{3s}

In the first subsection below we present the statements describing
the general case, whereas the second subsection is dedicated to the
case of $\nu=1$ and $J_{\ell \ell'}\geq 0$.

\subsection{Euclidean Gibbs measures in the general case}\label{ss3.1}

We begin by establishing existence of tempered Euclidean Gibbs
measures and compactness of their set $\mathcal{G}^{\rm t}$. For
models with non-compact spins, here they are even
infinite-dimensional, such a property is far from being evident.
\begin{theorem} \label{1tm}
For every $\beta>0$, the set of tempered Euclidean Gibbs measures
$\mathcal{G}^{\rm t}$ is non-void and $\mathcal{W}^{\rm t}$-
compact.
\end{theorem}
The next theorem gives an exponential moment estimate similar to
(\ref{16h}). Recall that the H\"{o}lder norm $|\cdot |_{C_{\beta
}^{\sigma }}$ was defined by (\ref{16z}).
\begin{theorem} \label{2tm}
For every $\sigma \in (0, 1/2)$ and $\varkappa >0$, there exists  a
positive constant $ C_{\ref{43}}$ such that, for any $\ell $ and for
all $\mu \in \mathcal{G}^{\rm t}$,
\begin{equation} \label{43}
\int_{\mathit{\Omega}} \exp\left(\lambda_\sigma |\omega_\ell
|_{C^\sigma_\beta}^2 + \varkappa |\omega_\ell |_{L^2_\beta}^2
\right)\mu({\rm d}\omega) \leq C_{\ref{43}},
\end{equation}
where $\lambda_\sigma$ is the same as in (\ref{16h}).
\end{theorem}
According to (\ref{43}), the one-site projections of each $\mu\in
\mathcal{G}^{\rm t}$ are sub-Gaussian. The bound $C_{\ref{43}}$ does
not depend on $\ell$ and is the same for all $\mu \in
\mathcal{G}^{\rm t}$, though it may depend on $\sigma $ and
$\varkappa$. Th estimate (\ref{43}) plays a crucial role in the
theory of the set $\mathcal{G}^{\rm t}$. Such estimates are also
important in the study of the Dirichlet operators $H_\mu$ associated
with the measures $\mu\in \mathcal{G}^{\rm t}$, see \cite{AKR,AKR1}.

 The set of tempered configurations
$\mathit{\Omega}^{\rm t}$ was introduced in (\ref{29}), (\ref{30})
by means of rather slack restrictions (c.f., (\ref{24})) imposed on
the $L^2_\beta$-norms of $\omega_\ell$. The elements of
$\mathcal{G}^{\rm t}$ are supported by this set, see (\ref{40a}). It
turns out that they have a much smaller support (a kind of the
Lebowitz-Presutti one). Given $b>0$ and $\sigma\in (0, 1/2)$, we
define
\begin{eqnarray} \label{47}
\mathit{\Xi} ( b , \sigma) &  = & \{ \xi \in \mathit{\Omega} \ |  \
(\forall \ell_0 \in \mathbb{L}) \ (\exists \Lambda_{\xi ,
\ell_0}\Subset
\mathbb{L}) \ (\forall \ell \in \Lambda_{\xi , \ell_0}^c): \\
& & \qquad \qquad \qquad \qquad  \qquad \qquad |\xi_\ell
|_{C^\sigma_\beta}^2 \leq b \log( 1 + |\ell - \ell_0|)\}, \nonumber
\end{eqnarray}
which in view of (\ref{25}) is a Borel subset of
$\mathit{\Omega}^{\rm t}$.
\begin{theorem}\label{3tm}
For every $\sigma \in (0, 1/2)$, there exists $b>0$, which depends
on $\sigma$ and on the parameters of the model only, such that for
all $\mu \in \mathcal{G}^{\rm t}$,
\begin{equation} \label{48}
\mu (\mathit{\Xi}(b , \sigma)) = 1.
\end{equation}
\end{theorem}
The last result in this group is a sufficient condition for
$\mathcal{G}^{\rm t}$ to be a singleton, which holds for high
temperatures (small $\beta$). It is obtained by controlling the
`non-convexity' of the potential energy (\ref{pe}). Let us decompose
\begin{equation}\label{decom}
V_{\ell} = V_{1, \ell} + V_{2, \ell},
\end{equation}
where $V_{1, \ell}\in C^2 (\mathbb{R}^\nu)$ is such that
\begin{equation} \label{dc1}
- a \leq b \ \stackrel{\rm def}{=} \ \inf_{\ell} \inf_{x, y \in
\mathbb{R}^\nu, \ y\neq 0}\left( V''_{1,\ell}(x)y, y \right)/|y|^2 <
\infty.
\end{equation}
As for the second term, we set
\begin{equation} \label{dc2}
0 \leq \delta \ \stackrel{\rm def}{=} \ \sup_{\ell} \left\{ \sup_{x
\in \mathbb{R}^\nu}V_{2, \ell}(x) - \inf_{x \in \mathbb{R}^\nu}V_{2,
\ell}(x) \right\} \leq \infty.
\end{equation}
Its role is to produce multiple minima of the potential energy
responsible for eventual phase transitions. Clearly, the
decomposition (\ref{decom}) is not unique; its optimal realizations
for certain types of $V_\ell$ are discussed in section 6 of
\cite{AKRT}.
\begin{theorem} \label{httm}
The set $\mathcal{G}^{\rm t}$ is a singleton if
\begin{equation}
\label{dc3} e^{\beta \delta} <(a + b)/\hat{J}_0.
\end{equation}
\end{theorem}
One observes that the latter condition surely holds at all $\beta$
if
\begin{equation} \label{si}
\delta = 0 \quad {\rm and} \quad \hat{J}_0 < a + b.
\end{equation}
In this case the potential energy $W_\Lambda$ given by (\ref{pe}) is
convex. The conditions (\ref{si}) and (\ref{dc3}) do not contain the
particle mass $m$; hence, the property stated holds also in the
quasi-classical limit\footnote{For details on this limit see
\cite{AKKR}.} $m \rightarrow + \infty$.

\subsection{Scalar ferromagnetic models}

Here we study in more detail the case $\nu=1$ and $J_{\ell \ell'}
\geq 0$, that is tacitly assumed in this subsection.
 Recall that the components $\omega_\ell$ of
$\omega\in \mathit{\Omega}$ are continuous functions on $S_\beta
\cong [0, \beta]$. For $\omega, \tilde{\omega} \in \mathit{\Omega}$,
we set $\omega \leq \tilde{\omega}$ if for all $\ell $ and $\tau \in
[0,\beta]$, one has $\omega_\ell (\tau) \leq \tilde{\omega}_\ell
(\tau)$. This allows one to define an order on
$\mathcal{P}(\mathit{\Omega})$, called stochastic domination or FKG
order, see \cite{Pr}. A function $f : \mathit{\Omega} \rightarrow
\mathbb{R}$ is called increasing if $\omega \leq \tilde{\omega}$
implies $f(\omega) \leq f(\tilde{\omega})$. Clearly, increasing
functions  $f\in C_{\rm b}(\mathit{\Omega}^{\rm t})$ constitute  a
measure determining class. For $\mu \in
\mathcal{P}(\mathit{\Omega}^{\rm t})$ and $f \in C_{\rm
b}(\mathit{\Omega}^{\rm t})$, we write (c.f., (\ref{f}))
\[
\mu (f) = \int_{\mathit{\Omega}^{\rm t}} f (\omega) \mu({\rm
d}\omega).
\]
Then for $\mu_1 , \mu_2 \in \mathcal{P}(\mathit{\Omega}^{\rm t})$,
we set
\begin{equation} \label{w1}
  \mu_1  \leq \mu_2 \ \  \ {\rm if} \ \ \
  \mu_1 (f) \leq \mu_2 (f), \quad \ {\rm for} \ \ {\rm all} \ \ {\rm increasing}
   \quad  f\in C_{\rm b}(\mathit{\Omega}^{\rm t}).
\end{equation}
A measure $\mu \in \mathcal{G}^{\rm t}$ is called {\em shift
invariant} if its Matsubara functions (\ref{kms}) have the property
(\ref{mats1}).
\begin{theorem} \label{4tm}
The set $\mathcal{G}^{\rm t}$  has unique maximal $\mu_{+}$ and
minimal $\mu_{-}$ elements in the sense of the order (\ref{w1}).
These elements are extreme and shift invariant; they are also
translation invariant if the model is so. If  $V_\ell (x ) = V_\ell
(-x)$ for all $\ell $, then $\mu_{+} (B) = \mu_{-} (- B)$ for all $B
\in \mathcal{B}(\mathit{\Omega})$.
\end{theorem}
\noindent Phase transitions correspond to $\mathcal{G}^{\rm t}$
possessing more than one element. In the underlying physical systems
phase transitions manifest themselves in the macroscopic
displacements of particles from their equilibrium positions (a
long-range order).
 For translation invariant ferromagnetic
models with $\nu =1$ and $V_\ell = V$ obeying certain conditions,
the appearance of the long-range order at low temperatures was
proven in \cite{BK1,DLP,Helf,Kond,Pastur}. Thus, one can expect that
also in our case $|\mathcal{G}^{\rm t}| > 1$ at big $\beta$. We
prove this under certain conditions imposed on $d$, $J_{\ell\ell'}$
and $V_\ell$. First we suppose that the interaction between the
nearest neighbors is uniformly nonzero
\begin{equation} \label{p1}
\inf_{|\ell -\ell'|=1} J_{\ell \ell'} \ \stackrel{\rm def}{=} \ J
>0.
\end{equation}
Next we suppose that $V_\ell$ are even continuous functions and
the upper bound in (\ref{3}) can be chosen as
\begin{equation} \label{ub}
V(x_\ell ) = \sum_{s=1}^r b^{(s)} x_\ell^{2s}; \quad 2 b^{(1)} < -
a ; \ \ \ b^{(s)}\geq 0, \  s\geq 2,
\end{equation}
where $a$ is the same as in (\ref{16e}) or in (\ref{pe}), and $r\geq
2$ is either a positive integer or infinite. For $r=+\infty$, we
assume that the series
\begin{equation} \label{p3}
\mathit{\Phi}(t) = \sum_{s=2}^{+\infty}
\frac{(2s)!}{2^{s-1}(s-1)!}{b}^{(s)} t^{s-1},
\end{equation} converges at some $t>0$.
Since $2b^{(1)} + a
<0$, the equation
\begin{equation} \label{p4}
a + 2b^{(1)} + \mathit{\Phi}(t) = 0,
\end{equation} has a unique solution $t_* >0$.
Finally, we suppose that for every $\ell$,
\begin{equation} \label{ub1}
V(x_\ell ) - V_\ell (x_\ell) \leq V(\tilde{x}_\ell )  - V_\ell
(\tilde{x}_\ell) , \quad {\rm whenever} \  \ x_\ell^2 \leq
\tilde{x}_\ell^2.
\end{equation}
By these assumptions all $V_\ell$ are `uniformly double-welled'. If
$V_\ell (x_\ell)= v_\ell (x_\ell^2)$ and $v_\ell$ are
differentiable, the condition (\ref{ub1}) may be formulated as an
upper bound for $v_\ell'$. Recall that $\mathbb{L} = \mathbb{Z}^d$;
for $d\geq 2$, we set
\begin{equation} \label{cj}
\theta_d = \frac{1}{(2\pi)^d}\int_{(- \pi, \pi]^d}\frac{{\rm
d}p}{\sqrt{E(p)}}, \quad E(p) = \sum_{j=1}^d [1-\cos p_j].
\end{equation}
\begin{theorem} \label{phtm}
Let $d\geq3$ and the above assumptions hold. Then under the
condition
\begin{equation} \label{cj1}
J > \theta_d^2 / 8 m t_*^2,
\end{equation}
there exists $\beta_*>0$ such that $|\mathcal{G}^{\rm t}|
>1$ whenever $\beta> \beta_*$. The bound $\beta_*$ is the unique
solution of the equation
\begin{equation} \label{cja}
t_* = \frac{1}{(2\pi)^d}\int_{(- \pi, \pi]^d}\frac{1}{\sqrt{8 m J
E(p)}}\coth\left( \frac{\beta^2 J E(p)}{2m}  \right)^{1/2}{\rm d}p.
\end{equation}
\end{theorem}
As was shown in \cite{AKK,AKKRprl,Koz}, strong quantum effects,
corresponding in particular to small values of the particle mass
$m$, can suppress abnormal fluctuations and hence phase transitions.
Therefore, one can expect that they can yield $|\mathcal{G}^{\rm
t}|=1$. In the probabilistic interpretation our model describes a
system of interacting diffusion processes, in which strong quantum
effects correspond to large diffusion intensity. The most general
result in this domain -- the uniqueness at all $\beta$ due to strong
quantum effects -- was proven in \cite{AKKR02b}. In the present
paper we essentially extend the class of self-interaction potentials
for which this result holds as well as make precise the bounds of
the uniqueness regime. Furthermore, unlike to the mentioned papers
we do not suppose that the interaction has finite range and the
model is translation invariant. In Theorem \ref{5tm} below we assume
that the potentials $V_\ell$ are even continuous functions
possessing the following property. There exists a convex function $v
:[0, +\infty) \rightarrow \mathbb{R}$ such that (c.f., (\ref{ub1}))
\begin{equation} \label{49}
V_\ell (x_\ell ) -  v (x_\ell^2) \leq V_\ell (\tilde{x}_\ell ) - v
(\tilde{x}_\ell^2) \quad {\rm whenever} \ \ x_\ell^2 <
\tilde{x}_\ell^2 .
\end{equation}
In typical cases of $V_\ell$, like (\ref{4}), as such a $v$ one can
take a convex polynomial of degree $r\geq 2$.
 Now let us introduce
 the following one-particle Schr\"odinger operator (c.f., (\ref{16e}), (\ref{sch}))
\begin{equation} \label{2}
\tilde{H}_\ell = - \frac{1}{2m}\left(\frac{\partial }{\partial
x_\ell }\right)^2 + \frac{a}{2} x_\ell^2 + v (x_\ell^2).
\end{equation}
It
 has purely discrete non-degenerate spectrum
$\{E_{n}\}_{n \in \mathbb{N}_0}$. Thus, one can define the parameter
\begin{equation} \label{51}
\mathit{\Delta} = \min_{n \in \mathbb{N}}\left(E_{n} - E_{n-1}
\right),
\end{equation}
which depends on $m$, $a$, and on the choice of $v$. Recall, that
$\hat{J}_0$ was defined by (\ref{6}).
\begin{theorem} \label{5tm}
Let the above assumptions regarding the potentials $V_\ell$ hold.
Then  the set of tempered Euclidean Gibbs measures  is a singleton
if
\begin{equation} \label{52}
m \mathit{\Delta}^2 > \hat{J}_0.
\end{equation}
\end{theorem}
\noindent Note that the above result holds for all $\beta>0$. Thus,
(\ref{52}) is a stability condition like (\ref{si}), where the
parameter $m \mathit{\Delta}^2$ appears as the oscillator rigidity
(or diffusion intensity). If it holds, a
stability-due-to-quantum-effects occurs, see
\cite{AKKRprl,KK,Koz,Ko}. If $v$ is a polynomial of degree $r \geq
2$, the rigidity $m \mathit{\Delta}^2$ is a continuous function of
the particle mass $m$; it gets small in the quasi-classical limit $m
\rightarrow +\infty$, see \cite{Ko}. At the same time, for $m
\rightarrow 0+$, one has $m \mathit{\Delta}^2 =
O(m^{-(r-1)/(r+1)})$, see \cite{AKK,Ko}. Hence, (\ref{52}) certainly
holds in the small mass limit, c.f., \cite{AKKR01,AKKR02b}. To
compare the latter result with Theorem \ref{phtm} let us assume that
$J_{\ell \ell'} = J$ iff $|\ell - \ell'|=1$ and all $V_\ell$
coincide with the function given by (\ref{ub}). Then the parameter
(\ref{51}) obeys the estimate $\mathit{\Delta} < 1 /2 m t_*$, see
\cite{Ko}, where $t_*$ is the same as in (\ref{cj1}), (\ref{cja}).
In this case the condition (\ref{52}) can be rewritten as
\begin{equation} \label{cj2}
J < 1 / 8 d m t_*^2.
\end{equation}
One can show that $\theta_d > 1/d$ and $d \theta_d^2 \rightarrow 1 $
as $d \rightarrow + \infty$, which indicates that the estimates
(\ref{cj1}) and (\ref{cj2}) become precise for sufficiently large
dimensions.

 Now we consider a translation invariant version of our model
and impose further conditions on the self-interaction potential. Set
\begin{equation} \label{m1}
\mathcal{F}_{\rm Laguerre} = \left\{ \varphi: \mathbb{R} \rightarrow
\mathbb{R} \ \left\vert \ \varphi(t) = \varphi_0 \exp( \gamma_0 t)
t^n \prod_{i=1}^{\infty} (1+ \gamma_i t) \right. \right\},
\end{equation}
where $\varphi_0 >0$, $n\in \mathbb{N}_0$, $\gamma_i \geq 0$ for all
$i \in \mathbb{N}_0$, and $\sum_{i=1}^\infty \gamma_i < \infty$.
Each $\varphi \in \mathcal{F}_{\rm Laguerre}$ can be extended to an
entire function $\varphi:\mathbb{C}\rightarrow \mathbb{C}$, which
has no zeros outside of $(-\infty, 0]$. These are Laguerre entire
functions, see \cite{Iliev,Koz1,KW}. In the next theorem the
parameter $a$ is the same as in (\ref{16e}).
\begin{theorem} \label{6tm}
Let the model we consider be translation invariant with
 the self-interaction potentials $V$ being of the form
\begin{equation} \label{m2}
V (x) = v (x^2) - h x, \quad h \in \mathbb{R},
\end{equation}
with $v(0) = 0$ and such that for a certain $b\geq a/2$, the
derivative $v'$ obeys the condition $b + v' \in \mathcal{F}_{\rm
Laguerre}$. Then the set $\mathcal{G}^{\rm t}$ is a singleton if $h
\neq 0$.
\end{theorem}

\subsection{Comments} \label{ss3.2} Here we comment the theorems and compare them
with the corresponding results known for
similar models. \vskip.1cm
\begin{itemize}
\item {\bf Theorem \ref{1tm}.} A classical tool for proving existence of Gibbs measures is
the celebrated Dobrushin criterion, Theorem 1 in \cite{Do}. To apply
it in our case one should find a compact positive function $h$
defined on the single-spin space ${C}_\beta$ such that for all
$\ell$ and $\xi \in \mathit{\Omega}$,
\begin{equation} \label{d1}
\int_{\mathit{\Omega}} h(\omega_\ell) \pi_\ell ({\rm d}\omega|\xi)
\leq A  + \sum_{\ell'} I_{\ell \ell'} h(\xi_{\ell'}),
\end{equation}
where
\[
 A >0 ; \ \quad I_{\ell \ell'} \geq 0 \quad  {\rm for} \ \ {\rm all} \ \ \ell ,
 \ell'
  , \ \quad {\rm and} \quad \  \sup_{\ell} \sum_{\ell'} I_{\ell \ell'}
< 1.
\]
 Here and in the sequel to simplify notations we
denote $\pi_{\{\ell\}}$ by $\pi_\ell$ . Then the estimate (\ref{d1})
would yield that for any $\xi \in \mathit{\Omega}$, such that
$\sup_{\ell} h(\xi_\ell ) < \infty$, the family
$\{\pi_\Lambda(\cdot|\xi)\}_{\Lambda \Subset \mathbb{L}}$ is
relatively compact in the weak topology on
$\mathcal{P}(\mathit{\Omega})$ (but not yet in $\mathcal{W}_\alpha$,
$\mathcal{W}^{\rm t}$). Next one would have to show that any
accumulation point of $\{\pi_\Lambda(\cdot|\xi)\}_{\Lambda \Subset
\mathbb{L}}$ is a Gibbs measure, which is much stronger than the
fact established by our Lemma \ref{3lm}. Such a scheme was used in
\cite{BH,COPP,Sin} where the existence of Gibbs measures for lattice
systems with the single-spin space $\mathbb{R}$ was proven. Those
proofs heavily utilized the specific properties of the models, e.g.,
attractiveness and translation invariance. The direct extension of
this scheme to quantum models seems to be impossible. The scheme we
employ to proving Theorem \ref{1tm} is based on compactness
arguments in the topologies $\mathcal{W}_\alpha$, $\mathcal{W}^{\rm
t}$. After obvious modifications it can be applied to models with
much more general types of inter-particle interaction potentials.
Additional comments on this matter follow Corollary \ref{1co}.

\item {\bf Theorem \ref{2tm}} gives a uniform
integrability estimate for tempered Euclidean Gibbs measures in
terms of model parameters, which in principle can be proven before
establishing the existence. For systems of classical unbounded
spins, the problem of deriving such estimates was first posed in
\cite{BH} (see the discussion following Corollary \ref{1co}). For
quantum anharmonic systems, similar estimates were obtained in the
so called analytic approach, which is an equivalent alternative to
the approach based on the DLR equation, see
\cite{AKPR01b,AKPR01c,AKPR04,AKRT2}. In this analytic approach
$\mathcal{G}^{\rm t}$ is defined as the set of probability measures
satisfying an integration-by-parts formula, determined by the model.
This gives additional tools for studying $\mathcal{G}^{\rm t}$ and
provides  a background for the stochastic dynamics method in which
the Gibbs measures are treated as invariant distributions for
certain infinite-dimensional stochastic evolution equations, see
\cite{AKRT3}. In both analytic and stochastic dynamics methods one
imposes a number of technical conditions on the interaction
potentials and uses advanced tools of stochastic analysis. The
method we employ to proving Theorem \ref{2tm} is much more
elementary. At the same time, Theorem \ref{2tm} gives an improvement
of the corresponding results of \cite{AKPR01c,AKPR04} because: (a)
the estimate (\ref{43}) gives a much stronger bound; (b) we do not
suppose that the functions $V_\ell$ are differentiable -- an
important assumption of the analytic approach.

\item {\bf Theorem \ref{3tm}.} As might be clear from the proof of
this theorem, every $\mu\in \mathcal{P}(\mathit{\Omega}^{\rm t})$
obeying the estimate (\ref{43}) possesses the support property
(\ref{48}). For Gibbs measures of classical lattice systems of
unbounded spins, a similar property  was first established in
\cite{LP}; hence, one can call $\mathit{\Xi} (b , \sigma)$ a
Lebowitz-Presutti type support. This result of \cite{LP} was
obtained by means of Ruelle's superstability estimates \cite{Ru1},
applicable to translation invariant models only. The generalization
to translation invariant quantum model was done in \cite{PY}, where
superstable Gibbs measures were specified by the following support
property
\[
\sup_{N\in \mathbb{N}} \left\{ (1 + 2N)^{-d}\sum_{\ell: |\ell|\leq
N} |\omega_\ell|^2_{L^2_\beta} \right\} \leq C(\omega) , \quad
\mu-{\rm a.s.}.
\]
Here we note that by the Birkhoff-Khinchine ergodic theorem, for any
translation invariant measure $\mu\in
\mathcal{P}(\mathit{\Omega}^{\rm t})$ obeying (\ref{43}), it follows
that for every $\sigma \in (0, 1/2)$, $\varkappa>0$, and
$\mu$-almost all $\omega$,
\[
\sup_{N\in \mathbb{N}} \left\{ (1 + 2N)^{-d}\sum_{\ell: |\ell|\leq
N} \exp\left(\lambda_\sigma |\omega_\ell|^2_{C_\beta^{\sigma}} +
\varkappa |\omega_\ell|^2_{L^2_\beta}\right)\right\} \leq C (\sigma
,\varkappa, \omega).
\]
In particular, every periodic Euclidean Gibbs measure constructed in
subsection \ref{sss5.2} below has the above property.

\item {\bf Theorem \ref{httm}} establishes a
sufficient uniqueness condition, holding at high-temperatures (small
$\beta$). Here we follow the papers \cite{AKRT,AKRT1}, where a
similar uniqueness statement was proven for translation invariant
ferromagnetic scalar version of our model. This was done by means of
another renown Dobrushin result, Theorem 4 in \cite{Do}, which gives
a sufficient condition for the uniqueness of Gibbs measures. The
main tool used in \cite{AKRT,AKRT1} to estimating the elements of
the Dobrushin matrix was the logarithmic Sobolev inequality for the
kernels $\pi_\ell$.

\item {\bf Theorem \ref{4tm}} is an extension of the
corresponding statement proven in \cite{BH} for classical lattice
models. The extreme elements mentioned in Theorem \ref{4tm} play an
important role in proving Theorems \ref{phtm}, \ref{5tm}, \ref{6tm}.

\item {\bf Theorem \ref{phtm}.} For translation invariant lattice
models, phase transitions are established by showing the existence
of nonergodic (with respect to the group of lattice translations)
Gibbs measures. This mainly was being done by means of the infrared
estimates, see \cite{BK1,DLP,Helf,Kond,Pastur}. Here we use a
version of the technique developed in those papers and the
corresponding correlation inequalities which allow us to compare the
model considered with its translation invariant version (reference
model).

\item {\bf Theorem \ref{5tm}.} For translation invariant models
with finite range interactions and with the anharmonic
self-interaction potential possessing special properties, the
uniqueness by strong quantum effects was proven in \cite{AKKR02b}
(see also \cite{AKKR01}). With the help of the extreme elements
$\mu_{\pm}\in \mathcal{G}^{\rm t}$ we essentially extend the results
of those papers. As in the case of Theorem \ref{phtm} we employ
correlation inequalities to compare the model with a proper
reference model.

\item {\bf Theorem \ref{6tm}.} For classical lattice models, the
uniqueness at nonzero $h$  was proven in \cite{BH,LML, LP} under the
condition that the potential (\ref{m2}) possesses the property which
we establish below in Definition \ref{lydf}. The novelty of Theorem
\ref{6tm} is that it describes a quantum model and gives an explicit
sufficient condition for $V$ to possess such a
property\footnote{Examples follow Proposition \ref{lypn}.}. This
theorem is valid also in the quasi-classical limit $m \rightarrow
+\infty$, in which it covers all the cases considered in
\cite{BH,LML, LP}. For $(\phi^4)_2$ Euclidean quantum fields, a
similar statement was proven in \cite{Gelerak}.
\end{itemize}

\section{Properties of the Local Gibbs Specification}
\label{3as} Here we develop our main tools  based on the
properties of the kernels (\ref{34}).

\subsection{Moment estimates}
\label{ss4.1} Moment estimates for the kernels (\ref{34}) allow one
to prove the $\mathcal{W}^{\rm t}$-relative compactness of the set
$\{\pi_\Lambda(\cdot|\xi)\}_{\Lambda \Subset \mathbb{L}}$, which by
Lemma \ref{3lm} guaranties that $\mathcal{G}^{\rm t} \neq
\emptyset$. Integrating them over $\xi\in\mathit{\Omega}^{\rm t}$ we
get by the DLR equation (\ref{40}) the corresponding estimates also
for the elements of $\mathcal{G}^{\rm t}$. Recall that $\pi_\ell$
stands for $\pi_{\{\ell\}}$.
\begin{lemma} \label{4lm}
For any $\varkappa$, $\vartheta >0$, and $\sigma \in (0, 1/2)$,
there exists $C_{\ref{53}}
>0$ such that for all $\ell \in \mathbb{L}$ and $\xi \in
\mathit{\Omega}^{\rm t}$,
\begin{equation} \label{53}
\int_{\mathit{\Omega}}\exp\left\{\lambda_\sigma |\omega_\ell
|^2_{C^\sigma_\beta} +  \varkappa |\omega_\ell |^2_{L^2_\beta}
\right\}\pi_\ell ({\rm d}\omega|\xi) \leq \exp\left\{ C_{\ref{53}} +
\vartheta \sum_{\ell'}
|J_{\ell\ell'}|\cdot|\xi_{\ell'}|_{L^2_\beta}^2\right\}.
\end{equation}
Here $\lambda_\sigma>0$ is the same as in (\ref{43}).
\end{lemma}
\begin{proof}
Note that by (\ref{llb1}) the left-hand side is finite and the
second term in $\exp \{ \cdot\}$ on the right-hand side is also
finite since $\xi\in \mathit{\Omega}^{\rm t}$. For any $\vartheta
>0$, one has (see (\ref{6}))
\begin{equation} \label{53z}
\left\vert \sum_{\ell'}J_{\ell \ell'} (\omega_\ell ,
\xi_{\ell'})_{L^2_\beta}\right\vert \leq \frac{ \hat{J}_0}{ 2
\vartheta}|\omega_\ell |^2_{L^2_\beta} + \frac{ \vartheta}{2}
\sum_{\ell'} |J_{\ell \ell'}|\cdot |\xi_{\ell'}|^2_{L^2_\beta},
\end{equation}
which holds for all $\omega, \xi \in \mathit{\Omega}^{\rm t}$.
 By these estimates and (\ref{19}), (\ref{23}), (\ref{33}),
 (\ref{34}) we get
\begin{eqnarray} \label{54}
& & {\rm LHS}(\ref{53}) \leq [1/ Y_\ell( \vartheta)]\cdot
\exp\left\{ \vartheta \sum_{\ell'}
|J_{\ell\ell'}|\cdot |\xi_{\ell'}|_{L^2_\beta}^2\right\} \\
& & \times  \int_{\mathit{\Omega}}\exp\left\{ \lambda_\sigma
|\omega_\ell |^2_{C^\sigma_\beta} + \left( \varkappa  + \hat{J}_0 /
2 \vartheta \right)|\omega_\ell |^2_{L^2_\beta} - \int_0^\beta
V_\ell (\omega_\ell (\tau)){\rm d}\tau \right\}\chi ({\rm
d}\omega_\ell), \nonumber
\end{eqnarray}
where
\[
Y_\ell ( \vartheta) = \int_{\mathit{\Omega}}\exp\left\{ -
\frac{\hat{J}_0 } {2 \vartheta}\cdot |\omega_\ell |^2_{L^2_\beta} -
\int_0^\beta V_\ell (\omega_\ell (\tau)){\rm d}\tau \right\}\chi
({\rm d}\omega_\ell).
\]
Now we use the upper bound (\ref{3}) to estimate $\inf_{\ell} Y_\ell
(\vartheta)$, the lower bound (\ref{3}) to estimate the integrand in
(\ref{54}), take into account Proposition \ref{1pn}, and arrive at
(\ref{53}).
\end{proof}
\noindent By Jensen's inequality we readily get from (\ref{53})
the following Dobrushin-like bound.
\begin{corollary} \label{1co}
For all $\ell$ and $\xi \in \mathit{\Omega}^{\rm t}$, the kernels
$\pi_\ell (\cdot|\xi)$,  obey the estimate
\begin{equation} \label{55}
\int_{\mathit{\Omega}} h(\omega_\ell) \pi_\ell ({\rm d}\omega |\xi)
\leq  C_{\ref{53}}  + (\vartheta/\varkappa) \sum_{\ell'}
|J_{\ell\ell'}|\cdot h(\xi_{\ell'}),
\end{equation}
with
\begin{equation} \label{56}
h(\omega_\ell ) = \lambda_\sigma |\omega_\ell |^2_{C^\sigma_\beta }
+ \varkappa |\omega_\ell |^2_{L^2_\beta },
\end{equation}
which is a compact function $h:C_\beta \rightarrow \mathbb{R}$.
\end{corollary}
\noindent For translation invariant lattice systems with the
single-spin space $\mathbb{R}$ and  ferromagnetic pair interactions,
integrability estimates like
\[
\log\left\{\int_{\mathbb{R}^{\mathbb{L}} }\exp (\lambda |x_\ell
|)\pi_\ell ({\rm d}x|y)\right\} < A + \sum_{\ell'} I_{\ell \ell'}
|y_{\ell'}|,
\]
were first obtained by J. Bellissard and R. H{\o}egh-Krohn, see
Proposition III.1 and Theorem III.2  in \cite{BH}. Dobrushin
 type estimates like (\ref{d1}) were also proven in \cite{COPP,Sin}.
The methods used there essentially employed the properties of the
model and hence cannot be of use in our situation. Our approach to
getting such estimates is much simpler; at the same time, it is
applicable in both cases -- classical and quantum. Its peculiarities
are: (a) first we prove the exponential integrability (\ref{53}) and
then derive the Dobrushin bound (\ref{55}) rather than prove it
directly; (b) the function (\ref{56}) consists of two additive
terms, the first of which is to guarantee the compactness while the
second one controls the inter-particle interaction.

Now by means of (\ref{53}) we obtain moment estimates for the
kernels $\pi_\Lambda$ with arbitrary $\Lambda \Subset \mathbb{L}$.
Let the parameters $\sigma$, $\varkappa$, and $\lambda_\sigma$ be
the same as in (\ref{53}). For $\ell \in \Lambda \Subset
\mathbb{L}$, we define
\begin{equation} \label{62}
n_\ell (\Lambda |\xi) = \log\left\{\int_{\mathit{\Omega}}\exp\left(
\lambda_\sigma |\omega_\ell |^2_{C^\sigma_\beta} + \varkappa
|\omega_\ell |^2_{L^2_\beta}\right) \pi_\Lambda ({\rm d}\omega|\xi)
\right\},
\end{equation}
which is finite by (\ref{llb1}).
\begin{lemma} \label{melm}
For every $\alpha \in \mathcal{I}$, there exists $C_{\ref{66a}}
(\alpha)
>0$ such that for all $\ell _{0}$ and $\xi \in \mathit{\Omega }^{
\mathrm{t}}$,
\begin{equation} \label{66a}
\limsup_{\Lambda \nearrow \mathbb{L}} \sum_{\ell \in \Lambda
}n_{\ell }(\Lambda |\xi )w_{\alpha }(\ell _{0},\ell ) \leq
C_{\ref{66a}}(\alpha);
\end{equation}%
hence,
\begin{equation} \label{66c}
\limsup_{\Lambda \nearrow \mathbb{L}} n_{\ell_0} (\Lambda |\xi) \leq
C_{\ref{66a}}(\alpha), \quad {for} \ \ {any} \ \ \alpha \in
\mathcal{I}.
\end{equation}
Thereby, for all $\xi\in \mathit{\Omega}^{\rm t}$, there exists
$C_{\ref{590}}(\ell, \xi)>0$ such that for all $\Lambda \Subset
\mathbb{L}$ containing $\ell$,
\begin{equation} \label{590}
n_{\ell} (\Lambda | \xi) \leq C_{\ref{590}}(\ell, \xi).
\end{equation}
\end{lemma}
\begin{proof}
Given $\varkappa >0$ and $\alpha \in \mathcal{I}$, we  fix
$\vartheta
>0$ such that
\begin{equation}
\vartheta \sum_{\ell ^{\prime }}|J_{\ell \ell ^{\prime }}|\leq \vartheta \hat{J}%
_{0}\leq \vartheta \hat{J}_{\alpha }<\varkappa .  \label{63b}
\end{equation}%
Then integrating both sides of the exponential bound (\ref{53}) with
respect to the measure $\pi _{\Lambda }(\mathrm{d}\omega |\xi )$  we
get
\begin{eqnarray}
\qquad \quad n_{\ell }(\Lambda |\xi ) &\leq &C_{\ref{53}} +\vartheta
\sum_{\ell ^{\prime }\in \Lambda ^{c}}|J_{\ell \ell ^{\prime
}}|\cdot |\xi _{\ell ^{\prime
}}|_{L_{\beta }^{2}}^{2}  \label{65} \\
&+&\log \left\{ \int_{\mathit{\Omega }}\exp \left( \vartheta
\sum_{\ell ^{\prime }\in \Lambda }|J_{\ell \ell ^{\prime }}|\cdot
 |\omega _{\ell ^{\prime }}|_{L_{\beta }^{2}}^{2}\right) \pi
_{\Lambda }(\mathrm{d}\omega
|\xi )\right\}  \notag \\
&\leq &C_{\ref{53}} +\vartheta \sum_{\ell ^{\prime }\in \Lambda
^{c}}|J_{\ell \ell ^{\prime }}|\cdot |\xi _{\ell ^{\prime
}}|_{L_{\beta }^{2}}^{2}  + \vartheta /\varkappa \sum_{\ell ^{\prime
}\in \Lambda }|J_{\ell \ell ^{\prime }}|\cdot n_{\ell ^{\prime
}}(\Lambda |\xi ).  \notag
\end{eqnarray}%
Here we have used (\ref{63b}) and the multiple H\"{o}lder inequality
\[
\int \left( \prod\nolimits_{i=1}^{n}\varphi _{i}^{\alpha
_{i}}\right)
\mathrm{d}\mu \leq \prod\nolimits_{i=1}^{n}\left( \int \varphi _{i}\mathrm{d}%
\mu \right) ^{\alpha _{i}},
\]
in which $\mu $ is a probability measure, $\varphi _{i}\geq 0$
(respectively,  $\alpha _{i}\geq 0$), $i=1 , \dots, n$, are
functions (respectively,  numbers such that $\sum_{i=1}^{n}\alpha
_{i}\leq 1$). Then (\ref{65}) yields
\begin{align}
& n_{\ell _{0}}(\Lambda |\xi )\leq \sum_{\ell \in \Lambda }n_{\ell
}(\Lambda
|\xi )w_{\alpha }(\ell _{0},\ell )  \label{66} \\
& \leq \frac{1}{1-\vartheta \hat{J}_{\alpha }/\varkappa }\left[
C_{\ref{53}} \sum_{\ell ^{\prime }\in \Lambda }w_{\alpha }(\ell
_{0},\ell ^{\prime })+\vartheta \hat{J}_{\alpha }\sum_{\ell
^{\prime }\in \Lambda ^{c}}|\xi _{\ell
^{\prime }}|_{L_{\beta }^{2}}^{2}w_{\alpha }(\ell _{0},\ell ^{\prime })%
\right] .  \notag
\end{align}%
Therefrom, for all $\xi \in \mathit{\Omega }^{\mathrm{t}}$, we get
\begin{eqnarray}
& &  \limsup_{\Lambda \nearrow \mathbb{L}}n_{\ell _{0}}(\Lambda |\xi
)\leq \limsup_{\Lambda \nearrow \mathbb{L}} \sum_{\ell \in \Lambda
}n_{\ell
}(\Lambda |\xi )w_{\alpha }(\ell _{0},\ell )  \label{67d} \\
& &  \quad \qquad \leq \frac{C_{\ref{53}} }{1-\vartheta
\hat{J}_{\alpha }/\varkappa }\sum_{\ell }w_{\alpha }(\ell _{0},\ell
)\ \overset{\mathrm{def}}{=}\text{ }C_{\ref{66a}} (\alpha), \notag
\end{eqnarray}
which gives (\ref{66a}) and (\ref{66c}). The proof of (\ref{590}) is
straightforward.
\end{proof}
\vskip.1cm \noindent Recall that  the norm $\|\cdot\|_\alpha$ was
defined by (\ref{29}). Given $\alpha \in \mathcal{I}$ and $\sigma
\in (0, 1/2)$, we set
\begin{equation} \label{59a}
\|\xi\|_{\alpha , \sigma} = \left[\sum_{\ell} |\xi_\ell
|^2_{C^\sigma_\beta}w_\alpha (0, l)\right]^{1/2}.
\end{equation}
\begin{lemma} \label{klm}
Let the assumptions of Lemma \ref{4lm} be satisfied. Then for every
 $\alpha \in \mathcal{I}$ and $\xi\in
 \mathit{\Omega}^{\rm t}$, one finds a positive
$C_{\ref{61q}}(\xi)$ such that for all  $\Lambda \Subset
\mathbb{L}$,
\begin{equation} \label{61q}
\int_{\mathit{\Omega}} \|\omega\|_{\alpha }^2 \pi_\Lambda ({\rm
d}\omega|\xi) \leq C_{\ref{61q}}(\xi).
\end{equation}
Furthermore, for every
 $\alpha \in \mathcal{I}$,
$\sigma \in (0, 1/2)$, and $\xi\in
 \mathit{\Omega}^{\rm t}$ for which the norm (\ref{59a}) is finite,
 one finds a $C_{\ref{61a}}(\xi)>0$ such that for
 all $\Lambda \Subset \mathbb{L}$,
 \begin{equation} \label{61a}
\int_{\mathit{\Omega}} \|\omega\|_{\alpha , \sigma}^2 \pi_\Lambda
({\rm d}\omega|\xi) \leq C_{\ref{61a}}(\xi).
\end{equation}
\end{lemma}
\begin{proof}
For any fixed $\xi \in\mathit{\Omega}^{\rm t}$, by the Jensen
inequality and (\ref{66}) one has
\begin{eqnarray} \label{59s}
& & \lim\sup_{\Lambda \nearrow \mathbb{L}}\int_{\mathit{\Omega}}
\|\omega\|^2_{\alpha } \pi_\Lambda ({\rm d}\omega|\xi) \\ & & \quad
\leq \lim\sup_{\Lambda \nearrow
\mathbb{L}}\left[\frac{1}{\varkappa}\sum_{\ell \in \Lambda } n_\ell
(\Lambda|\xi) w_\alpha (0, \ell) + \sum_{\ell \in
\Lambda^c}|\xi_\ell |_{L_\beta^2}^2 w_\alpha (0, \ell) \right] \nonumber \\
& & \quad \leq C_{\ref{66a}}(\alpha )/ \varkappa . \nonumber
\end{eqnarray}
Hence, the set consisting of the left-hand sides of (\ref{61q})
indexed by $\Lambda \Subset \mathbb{L}$ is bounded in $\mathbb{R}$.
The proof of (\ref{61a}) is analogous.
\end{proof}

\subsection{Weak convergence of tempered measures}
\label{wcss}

Recall that $f:\mathit{\Omega}\rightarrow \mathbb{R}$ is called a
local function if it is measurable with respect to
$\mathcal{B}(\mathit{\Omega}_\Lambda)$ for a certain $\Lambda\Subset
\mathbb{L}$.
\begin{lemma} \label{tanlm}
Let a sequence $\{\mu_n\}_{n \in \mathbb{N}}\subset
\mathcal{P}(\mathit{\Omega}^{\rm t})$ have the following properties:
(a) for every $\alpha \in \mathcal{I}$, each its element obeys the
estimate
\begin{equation} \label{ta1}
\int_{\mathit{\Omega}^{\rm t}} \|\omega\|_\alpha^2  \mu_n ({\rm
d}\omega) \leq C_{\ref{ta1}} (\alpha),
\end{equation}
with one and the same $C_{\ref{ta1}}(\alpha)$; (b) for every local
$f\in C_{\rm b}(\mathit{\Omega}^{\rm t})$, the sequence $\{\mu_n
(f)\}_{n \in \mathbb{N}}\subset \mathbb{R}$ is fundamental. Then
$\{\mu_n\}_{n \in \mathbb{N}}$ converges in $\mathcal{W}^{\rm t}$ to
a certain $\mu \in \mathcal{P}(\mathit{\Omega}^{\rm t})$.
\end{lemma}
\begin{proof}
The topology of the Polish space $\mathit{\Omega}^{\rm t}$ is
consistent with the following metric (c.f., (\ref{met}))
\begin{eqnarray} \label{tan2}
\rho (\omega , \tilde{\omega}) = \sum_{k=1}^\infty
2^{-k}\frac{\|\omega - \tilde{\omega}\|_{\alpha_k} }{1 + \|\omega -
\tilde{\omega}\|_{\alpha_k} } + \sum_{\ell} 2^{-|\ell|}
\frac{|\omega_\ell - \tilde{\omega}_\ell|_{C_\beta}}{1 +
|\omega_\ell - \tilde{\omega}_\ell|_{C_\beta}},
\end{eqnarray}
where $\{\alpha_k\}_{k \in \mathbb{N}}\subset
\mathcal{I}=(\underline{\alpha}, \overline{\alpha})$ is a monotone
strictly decreasing sequence converging to $\underline{\alpha}$. Let
us denote by $C^{\rm u}_{\rm b}(\mathit{\Omega}^{\rm t}; \rho)$ the
set of all bounded functions $f: \mathit{\Omega}^{\rm t} \rightarrow
\mathbb{R}$, which are uniformly continuous with respect to
(\ref{tan2}). Thus, to prove the lemma it suffices to show that
under its conditions the sequence $\{\mu_n (f)\}_{n \in \mathbb{N}}$
is fundamental for every $f \in C^{\rm u}_{\rm
b}(\mathit{\Omega}^{\rm t}; \rho)$. Given $\delta
>0$, we choose $\Lambda_\delta \Subset \mathbb{L}$ and $k_\delta
\in \mathbb{N}$ such that
\begin{equation} \label{tan3}
\sum_{\ell \in \Lambda_\delta^c} 2^{-|\ell|} < \delta/3, \qquad \
\sum_{k =k_\delta}^\infty 2^{-k} = 2^{- k_\delta + 1} < \delta/3.
\end{equation}
For this $\delta$ and a certain $R>0$, we choose $\Lambda_\delta
(R)\Subset \mathbb{L}$ to obey
\begin{equation} \label{tan4}
\sup_{\ell \in \mathbb{L} \setminus \Lambda_\delta (R)}\left\{
w_{\alpha_{k_\delta
-1}}(0,\ell)/w_{\alpha_{k_\delta}}(0,\ell)\right\} <
\frac{\delta}{3 R^2},
\end{equation}
which is possible in view of (\ref{te1}). Finally, for $R>0$, we set
\begin{equation} \label{tan40}
B_R = \{ \omega \in \mathit{\Omega}^{\rm t} \ | \
\|\omega\|_{\alpha_{k_\delta}} \leq R\}.
\end{equation}
By (\ref{ta1}) and the Chebyshev inequality, one has that for all $n
\in \mathbb{N}$,
\begin{equation} \label{tan5}
\mu_n \left(\mathit{\Omega}^{\rm t} \setminus B_R \right) \leq
C_{\ref{ta1}}(\alpha_{k_\delta})/ R^2 .
\end{equation}
Now for $f \in C^{\rm u}_{\rm b}(\mathit{\Omega}^{\rm t}; \rho)$,
$\Lambda \Subset \mathbb{L}$, and $n , m\in \mathbb{N}$, we have
\begin{eqnarray} \label{tan6}
\left\vert \mu_n (f) - \mu_m (f)  \right\vert & \leq & \left\vert
\mu_n (f_\Lambda ) - \mu_m (f_\Lambda)  \right\vert \\ & + & 2
\max\{\mu_n ( |f - f_\Lambda|); \mu_m ( |f - f_\Lambda|) \},
\nonumber
\end{eqnarray}
where we set $f_\Lambda (\omega ) = f(\omega_\Lambda \times
0_{\Lambda^c})$. By (\ref{tan5}),
\begin{eqnarray} \label{tan7}
\mu_n ( |f - f_\Lambda|) & \leq & 2
C_{\ref{ta1}}(\alpha_{k_\delta}) \|f\|_\infty
/ R^2\\
& + & \int_{B_R} \left\vert f(\omega) - f(\omega_\Lambda \times
0_{\Lambda^c} )\right\vert \mu_n ({\rm d} \omega). \nonumber
\end{eqnarray}
For chosen $f \in C^{\rm u}_{\rm b}(\mathit{\Omega}^{\rm t}; \rho)$
and $\varepsilon >0$, one finds $\delta >0$ such that for all
$\omega, \tilde{\omega} \in \mathit{\Omega}^{\rm t}$,
\[
\left\vert f (\omega) - f(\tilde{\omega})\right\vert <
\varepsilon/6, \quad {\rm whenever} \ \  \rho(\omega ,
\tilde{\omega}) < \delta.
\]
For these $f$, $\varepsilon$, and $\delta$, one picks up
$R(\varepsilon, \delta)>0$ such that
\begin{equation} \label{tan8}
C_{\ref{ta1}}(\alpha_{k_\delta})\|f\|_{\infty}/ \left[R(\varepsilon,
\delta)\right]^2 < {\varepsilon}/{12}.
\end{equation}
Now one takes $\Lambda \Subset \mathbb{L}$, which contains both
$\Lambda_\delta$ and $\Lambda_\delta [R(\varepsilon, \delta)]$
defined by (\ref{tan3}), (\ref{tan4}). For this $\Lambda$, $\omega
\in B_{R(\varepsilon, \delta)}$, and $k = 1 , 2 , \dots , k_\delta
-1$, one has
\begin{eqnarray} \label{tan9}
\qquad \|\omega - \omega_\Lambda \times 0_{\Lambda^c}\|_{\alpha_k}^2
& = & \sum_{\ell \in \Lambda^c} |\omega_\ell|_{L^2_\beta}^2
w_{\alpha_{k_\delta}}(0,\ell)\left[w_{\alpha_{k}}(0,\ell)/w_{\alpha_{k_\delta}}(0,\ell)
\right]\\ & \leq & \frac{\delta}{3\left[R(\varepsilon,
\delta)\right]^2} \sum_{\ell \in
\Lambda^c}|\omega_\ell|_{L^2_\beta}^2 w_{\alpha_{k_\delta}}(0,\ell)
< \frac{\delta}{3}, \nonumber
\end{eqnarray}
where (\ref{tan4}), (\ref{tan40}) have been used. Then by
(\ref{tan2}), (\ref{tan3}), it follows that
\begin{equation} \label{tan10}
\forall \omega\in B_{R(\varepsilon, \delta)}: \quad \rho (\omega ,
\omega_\Lambda \times 0_{\Lambda^c}) < \delta,
\end{equation}
which together with (\ref{tan8}) yields in (\ref{tan7})
\[
\mu_n ( |f - f_\Lambda|) < \frac{\varepsilon}{6} +
\frac{\varepsilon}{6}\mu_n \left(B_{R(\varepsilon, \delta)}
\right)\leq \frac{\varepsilon}{3}.
\]
By assumption (b) of the lemma, one finds $N_\varepsilon$ such that
for all $n , m >N_\varepsilon$,
\[
\left\vert \mu_n (f_\Lambda ) - \mu_m (f_\Lambda ) \right\vert <
\frac{\varepsilon}{3}.
\]
Applying the latter two estimates in (\ref{tan6}) we get that the
sequence $\{\mu_n\}_{n\in \mathbb{N}}$ is fundamental in the
topology $\mathcal{W}^{\rm t}$ in which the space
$\mathcal{P}(\mathit{\Omega}^{\rm t})$ is complete.
\end{proof}

\section{Proof of Theorems \ref{1tm} -- \ref{httm}}
\label{4s}

The existence of Euclidean Gibbs measures and the integrability
estimate (\ref{43}) can be proven independently. To establish the
compactness of $\mathcal{G}^{\rm t}$ we will need the estimate
(\ref{43}). Thus, we first prove  Theorem \ref{2tm}. \noindent

 \textbf{Proof of Theorem
\ref{2tm}:} \ Let us show that every $\mu \in \mathcal{P}(\Omega)$
which solves the DLR equation (\ref{40}) ought to obey the estimate
(\ref{43}) with one and the same $C_{\ref{43}}$. To this end we
apply the bounds for the kernels $\pi_\Lambda (\cdot|\xi)$ obtained
above. Consider the functions
\[
G_N (\omega_\ell) \ \stackrel{\rm def}{=} \ \exp\left(\min\left\{
\lambda_\sigma |\omega_\ell|_{C_\beta^\sigma}^2 + \varkappa
|\omega_\ell|_{L_\beta^2}^2; N\right\} \right), \quad N\in
\mathbb{N}.
\]
By (\ref{40}), Fatou's lemma, and the estimate (\ref{66c}) with an
arbitrarily chosen
 $\alpha \in \mathcal{I}$ we get
\begin{eqnarray*}
& & \int_{\mathit{\Omega}} G_N (\omega_\ell)\mu({\rm d}\omega)
=\limsup_{\Lambda \nearrow \mathbb{L}}\int_{\mathit{\Omega}} \left[
\int_{\mathit{\Omega}} G_N (\omega_\ell)\pi_\Lambda ({\rm
d}\omega|\xi)\right]\mu({\rm d}\xi) \\ & & \quad \leq
\limsup_{\Lambda \nearrow \mathbb{L}}\int_{\mathit{\Omega}}
\left[\int_{\mathit{\Omega}}\exp\left( \lambda_\sigma
|\omega_\ell|_{C_\beta^\sigma}^2 + \varkappa
|\omega_\ell|_{L_\beta^2}^2 \right)\pi_\Lambda ({\rm
d}\omega|\xi)\right]\mu({\rm d}\xi) \\ & & \quad \leq
\int_{\mathit{\Omega}}\left[ \limsup_{\Lambda \nearrow
\mathbb{L}}\int_{\mathit{\Omega}}\exp\left( \lambda_\sigma
|\omega_\ell|_{C_\beta^\sigma}^2 + \varkappa
|\omega_\ell|_{L_\beta^2}^2 \right)\pi_\Lambda ({\rm
d}\omega|\xi)\right]\mu({\rm d}\xi)\\ & & \quad \leq \exp
C_{\ref{66a}}(\alpha)  \ \stackrel{\rm def}{=} \ C_{\ref{43}} .
\end{eqnarray*}
In view of the support property (\ref{40b}) of any measure solving
the equation (\ref{40}) we can pass here to the limit $N \rightarrow
+\infty$ and get (\ref{43}).
 $\square$
\begin{corollary} \label{tanco}
For every $\alpha \in \mathcal{I}$, the topologies induced on
$\mathcal{G}^{\rm t}$ by $\mathcal{W}_\alpha$ and $\mathcal{W}^{\rm
t}$ coincide.
\end{corollary}
\begin{proof}
Follows immediately from Lemma \ref{tanlm} and the estimate
(\ref{43}).
\end{proof}
\vskip.1cm \noindent
 \textbf{Proof of Theorem
\ref{1tm}:} \ Let us introduce the next scale of Banach spaces
(c.f., (\ref{29}))
\begin{equation} \label{57}
\mathit{\Omega}_{\alpha, \sigma} = \left\{ \omega \in
\mathit{\Omega} \ \left\vert \ \|\omega\|_{\alpha , \sigma}  <
\infty \right. \right\}, \quad \sigma \in (0, 1/2), \ \ \alpha \in
\mathcal{I},
\end{equation}
where the norm $\|\cdot\|_{\alpha , \sigma}$ was defined by
(\ref{59a}).  For any pair $\alpha , \alpha'\in \mathcal{I}$ such
that $\alpha < \alpha'$, the embedding $\mathit{\Omega}_{\alpha ,
\sigma} \hookrightarrow \mathit{\Omega}_{\alpha'}$ is compact, see
Remark \ref{1rm}. This fact and the estimate (\ref{61a}), which
holds for any $\xi \in \mathit{\Omega}_{\alpha, \sigma}$, imply by
Prokhorov's criterion the relative compactness of the set
$\{\pi_{\Lambda}(\cdot|\xi)\}_{\Lambda \Subset \mathbb{L}}$ in
$\mathcal{W}_{\alpha'}$. Therefore, the sequence
$\{\pi_{\Lambda}(\cdot|0)\}_{\Lambda \Subset \mathbb{L}}$ is
relatively compact in every $\mathcal{W}_\alpha$, $\alpha \in
\mathcal{I}$. Then Lemma \ref{3lm} yields that $\mathcal{G}^{\rm t}
\neq \emptyset$. By the same Prokhorov criterion and the estimate
(\ref{43}), we get the $\mathcal{W}_\alpha$-relative compactness of
$\mathcal{G}^{\rm t}$. Then in view of the Feller property (Lemma
\ref{2lm}), the set $\mathcal{G}^{\rm t}$ is closed and hence
compact in every $\mathcal{W}_\alpha$, $\alpha\in \mathcal{I}$,
which by Corollary \ref{tanco} completes the proof. $\square$.
\vskip.1cm
 \noindent
 \textbf{Proof of Theorem \ref{3tm}:} \
To some extent we shall  follow the line of arguments used in the
proof of Lemma 3.1 in \cite{LP}. Given $\ell, \ell_0$, $b>0$,
$\sigma \in (0, 1/2)$, and $\Lambda \subset \mathbb{L}$, we
introduce
\begin{eqnarray} \label{44}
\qquad \mathit{\Xi}_\ell (\ell_0 , b , \sigma) & = & \{ \xi \in
\mathit{\Omega} \ | \ |\xi_\ell |_{C^\sigma_\beta}^2 \leq b \log ( 1
+ |\ell - \ell_0|) \}, \\ \qquad \mathit{\Xi}_\Lambda (\ell_0 , b ,
\sigma) & = & \bigcap_{\ell \in \Lambda} \Xi_\ell (\ell_0 , b ,
\sigma) . \nonumber
\end{eqnarray}
For a cofinal sequence $\mathcal{L}$, we set
\begin{equation} \label{45}
\mathit{\Xi} (\ell_0 , b , \sigma)  = \bigcup_{\Lambda \in
\mathcal{L}}\mathit{\Xi}_{\Lambda^c} (\ell_0 , b , \sigma), \quad
\mathit{\Xi}( b , \sigma)  = \bigcap_{\ell_0 \in
\mathbb{L}}\mathit{\Xi} (\ell_0 , b , \sigma).
\end{equation}
The latter $\mathit{\Xi} (b, \sigma)$ is a subset of
$\mathit{\Omega}^{\rm t}$ and is the same as the one given by
(\ref{47}). To prove the theorem let us show that for any $\sigma
\in (0, 1/2)$, there exists $b>0$ such that for all $\ell_0$ and
$\mu \in \mathcal{G}^{\rm t}$,
\begin{equation} \label{69}
\mu \left( \mathit{\Omega} \setminus \mathit{\Xi}(\ell_0 , b ,
\sigma)\right) = 0.
\end{equation}
By (\ref{44}) we have
\begin{eqnarray} \label{mar}
\qquad \quad \mathit{\Omega} \setminus \mathit{\Xi}_{\Lambda^c}
(\ell_0 , b, \sigma)  & = & \{ \xi \in \mathit{\Omega} \ | \
(\exists \ell \in \Lambda^c):\ \ |\xi_\ell |_{C^\sigma_\beta}^2 > b
\log (1 + |\ell - \ell_0|)\} \\ &  \subset &  \{ \xi \in
\mathit{\Omega} \ | \ (\exists \ell \in \Delta^c):\ \ |\xi_\ell
|_{C^\sigma_\beta}^2 > b \log (1 + |\ell - \ell_0|)\}, \nonumber
\end{eqnarray}
for any $\Delta \subset \Lambda$. Therefore,
\begin{equation} \label{70}
\mu \left(\bigcap_{\Lambda \in \mathcal{L}} \left[  \mathit{\Omega}
\setminus \mathit{\Xi}_{\Lambda^c} (\ell_0 , b, \sigma)\right]
\right) = \lim_{\mathcal{L}}\mu \left(  \mathit{\Omega} \setminus
\mathit{\Xi}_{\Lambda^c} (\ell_0 , b, \sigma)\right),
\end{equation}
which holds for any cofinal sequence $\mathcal{L}$. By (\ref{mar}),
\begin{eqnarray*}
\mu \left(  \mathit{\Omega} \setminus \mathit{\Xi}_{\Lambda^c}
(\ell_0 , b, \sigma) \right) & = & \mu \left(\bigcup_{\ell \in
\Lambda^c} \left[\mathit{\Omega} \setminus \mathit{\Xi}_\ell (\ell_0
, b , \sigma) \right] \right)\\ & \leq & \sum_{\ell \in \Lambda^c}
\mu\left( \left\{ \xi \ | \ |\xi_\ell|^2_{C^\sigma_\beta}
> b \log ( 1 + |\ell - \ell_0|) \right\} \right)\\ & = & \sum_{\ell \in
\Lambda^c} \mu\left( \left\{ \xi \ | \ \exp\left( \lambda_\sigma
|\xi_\ell|^2_{C^\sigma_\beta} \right) > (1 + |\ell - \ell_0|)^{b
\lambda_\sigma} \right\} \right).
\end{eqnarray*}
Applying here the Chebyshev inequality and the estimate (\ref{43})
we get
\[
\mu \left( \mathit{\Omega} \setminus \mathit{\Xi}_{\Lambda^c}
(\ell_0 , b, \sigma) \right) \leq C_{\ref{43}} \sum_{\ell \in
\Lambda^c} ( 1 + |\ell - \ell_0|)^{ - b \lambda_\sigma}.
\]
As $\mathbb{L} = \mathbb{Z}^d$, the latter series converges for any
$b > d /\lambda_\sigma$. In this case by (\ref{70})
\begin{eqnarray*}
\mu \left(  \mathit{\Omega} \setminus \mathit{\Xi} (\ell_0 , b,
\sigma) \right)  =  \lim_{\mathcal{L}}\mu \left(\left[
\mathit{\Omega} \setminus \mathit{\Xi}_{\Lambda^c} (\ell_0 , b,
\sigma)\right] \right) = 0,
\end{eqnarray*}
which yields (\ref{69}). $\square$ \vskip.1cm \noindent
 By $\mathcal{E}(\mathit{\Omega}^{\rm t})$ we denote the set of all
 continuous local functions $f:\mathit{\Omega}^{\rm t} \rightarrow \mathbb{R}$,
 for each of which there exist $\sigma\in (0,1/2)$,
$\Delta_f \Subset \mathbb{L}$, and $D_f >0$, such that
\begin{equation} \label{loc}
|f(\omega) |^2 \leq D_f \sum_{\ell \in \Delta_f} \exp\left(
\lambda_\sigma |\omega_\ell|_{C^\sigma_\beta}^2 \right), \quad
{\rm for \ all} \ \  \omega\in \mathit{\Omega}^{\rm t},
\end{equation}
where $\lambda_\sigma$ is as in (\ref{43}). Let ${\rm
ex}(\mathcal{G}^{\rm t})$ stand for the set of all extreme elements
of $\mathcal{G}^{\rm t}$.
\begin{lemma} \label{locco}
For every $\mu\in{\rm ex}(\mathcal{G}^{\rm t})$ and any cofinal
sequence $\mathcal{L}$, it follows that:
 (a) the sequence $\{\pi_\Lambda
(\cdot|\xi)\}_{\Lambda \in \mathcal{L}}$ converges in
$\mathcal{W}^{\rm t}$ to this $\mu$ for $\mu$-almost all $\xi \in
\mathit{\Omega}^{\rm t}$; (b) for every $f \in
\mathcal{E}(\mathit{\Omega}^{\rm t})$, one has
$\lim_{\mathcal{L}}\pi_\Lambda (f|\xi)= \mu(f)$ for $\mu$-almost all
$\xi \in \mathit{\Omega}^{\rm t}$.
\end{lemma}
\begin{proof}
Claim (c) of Theorem 7.12, page 122 in \cite{Ge}, implies that for
any local $f\in C_{\rm b}(\mathit{\Omega}^{\rm t})$,
\begin{equation} \label{a1}
\lim_{\mathcal{L}} \pi_\Lambda (f|\xi) = \mu(f), \quad {\rm for} \
\mu{\rm -almost} \ {\rm all} \ \xi\in\mathit{\Omega}^{\rm t}.
\end{equation}
Then the convergence stated in claim (a) follows from Lemmas
\ref{klm} and \ref{tanlm}. Given $f \in
\mathcal{E}(\mathit{\Omega}^{\rm t})$ and $N \in \mathbb{N}$, we set
$\mathit{\Omega}_N = \{\omega \in \mathit{\Omega} \ | \ |f(\omega)|>
N\}$ and
\[
f_N (\omega) = \left\{ \begin{array}{ll} f(\omega)  \ \ &{\rm if}
\ \ |f(\omega)| \leq N; \\ N f(\omega)/|f(\omega)| \ \ &{\rm
otherwise}.
\end{array} \right.
\]
Each $f_N$ belongs to $C_{\rm b}(\mathit{\Omega}^{\rm t})$ and $f_N
\rightarrow f$ point-wise as $N \rightarrow +\infty$. Then by
(\ref{a1}) there exists  a Borel set $\mathit{\Xi}_\mu \subset
\mathit{\Omega}^{\rm t}$, such that $\mu(\mathit{\Xi}_\mu) = 1$ and
for every $N \in \mathbb{N}$,
\begin{equation} \label{1aa}
\lim_{\mathcal{L}} \pi_\Lambda (f_N |\xi) = \mu(f_N), \quad {\rm
for } \ {\rm all} \ \xi \in \mathit{\Xi}_\mu.
\end{equation}
Note that by (\ref{62}), (\ref{590}), and (\ref{loc}), for any $\xi
\in \mathit{\Xi}_\mu$ one finds a positive $C_{\ref{591}}(f, \xi)$
such that for all $\Lambda \Subset \mathbb{L}$, which contain
$\Delta_{ f}$, it follows that
\begin{equation} \label{591}
\int_{\mathit{\Omega}} |f(\omega)|^2 \pi_\Lambda ({\rm d}\omega|\xi)
\leq {C}_{\ref{591}} (f , \xi).
\end{equation}
Hence
 \begin{eqnarray*}
& & |\pi_\Lambda (f |\xi) - \pi_\Lambda (f_N|\xi)| \leq 2
\int_{\mathit{\Omega}_N}  |f(\omega)|\pi_\Lambda ({\rm
d}\omega|\xi)\\
& & \qquad \leq \frac{2}{N} \cdot\int_{\mathit{\Omega}}|f(\omega)|^2
\pi_\Lambda ({\rm d}\omega|\xi) \leq \frac{2}{N} \cdot
{C}_{\ref{591}} (f , \xi).
\end{eqnarray*}
 Similarly, by means of (\ref{loc}) and Theorem
\ref{2tm}, one gets
\[
|\mu (f) - \mu (f_N)| \leq \frac{2}{N}\cdot  D_f {C}_{\ref{43}}.
\]
The latter two inequalities  and (\ref{1aa}) allow us to estimate
$|\pi_\Lambda (f|\xi) - \mu(f)|$ and thereby to complete the
proof.
\end{proof}
\vskip.1cm \noindent \textbf{Proof of Theorem \ref{httm}:} For the
scalar translation invariant version of the model considered here,
the high-temperature uniqueness was proven in \cite{AKRT,AKRT1}. The
proof given below is a modification of the arguments used there.
Thus, we can be brief.

 Given $\ell$ and
$\xi, \xi' \in \mathit{\Omega}^{\rm t}$, we define the distance
\[
   R[\pi_\ell (\cdot|\xi); \pi_\ell (\cdot|\xi')]
 =  \sup_{f \in {\rm Lip}_1 (L^2_\beta)} \left\vert
\int_{\mathit{\Omega}} f(\omega_\ell)\pi_\ell({\rm d}\omega|\xi) -
\int_{\mathit{\Omega}} f(\omega_\ell)\pi_\ell({\rm
d}\omega|\xi')\right\vert,
\]
where ${\rm Lip}_1 (L^2_\beta)$ stands for the set of
Lipschitz-continuous functions $f:L^2_\beta\rightarrow \mathbb{R}$
 with the Lipschitz constant equal to one. The
proof is based upon the Dobrushin criterium (Theorem 4 in
\cite{Do}), which employs the matrix
\begin{equation} \label{ta21}
C_{\ell\ell'} = \sup\left\{ \frac{R[\pi_\ell(\cdot|\xi);
\pi_\ell(\cdot|\xi')]}{|\xi_\ell -
\xi_{\ell'}|_{L^2_\beta}}\right\}, \quad \ell \neq \ell', \ \ \ell,
\ell'\in \mathbb{L},
\end{equation}
where the supremum is taken over all $\xi, \xi' \in
\mathit{\Omega}^{\rm t}$ which differ only at $\ell'$. According
to this criterium the uniqueness stated will follow from the fact
\begin{equation} \label{ta22}
\sup_{\ell} \sum_{\ell' \in \mathbb{L}\setminus\{\ell\}} C_{\ell
\ell'} < 1.
\end{equation}
In view of (\ref{llb1}) the map
\begin{equation} \label{ta23}
L^2_\beta \ni \xi_{\ell'} \mapsto \mathit{\Upsilon}(\xi_{\ell'}) \
\stackrel{\rm def}{=} \ \int_{\mathit{\Omega}} f(\omega_\ell)
\pi_\ell ({\rm d}\omega |\xi)
\end{equation}
has the following derivative in direction $\zeta \in L^2_\beta$
\[
\left(\nabla \mathit{\Upsilon} (\xi_{\ell'}) ,
\zeta\right)_{L^2_\beta} =  - J_{\ell \ell'} \left[\pi_\ell \left(f
\cdot (\omega_\ell , \zeta)_{L^2_\beta} \left\vert \xi
\right.\right) - \pi_\ell \left(f|\xi\right) \cdot \pi_\ell \left(
(\omega_\ell , \zeta)_{L^2_\beta}\left\vert \xi \right.
\right)\right].
\]
By Theorem 5.1 of \cite{AKRT} the measures $\pi_\ell (\cdot|\xi)$
obey the logarithmic Sobolev inequality with the constant
\begin{equation}\label{ta25}
C_{\rm LS} = e^{\beta \delta}/ (a+b),
\end{equation}
which is independent of $\xi$. By standard arguments this yields the
estimate
\begin{equation}\label{ta26}
\left\vert \left(\nabla \mathit{\Upsilon} (\xi_{\ell'}) ,
\zeta\right)_{L^2_\beta}\right\vert \leq C_{\rm LS} |J_{\ell
\ell'}|\cdot |\zeta|^2_{L^2_\beta}.
\end{equation}
Then with the help of the mean value theorem from (\ref{ta21}) and
(\ref{ta25}) we get
\[
C_{\ell \ell'} \leq |J_{\ell \ell'}| \cdot  e^{\beta \delta}/ (a+b).
\]
Thereby, the validity of the uniqueness condition (\ref{ta22}) is
ensured by (\ref{dc3}). $\square$

\section{Proof of Theorems \ref{4tm}, \ref{phtm}}
\label{5s}

\subsection{Correlation Inequalities}

\label{ss5.1}

Recall that Theorems \ref{4tm} -- \ref{6tm} describe ferromagnetic
scalar models, i.e., the ones with $\nu = 1$ and $J_{\ell \ell'}
\geq 0$. Thus, all the statements below refer to such models only.
Their proofs are mainly based on correlation inequalities. For the
Gibbs measures considered here, such inequalities were derived in
\cite{AKKR} in the framework of the lattice approximation technique,
analogous to that of Euclidean quantum fields \cite{Si74}. We begin
with the FKG inequality, Theorem 6.1 in \cite{AKKR}.
\begin{proposition} \label{fkg}
For all $\Lambda \Subset \mathbb{L}$, $\xi \in \mathit{\Omega}^{\rm
t}$ and any continuous increasing $\pi_\Lambda
(\cdot|\xi)$-integrable functions $f$ and $g $, it follows that
\begin{equation} \label{w2}
\pi_\Lambda ( f \cdot g |\xi) \geq \pi_\Lambda ( f |\xi) \cdot
\pi_\Lambda ( g |\xi),
\end{equation}
which yields that for all such functions,
\begin{equation} \label{w3}
 \xi \leq \tilde{\xi} \ \
\Longrightarrow \pi_\Lambda ( f |\xi) \leq \pi_\Lambda (
f|\tilde{\xi}).
\end{equation}
\end{proposition}
Next, there follow the GKS inequalities, Theorem 6.2 in \cite{AKKR}.
\begin{proposition} \label{gks}
Let the self-interaction potential have the form
\begin{equation} \label{w3a}
V_\ell (x) = v_\ell (x^2) - h_\ell x, \quad h_\ell \geq 0 \ \ {\rm
for \ all} \ \ell\in \mathbb{L},
\end{equation}
with $v_\ell$ being continuous. Let also the continuous functions
$f_1, \dots , f_{n+m} : \mathbb{R}\rightarrow \mathbb{R}$ be
polynomially bounded and such that every $f_i$ is either an odd
increasing function on $\mathbb{R}$ or an even positive function,
increasing on $[0, +\infty)$. Then the following inequalities hold
for all $\tau_1 , \dots , \tau_{n+m} \in [0, \beta]$, and all
$\ell_1 , \dots , \ell_{n+m}\in \Lambda$,
\begin{equation} \label{w3b}
\int_{\mathit{\Omega}} \left(\prod_{i=1}^n f_i (\omega_{\ell_i}
(\tau_i))\right) \pi_\Lambda \left({\rm d}\omega |0\right) \geq 0;
\end{equation}
\begin{eqnarray} \label{w3c}
& & \int_{\mathit{\Omega}}\left(\prod_{i=1}^n f_i (\omega_{\ell_i}
(\tau_i))\right) \cdot \left(\prod_{i=n+1}^{n+m}f_i
(\omega_{\ell_i}(\tau_i)) \right)\pi_\Lambda \left({\rm d
}\omega|0\right)
\\ & & \qquad \geq  \int_{\mathit{\Omega}}\left(\prod_{i=1}^n f_i (\omega_{\ell_i}
(\tau_i))\right) \pi_\Lambda \left({\rm d }\omega|0\right) \cdot
\int_{\mathit{\Omega}}\left(\prod_{i=n+1}^{n+m}f_i
(\omega_{\ell_i}(\tau_i)) \right)\pi_\Lambda \left({\rm d
}\omega|0\right). \nonumber
\end{eqnarray}
\end{proposition}
 Given $\xi \in \mathit{\Omega}^{\rm t}$, $\Lambda \Subset
\mathbb{L}$, and $\ell, \ell'$, $\tau , \tau' \in [0, \beta]$,  the
pair correlation function is
\begin{eqnarray}\label{w4}
K^\Lambda_{\ell \ell'}(\tau , \tau' |\xi)& = &
\int_{\mathit{\Omega}} \omega_\ell (\tau) \omega_{\ell'} (\tau')
\pi_\Lambda ({\rm d}\omega|\xi) \\ & - & \int_{\mathit{\Omega}}
\omega_\ell (\tau)  \pi_\Lambda ({\rm d}\omega|\xi) \cdot
\int_{\mathit{\Omega}}
 \omega_{\ell'} (\tau') \pi_\Lambda ({\rm
d}\omega|\xi). \nonumber
\end{eqnarray}
Then, by (\ref{w3}),
\begin{equation} \label{w4b}
K^\Lambda_{\ell \ell'}(\tau , \tau' |\xi) \geq 0,
\end{equation}
which holds for all $\ell, \ell'$, $\tau, \tau'$, and
$\xi\in\mathit{\Omega}^{\rm t}$.
 The following result is a version of the estimate (12.129), page 254 of \cite{FFS},
 which for the Euclidean Gibbs measures may be proven by means of
 the lattice approximation.
\begin{proposition} \label{ffs}
Let $V_\ell$ be of the form (\ref{w3a}) with $h_\ell =0$ and the
functions $v_\ell$ being convex. Then for all $\ell, \ell'$, $\tau,
\tau'$ and for any $\xi \in \mathit{\Omega}^{\rm t}$ such that $\xi
\geq 0$, it follows that
\begin{equation} \label{w5}
K^\Lambda_{\ell \ell'}(\tau , \tau'|\xi) \leq K^\Lambda_{\ell
\ell'}(\tau , \tau'|0).
\end{equation}
\end{proposition}
Let us consider
\begin{eqnarray} \label{v1}
& & U_{\ell_1  \ell_2  \ell_3 \ell_4}^\Lambda (\tau_1 , \tau_2 ,
\tau_3 , \tau_4) =
\int_{\mathit{\Omega}}\omega_{\ell_1}(\tau_1)\omega_{\ell_2}(\tau_2)
\omega_{\ell_3}(\tau_3) \omega_{\ell_4}(\tau_4) \pi_\Lambda ({\rm
d}\omega|0)\\ & &  \qquad \quad  \quad \quad - K_{\ell_1
\ell_2}^\Lambda(\tau_1 , \tau_2|0)K_{\ell_3 \ell_4}^\Lambda(\tau_3 ,
\tau_4|0) - K_{\ell_1 \ell_3}^\Lambda(\tau_1 , \tau_3|0)K_{\ell_2
\ell_4}^\Lambda(\tau_2 , \tau_4|0) \nonumber \\ & &  \qquad \quad
\quad \quad - K_{\ell_1 \ell_4}^\Lambda(\tau_1 , \tau_4|0)K_{\ell_2
\ell_3}^\Lambda(\tau_2 , \tau_3|0), \nonumber
\end{eqnarray}
which is the Ursell function for the measure $\pi_\Lambda
(\cdot|0)$. The next statement gives the Gaussian domination and
Lebowitz inequalities, see \cite{AKKR}.
\begin{proposition} \label{le}
Let $V_\ell$ be of the form (\ref{w3a}) with $h_\ell =0$ and
 the functions $v_\ell$ being convex. Then for all $n
\in \mathbb{N}$, $\ell_1, \dots, \ell_{2n}
 \in \Lambda\Subset \mathbb{L}$, $\tau_1, \dots, \tau_{2n} \in [0, \beta]$, it follows
 that
\begin{eqnarray} \label{vv1}
& &
\int_{\mathit{\Omega}}\omega_{\ell_1}(\tau_1)\omega_{\ell_2}(\tau_2)
\cdots \omega_{\ell_{2n}}(\tau_{2n}) \pi_\Lambda ({\rm d}\omega|0)
\nonumber\\ & & \qquad \quad \leq \sum_{\sigma} \prod_{j=1}^n
\int_{\mathit{\Omega}}\omega_{\ell_{\sigma(2j-1)}}
(\tau_{\sigma(2j-1)})\omega_{\ell_{\sigma(2j)}}(\tau_{\sigma(2j)})\pi_\Lambda
({\rm d}\omega|0),
\end{eqnarray}
where the sum runs through the set of all partitions of $\{1,
\dots , 2n\}$ onto unordered pairs. In particular,
\begin{equation}\label{v2}
U_{\ell_1  \ell_2  \ell_3 \ell_4}^\Lambda (\tau_1 , \tau_2 , \tau_3
, \tau_4) \leq 0.
\end{equation}
\end{proposition}

\subsection{Proof of Theorem \ref{4tm}}

\label{sdss}  Given $\ell_0$ and $b>0$, let $\xi^{\ell_0}
=(\xi^{\ell_0}_\ell)_{\ell \in \mathbb{L}}$ be the following
constant (with respect to $\tau \in S_\beta$) configuration
\begin{equation} \label{v3}
 \xi^{\ell_0}_{\ell} (\tau) = [ b
\log(1 + |\ell - \ell_0|)]^{1/2}.
\end{equation}
 \textbf{Proof of Theorem \ref{4tm}:} \ The existence and
uniqueness of $\mu_{\pm}$ can be proven by literal repetition of the
arguments used in \cite{BH} to proving Theorem IV.3. These measures
are extreme, see the proof of Proposition V.1 in \cite{BH}. Now let
us show how to construct $\mu_{\pm}$. Fix $\sigma \in (0, 1/2)$ and
$b$ obeying the condition $b> d/\lambda_\sigma$ (see the proof of
Theorem \ref{3tm}). In view of (\ref{25}), for any $\ell_0$,
$\xi^{\ell_0}$ belongs to $\mathit{\Omega}^{\rm t}$. It also belongs
to  $\mathit{\Xi}(\ell_0 , b, \sigma)$ and for all $\xi\in
\mathit{\Xi}( b, \sigma)$, one finds $\Delta \Subset \mathbb{L}$
such that $\xi_\ell (\tau) \leq \xi^{\ell_0}_{\ell} (\tau)$ for all
$\tau\in [0, \beta]$ and $\ell \in \Delta^c$. Therefore, for any
cofinal sequence $\mathcal{L}$ and $\xi \in \mathit{\Xi}( b,
\sigma)$, one finds $\Delta \in \mathcal{L}$ such that for all
$\Lambda \in \mathcal{L}$, $\Delta \subset \Lambda$, one has
$\pi_\Lambda (\cdot |\xi) \leq \pi_\Lambda (\cdot |\xi^{\ell_0})$.
As was established in the proof of Theorem \ref{1tm}, every sequence
$\{\pi_\Lambda (\cdot | \xi)\}_{\Lambda \in \mathcal{L}}$, $\xi \in
\mathit{\Xi}(b ,\sigma)\subset \mathit{\Omega}^{\rm t}$, is
relatively compact in any $\mathcal{W}_\alpha$, $\alpha \in
\mathcal{I}$, which by Lemmas \ref{klm}, \ref{tanlm} yields its
$\mathcal{W}^{\rm t}$-relative compactness. For a cofinal sequence
$\mathcal{L}$, let $\mu^{\ell_0}$ be any of the accumulating points
of $\{\pi_\Lambda (\cdot | \xi^{\ell_0})\}_{\Lambda \in
\mathcal{L}}$. By Lemma \ref{3lm} $\mu^{\ell_0} \in\mathcal{G}^{\rm
t}$ and by Lemma \ref{locco} $\mu^{\ell_0}$ dominates every element
of ${\rm ex}(\mathcal{G}^{\rm t})$. Hence, $\mu^{\ell_0} = \mu_+$
since the maximal element is unique. The same is true for the
remaining accumulation points of $\{\pi_\Lambda (\cdot |
\xi)\}_{\Lambda \in \mathcal{L}}$; thus, for every cofinal sequence
$\mathcal{L}$ and for every $\ell_0$, we have
\begin{equation} \label{rf41}
\lim_{\mathcal{L}} \pi_\Lambda (\cdot |\xi^{\ell_0}) = \mu_{+}.
\end{equation}
 As the configuration
(\ref{v3}) is constant with respect to $\tau\in S_\beta$, the kernel
$\pi_\Lambda (\cdot|\xi^{\ell_0})$ may be considered as the one
$\tilde{\pi}_\Lambda (\cdot|0)$ corresponding to the Schr\"odinger
operator (\ref{sch}) with an additional  site-dependent `external
field', i.e., to the operator
\[
H_\Lambda - \sum_{\ell\in \Lambda} x_\ell \cdot [b \log(1 + |\ell -
\ell_0|)]^{1/2}.
\]
Then the Matsubara functions have the property (\ref{mats1}), which
yields by (\ref{rf41}) the same property for the functions
$\mathit{\Gamma}^{\mu_+}$. Analogously, one sets $\mu_{-} =
\lim_{\mathcal{L}}\pi_\Lambda (\cdot |- \xi^{\ell_0}) \in {\rm
ex}(\mathcal{G}^{\rm t})$ and proves its shift invariance. The
remaining properties can be established by repetition of the
arguments used in \cite{BH} to prove Proposition V.3. $\square$.
\begin{lemma} \label{bhpn}
The set $\mathcal{G}^{\rm t}$ is a singleton if and only if
\begin{equation} \label{w9}
\int_{\mathit{\Omega}} \omega_{\ell} (0) \mu_{+} ({\rm d}\omega) =
\int_{\mathit{\Omega}} \omega_{\ell} (0) \mu_{-} ({\rm d}\omega),
\quad {for} \ \ {any} \ \ \ell.
\end{equation}
\end{lemma}
\begin{proof}
Since the measures $\mu_{\pm}$ are shift invariant, we have
\begin{equation} \label{soc2}
\int_{\mathit{\Omega}} \omega_{\ell} (\tau) \mu_{\pm} ({\rm
d}\omega) = \int_{\mathit{\Omega}} \omega_{\ell} (0) \mu_{\pm} ({\rm
d}\omega),\quad \ {\rm for} \ {\rm all} \ \ \tau\in [0, \beta].
\end{equation}
 Certainly, (\ref{w9}) holds if $|\mathcal{G}^{\rm t}|=1$. Let us
 prove the converse.
 One observes that each local bounded continuous
function on $\mathit{\Omega}$ may be written as
\[
\mathit{\Omega} \ni \omega \mapsto f(\omega_{\ell_1} (\tau_1) ,
\dots , \omega_{\ell_n} (\tau_n)),
\]
with certain $n\in \mathbb{N}$, $\ell_1 , \dots , \ell_n \in
\mathbb{L}$, $\tau_1 , \dots , \tau_n \in [0, \beta]$, and $f\in
C_{\rm b} (\mathbb{R}^n)$. Obviously, the set of all local bounded
continuous functions is a defining class for
$\mathcal{P}(\mathit{\Omega}^{\rm t})$. Thus, the proof will be done
if we show that all finite-dimensional projections
\[
\nu_{\ell_1 , \dots , \ell_n}^{\tau_1 , \dots , \tau_n; \pm} \
\stackrel{\rm def}{=} \ \nu_{n; \pm},
\]
of the measures $\mu_{\pm}$ coincide whenever (\ref{w9}) holds.
Hence, we have to prove that $\nu_{1;+} = \nu_{1;-}$ implies
$\nu_{n;+} = \nu_{n;-}$ for all $n \in \mathbb{N}$. To this end we
consider the Wasserstein distance between the pairs $\nu_{n;\pm}$
\begin{equation} \label{u1}
R\left[\nu_{n ,+}, \nu_{n, -}\right] = \inf \int_{\mathbb{R}^{2n}}
|x - x'| P({\rm d}x, {\rm d}x'),
\end{equation}
where the infimum is taken over the set $\Pi^{(2n)}_{\pm}$ of all
probability measures on $(\mathbb{R}^{2n},
\mathcal{B}(\mathbb{R}^{2n}))$ which marginal distributions are
$\nu_{n ,\pm}$, see \cite{Du}. The relation (\ref{w3}) defines an
order on $\mathcal{P}(\mathit{\Omega}^{\rm t})$ and hence on
$\mathcal{P}(\mathbb{R}^n)$. In this sense $\nu_{n , -}$ is
dominated by its counterpart. Then by Strassen's theorem (Theorem
2.4 in \cite{Li}), there exists $\tilde{P}\in \Pi^{(2n)}_{\pm}$ such
that
\[
\tilde{P} \left(\left\{ (x , x') \in \mathbb{R}^{2n} \ | \ x \geq x'
\right\}\right) = 1.
\]
Thereby,
\begin{eqnarray*}
R\left[\nu_{n ,+ }, \nu_{n ,-}\right] & \leq &
\int_{\mathbb{R}^{2n}} |x - x'| \tilde{P}({\rm d}x, {\rm d}x'),\\
&\leq & \sum_{i =1}^n \int_{\mathbb{R}^n} x^{(i)}\left[ \nu_{n
,+}({\rm d}x) - \nu_{n ,-}({\rm d}x)\right] \\ & = & \sum_{i =1}^n
\int_{\mathbb{R}^n} x^{(i)}\left[ \nu_{1 ,+}({\rm d}x) - \nu_{1
,-}({\rm d}x)\right],
\end{eqnarray*}
which completes the proof.
\end{proof}
 \vskip.1cm \noindent
Theorem \ref{4tm} and the lemma just proven have the following
\begin{corollary} \label{domco}
If $V_\ell (x) = V_\ell (-x)$ for all $\ell$, the set
$\mathcal{G}^{\rm t}$ is a singleton if and only if $\mu_+
(\omega_\ell (0)) = 0$ for all $\ell$.
\end{corollary}

\subsection{Reference models}

\label{rfss} We shall prove Theorems \ref{phtm}, \ref{5tm} by
comparing our model with two reference models, defined as follows.
Let $J$ and $V$ be the same as in (\ref{p1}) and (\ref{ub})
respectively. For $\Lambda \Subset \mathbb{L}$, we set (c.f.,
(\ref{sch}))
\begin{equation} \label{rf1}
H^{\rm low}_{\Lambda} = \sum_{\ell \in \Lambda}\left[ H_\ell^{\rm
har} + V(x_\ell)\right] - \frac{1}{2} \sum_{\ell, \ell'\in \Lambda}
J \epsilon_{\ell \ell'} x_\ell x_{\ell'},
\end{equation}
where $H_\ell^{\rm har}$ is given by (\ref{16e}) and $\epsilon_{\ell
\ell'} = 1$ if $|\ell - \ell'|=1$ and $\epsilon_{\ell \ell'} = 0$
otherwise. Next, for $\Lambda \Subset \mathbb{L}$, we set
\begin{equation} \label{rf2}
H^{\rm upp}_{\Lambda} = \sum_{\ell \in \Lambda} \left[H_\ell^{\rm
har} + v(x_\ell^2) \right]- \frac{1}{2} \sum_{\ell , \ell'} J_{\ell
\ell'} x_\ell x_{\ell'} = \sum_{\ell \in \Lambda} \tilde{H}_\ell -
\frac{1}{2} \sum_{\ell , \ell'} J_{\ell \ell'} x_\ell x_{\ell'},
\end{equation}
where $\tilde{H}_\ell$ is defined by (\ref{2}) and the interaction
intensities $J_{\ell \ell'}$ are the same as in (\ref{sch}). Since
both these models are particular cases of the model we consider,
their sets of Euclidean Gibbs measures have the properties
established by Theorems \ref{1tm} -- \ref{4tm}. By $\mu^{\rm
low}_{\pm}$, $\mu^{\rm upp}_{\pm}$ we denote the corresponding
extreme elements.
\begin{remark} \label{convexrm}
The self-interaction potentials of both reference models have the
form (\ref{w3a}) with the zero external field $h_\ell =0$ and the
functions $v_\ell$ being convex. Hence, they obey the conditions of
all the statements of subsection \ref{ss5.1}. The $low$-reference
model is translation invariant. The $upp$-reference model is
translation invariant if $J_{\ell \ell'}$ are invariant with respect
to the translations of $\mathcal{L}$.
\end{remark}
\begin{lemma} \label{rflm}
For every $\ell$, it follows that
\begin{equation} \label{rf3}
\mu^{\rm low}_{+} (\omega_\ell (0)) \leq \mu_{+} (\omega_\ell (0))
\leq \mu^{\rm upp}_{+} (\omega_\ell (0)).
\end{equation}
\end{lemma}
\begin{proof}
By (\ref{rf41}) we have that for any $\mathcal{L}$,
\begin{equation} \label{soc1}
\int_{\mathit{\Omega}}\omega_\ell (\tau)\mu_{\pm}({\rm d}\omega) =
\lim_{ \mathcal{L}}\int_{\mathit{\Omega}}\omega_\ell
(\tau)\pi_\Lambda ({\rm d}\omega|\pm \xi^{\ell_0}), \ \ {\rm for} \
{\rm all} \ \tau\in [0,\beta].
\end{equation}
Thus, the proof will be done if we show that for all $\Lambda
\Subset \mathbb{L}$ and $\ell \in \Lambda$,
\begin{equation} \label{rf4}
\pi_{\Lambda}^{\rm low} (\omega_\ell (0)|\xi^{\ell_0}) \leq
\pi_{\Lambda} (\omega_\ell (0)|\xi^{\ell_0}) \leq \pi_{\Lambda}^{\rm
upp} (\omega_\ell (0)|\xi^{\ell_0}),
\end{equation}
where $\pi_{\Lambda}^{\rm low}$ , $\pi_{\Lambda}^{\rm upp}$ are the
kernels for the reference models. First we prove the left-hand
inequality in (\ref{rf4}). For given $\Lambda \Subset \mathbb{L}$
and $t, s \in [0, 1]$, we introduce
\begin{eqnarray} \label{rf5}
& & \varpi^{(t, s)}_\Lambda ({\rm d}\omega_\Lambda ) = \frac{1}{Y(t,
s)} \exp\left( \frac{1}{2} \sum_{\ell , \ell' \in \Lambda} J
\epsilon_{\ell \ell'} (\omega_\ell , \omega_{\ell'})_{L^2_\beta} +
\sum_{\ell \in \Lambda}(\omega_\ell, \eta^{\ell_0, s}_\ell)_{L^2_\beta} \right. \\
& & \qquad - \sum_{\ell \in \Lambda} \int_0^\beta V(\omega_\ell
(\tau)){\rm d}\tau + \frac{s}{2} \sum_{\ell , \ell' \in
\Lambda}\left[J_{\ell \ell'} -   J \epsilon_{\ell
\ell'}\right](\omega_\ell , \omega_{\ell'})_{L^2_\beta} \nonumber
\\& & \qquad - \left. t \sum_{\ell \in \Lambda} \int_0^\beta \left[V_\ell ( \omega_\ell
(\tau)) - V ( \omega_\ell (\tau)) \right]{\rm d}\tau \right)
\chi_\Lambda ({\rm d}\omega_\Lambda), \nonumber
\end{eqnarray}
where, see (\ref{v3}),
\begin{eqnarray} \label{rf6}
\eta^{\ell_0, s}_\ell (\tau) & \stackrel{\rm def}{=} & \sum_{\ell'
\in \Lambda^c} J \epsilon_{\ell \ell'} \xi^{\ell_0}_{\ell'} (\tau) \\
& + & s \sum_{\ell' \in \Lambda^c}\left[ J_{\ell \ell'} - J
\epsilon_{\ell \ell'}\right] \xi^{\ell_0}_{\ell'} (\tau) \geq
\sum_{\ell' \in \Lambda^c} J \epsilon_{\ell \ell'}
\xi^{\ell_0}_{\ell'} (\tau)
> 0, \nonumber
\end{eqnarray}
which in fact is independent of $\tau$, and
\begin{eqnarray*}
& & Y(t, s) = \int_{\mathit{\Omega}_\Lambda} \exp\left( \frac{1}{2}
\sum_{\ell , \ell' \in \Lambda} J \epsilon_{\ell \ell'} (\omega_\ell
, \omega_{\ell'})_{L^2_\beta} +
\sum_{\ell \in \Lambda}(\omega_\ell, \eta^{\ell_0, s}_\ell)_{L^2_\beta} \right. \\
& & \qquad - \sum_{\ell \in \Lambda} \int_0^\beta V(\omega_\ell
(\tau)){\rm d}\tau + \frac{s}{2} \sum_{\ell , \ell' \in
\Lambda}\left[J_{\ell \ell'} -   J \epsilon_{\ell
\ell'}\right](\omega_\ell , \omega_{\ell'})_{L^2_\beta} \nonumber
\\& & \qquad - \left. t \sum_{\ell \in \Lambda} \int_0^\beta \left[V_\ell ( \omega_\ell
(\tau)) - V ( \omega_\ell (\tau)) \right]{\rm d}\tau \right)
\chi_\Lambda ({\rm d}\omega_\Lambda). \nonumber
\end{eqnarray*}
Since the site-dependent `external field' (\ref{rf6}) is positive,
the moments of the measure (\ref{rf5}) obey the GKS inequalities.
Therefore, for any $\ell \in \Lambda$, the function
\begin{equation} \label{rf7}
\phi (t, s) = \varpi_\Lambda^{(t, s)} (\omega_\ell (0)), \quad t, s
\in [0, 1],
\end{equation}
is continuous and increasing in  both variables. Indeed, taking into
account (\ref{p1}), (\ref{ub}), and (\ref{ub1}), we get
\begin{eqnarray*}
 \frac{\partial }{\partial s} \phi (t, s) & = &
\sum_{\ell'\in\Lambda} \left[J_{\ell \ell'} - J \epsilon_{\ell
\ell'}  \right]\xi^{\ell_0}_{\ell'} (0) \\ & \times & \int_0^\beta
\left\{ \varpi_\Lambda^{(t, s)} \left[ \omega_\ell (0)
\omega_{\ell'}(\tau)\right] - \varpi_\Lambda^{(t,s)} \left[
\omega_\ell (0) \right] \cdot \varpi_\Lambda^{(t, s)} \left[
\omega_{\ell'}(\tau)\right] \right\} {\rm d}\tau  \\
  & + &
 \frac{1}{2}\sum_{\ell_1 , \ell_2 \in \Lambda} \left[J_{\ell_1 \ell_2} - J
 \epsilon_{\ell_1
\ell_2}  \right]\left\{\varpi_\Lambda^{(t, s)} \left[
\omega_{\ell}(0) (\omega_{\ell_1}, \omega_{\ell_2})_{L^2_\beta}
\right] \right. \\ & - & \left. \varpi_\Lambda^{(t, s)} \left[
\omega_{\ell}(0)  \right] \cdot \varpi_\Lambda^{(t, s)} \left[
(\omega_{\ell_1}, \omega_{\ell_2})_{L^2_\beta} \right] \right\} \geq
0, \\ \frac{\partial }{\partial t} \phi (t, s) & = & \sum_{\ell' \in
\Lambda} \int_0^\beta \left\{\varpi_\Lambda^{(t, s)} \left(
\omega_{\ell}(0) \cdot \left[ V(\omega_{\ell'} (\tau)) - V_{\ell'}
(\omega_{\ell'} (\tau))\right] \right) \right.
\\ & - & \left. \varpi_\Lambda^{(t, s)} \left[
\omega_{\ell}(0)  \right] \cdot \varpi_\Lambda^{(t, s)} \left[
V(\omega_{\ell'} (\tau)) - V_{\ell'} (\omega_{\ell'} (\tau)) \right]
\right\}{\rm d}\tau \geq 0.
\end{eqnarray*}
But by (\ref{rf5}) and (\ref{rf7})
\[ \phi (0,0) = \pi_\Lambda^{\rm low} (\omega_\ell (0)), \quad \phi (1,1)
 = \pi_\Lambda (\omega_\ell (0)),
 \]
 which proves the left-hand inequality in (\ref{rf4}). To prove
 the right-hand one we have to take the measure (\ref{rf5}) with $s=1$ and
 $v(x^2_\ell)$ instead of $V(x_\ell)$ and repeat the above steps
 taking into account
 (\ref{49}).
\end{proof}
\begin{corollary}[Comparison Criterion]  \label{compco}
The model considered undergoes a phase transition if the ${\rm
low}$-reference model does so. The uniqueness of tempered Euclidean
Gibbs measures of the ${\rm upp}$-reference model implies that
$|\mathcal{G}^{\rm t}|=1$.
\end{corollary}

\subsection{Estimates for pair correlation functions}
\label{sss5.3}

 For $\Delta \subset \Lambda$, $\ell , \ell'\in \Lambda$,
$\tau , \tau' \in [0, \beta]$, and $t \in [ 0, 1]$, let us consider
\begin{equation} \label{w4zz}
Q^\Lambda_{\ell \ell'} (\tau , \tau'|\Delta, t) =
\int_{\mathit{\Omega}_\Lambda} \omega_\ell (\tau) \omega_{\ell'}
(\tau') \varpi_{\Lambda, \Delta}^{(t)} ({\rm d}\omega_\Lambda),
\end{equation}
where this time we have set
\begin{eqnarray} \label{w4z}
& & \varpi_{\Lambda, \Delta}^{(t)} ({\rm d}\omega_\Lambda)
  = \frac{1}{Y_{\Lambda, \Delta}(t)} \exp \left\{ \frac{1}{2} \sum_{\ell_1 ,
\ell_2 \in \Lambda \setminus \Delta}J_{\ell_1 \ell_2}
(\omega_{\ell_1}, \omega_{\ell_2})_{L^2_\beta}\right.
\\
& & \quad   + t \left( \sum_{\ell_1 \in \Delta} \sum_{\ell_2 \in
\Lambda \setminus \Delta}J_{\ell_1 \ell_2} (\omega_{\ell_1},
\omega_{\ell_2})_{L^2_\beta} +  \frac{1}{2} \sum_{\ell_1 , \ell_2
\in \Delta}J_{\ell_1 \ell_2} (\omega_{\ell_1},
\omega_{\ell_2})_{L^2_\beta} \right) \nonumber \\& & \quad \left.  -
\sum_{\ell \in \Lambda}\int_0^\beta V_\ell (\omega_\ell (\tau)){\rm
d}\tau \right\}\chi_\Lambda
({\rm d}\omega_\Lambda), \nonumber\\
 & &  Y_{\Lambda , \Delta}(t) =
 \int_{\mathit{\Omega}_\Lambda} \exp \left\{ \frac{1}{2} \sum_{\ell_1 ,
\ell_2 \in \Lambda \setminus \Delta}J_{\ell_1 \ell_2}
(\omega_{\ell_1}, \omega_{\ell_2})_{L^2_\beta}\right.
\nonumber \\
& & \quad   + t \left( \sum_{\ell_1 \in \Delta} \sum_{\ell_2 \in
\Lambda \setminus \Delta}J_{\ell_1 \ell_2} (\omega_{\ell_1},
\omega_{\ell_2})_{L^2_\beta} +  \frac{1}{2} \sum_{\ell_1 , \ell_2
\in \Delta}J_{\ell_1 \ell_2} (\omega_{\ell_1},
\omega_{\ell_2})_{L^2_\beta} \right) \nonumber \\& & \quad \left.  -
\sum_{\ell \in \Lambda}\int_0^\beta V_\ell (\omega_\ell (\tau)){\rm
d}\tau \right\}\chi_\Lambda ({\rm d}\omega_\Lambda). \nonumber
\end{eqnarray}
By literal repetition of the arguments used for proving Lemma
\ref{rflm} one proves the following
\begin{proposition} \label{wzlm}
The above $Q^\Lambda_{\ell \ell'} (\tau , \tau'|\Delta , t)$ is an
increasing continuous function of $t\in [0,1]$.
\end{proposition}
\begin{corollary} \label{wzco}
Let the conditions of Proposition \ref{gks} be satisfied. Then for
any pair $\Lambda \subset \Lambda ' \Subset \mathbb{L}$ and for all
$\tau$ and $\ell$, the functions (\ref{w3}) obey the estimate
\begin{equation} \label{w4zy}
K_{\ell \ell'}^\Lambda (\tau , \tau'|0) \leq K_{\ell
\ell'}^{\Lambda'} (\tau , \tau'|0),
\end{equation}
which holds for all $\ell , \ell' \in \Lambda$ and $\tau , \tau'
\in [0,\beta]$.
\end{corollary}
\noindent Now we obtain bounds for the correlation functions of the
reference models for a one-point $\Lambda = \{\ell\}$. Set
\begin{equation} \label{rf8}
{K}^{\rm upp}_\ell (\tau , \tau') = \pi_{\ell}^{\rm upp}
(\omega_\ell (\tau) \omega_\ell (\tau')|0), \quad {K}^{\rm low}_\ell
(\tau , \tau') = \pi_{\ell}^{\rm low} (\omega_\ell (\tau)
\omega_\ell (\tau')|0),
\end{equation}
Both these functions are independent of $\ell$.
 We recall that the parameter $\mathit{\Delta}$
was defined by (\ref{51}).
\begin{lemma} \label{dtplm}
For every $\beta$, it follows that
\begin{equation} \label{w70}
K^{\rm upp}_\ell \ \stackrel{\rm def}{=} \ \int_0^\beta {K}^{\rm
upp}_\ell (\tau , \tau') {\rm d}\tau \leq 1/ m \mathit{\Delta}^2.
\end{equation}
\end{lemma}
\begin{proof}
In view of (\ref{mats1}) the above integral in independent of
$\tau$. By (\ref{mats}) and (\ref{mul10})
\begin{equation} \label{z4}
{K}^{\rm upp}_\ell = \frac{1}{\tilde{Z}_\ell} \int_0^\beta {\rm
trace} \left\{ x_\ell e^{-\tau \tilde{H}_\ell} x_\ell e^{-(\beta
-\tau) \tilde{H}_\ell}\right\}{\rm d}\tau , \quad \tilde{Z}_\ell =
{\rm trace}[ e^{- \beta\tilde{H}_\ell}],
\end{equation}
where the Schr\"odinger operator $\tilde{H}$ was defined in
(\ref{2}). Its spectrum $\{{E}_n\}_{n \in \mathbb{N}}$ determines by
(\ref{51}) the parameter $\mathit{\Delta}$. Integrating in
(\ref{z4}) we get
\begin{eqnarray} \label{w8}
{K}^{\rm upp}_\ell & = &\frac{1}{\tilde{Z}_\ell} \sum_{n, n'\in
\mathbb{N}_0, \ n\neq n'} \left\vert (\psi_n , x_\ell
\psi_{n'})_{L^2(\mathbb{R})} \right\vert^2 \frac{(E_{n} -
E_{n'})(e^{-\beta E_{n'}}- e^{-\beta E_{n}})}{(E_{n} - E_{n'})^2}
\nonumber \\ &\leq & \frac{1}{\tilde{Z}_\ell}\cdot
\frac{1}{\mathit{\Delta}^2}\sum_{n, n'\in \mathbb{N}_0} \left\vert
(\psi_n , x_\ell \psi_{n'})_{L^2(\mathbb{R})} \right\vert^2(E_{n} -
E_{n'})(e^{-\beta E_{n'}}- e^{-\beta E_{n}})\nonumber
\\ & = &\frac{1}{\mathit{\Delta}^2} \cdot\frac{1}{\tilde{Z}_\ell} {\rm
trace} \left\{\left[x_\ell , \left[\tilde{H}_\ell, x_\ell\right]
\right] e^{-\beta \tilde{H}_\ell}\right\} = \frac{1}{m
\mathit{\Delta}^2} ,
\end{eqnarray}
where $\psi_n$, $n \in \mathbb{N}_0$ are the eigenfunctions of
$\tilde{H}_\ell$ and $[\cdot, \cdot]$ stands for commutator.
\end{proof}
\noindent For the functions $K_{\ell}^{\rm low}$, a representation
like (\ref{z4}) is obtained by means of the following Schr\"odinger
operator
\begin{equation} \label{z4z}
\hat{H}_\ell =  H_\ell^{\rm har} + V(x_\ell) = - \frac{1}{2m}
\left(\frac{\partial}{\partial x_\ell} \right)^2 + \frac{a}{2}
x_\ell^2 + V(x_\ell),
\end{equation}
where $m$ and $a$ are the same as in (\ref{2}) but $V$ is given by
(\ref{ub}). Thereby,
\begin{equation} \label{z4y}
{K}^{\rm low}_\ell (0,0) = {\rm trace}[ x_\ell^2 \exp(-\beta
\hat{H}_\ell)]/{\rm trace}[\exp(-\beta \hat{H}_\ell)] \
\stackrel{\rm def}{=} \ \hat{\varrho} (x_\ell^2).
\end{equation}
\begin{lemma} \label{lblm}
Let  $t_*$ be the solution of (\ref{p4}). Then ${K}^{\rm low}_\ell
(0,0) \geq t_*$.
\end{lemma}
\begin{proof}
By Bogoliubov's inequality (see e.g., \cite{Simonl}), it follows
that
\[
\hat{\varrho}_\ell \left([p_\ell , [\hat{H}_\ell , p_\ell]]\right)
\geq 0, \quad p_\ell = - \sqrt{-1} \frac{\partial}{\partial x_\ell},
\]
which by (\ref{ub}), (\ref{p3}) yields
\begin{eqnarray*}
& & a + 2 b^{(1)} + \sum_{s=2}^r 2 s (2 s-1)b^{(s)}
\hat{\varrho}\left[ x^{2(s-1)}_\ell\right] \\ & & \quad =  a + 2
b^{(1)} + \sum_{s=2}^r 2 s (2 s-1)b^{(s)} \pi^{\rm low}_{\ell}
\left[\left(\omega_\ell (0)\right)^{2(s-1)}\right] \geq 0. \nonumber
\end{eqnarray*}
Now we use  the Gaussian domination inequality (\ref{vv1}) and
obtain  ${K}^{\rm low}_\ell \geq t_*$.
\end{proof}

\subsection{Proof of Theorem \ref{phtm}} \label{sss5.2}

In view of Corollary \ref{compco} we show that
\begin{equation} \label{rf10}
\mu_{+}^{\rm low}(\omega_\ell (0)) >0,
\end{equation}
if the conditions of Theorem \ref{phtm} are satisfied. Note that the
left-hand side of (\ref{rf10}) is independent of $\ell$ as the
measure is translation invariant. The proof of (\ref{rf10}) will be
based on  \cite{Kond}, see also Theorem 8.1 in \cite{AKKR}. Here the
translation invariance and reflection positivity of the reference
model are used to show that for $\beta>\beta_*$ the set of its
Euclidean Gibbs measures contains a nonergodic element with respect
to the group of translations of $\mathbb{L}$. This gives
non-uniqueness and hence (\ref{rf10}). To follow this line  we
construct periodic Euclidean Gibbs states by ntroducing (c.f.,
(\ref{19}))
\begin{equation} \label{mm7}
{I}^{\rm per}_\Lambda (\omega_\Lambda) = - \frac{J}{2} \sum_{\ell ,
\ell' \in \Lambda}\epsilon_{\ell\ell'}^\Lambda (\omega_\ell ,
\omega_{\ell'})_{L^2_\beta} + \sum_{\ell\in \Lambda} \int_0^\beta V
\left(\omega_\ell (\tau) \right) {\rm d}\tau,
\end{equation}
where
\begin{equation} \label{mm8}
 \Lambda= (-L, L]^d\bigcap
\mathbb{L}, \ \ L\in \mathbb{N},
\end{equation}
and $\epsilon_{\ell\ell'}^\Lambda = 1$ if $|\ell - \ell'|_\Lambda
=1$ and $\epsilon_{\ell\ell'}^\Lambda = 0$ otherwise. Here $|\ell -
\ell'|_\Lambda= [|\ell_1 - \ell'_1|^2_L + \cdots + |\ell_d -
\ell'_d|^2_L]^{1/2}$ and $|\ell_j - \ell'_j|_L = \min\{|\ell_j -
\ell'_j|; L - |\ell_j - \ell'_j| \}$, $j = 1 , \dots , d$. Clearly,
${I}^{\rm per}_\Lambda$ is invariant with respect to the
translations of the torus which one obtains by identifying the
opposite walls of the box (\ref{mm8}). The energy functional
${I}^{\rm per}_\Lambda$ corresponds to the following periodic
Schr\"odinger operator
\begin{equation} \label{schp}
{H}^{\rm per}_\Lambda =  \sum_{\ell \in \Lambda} \left[H^{\rm
har}_\ell + V( x_\ell)\right]- \frac{J}{2} \sum_{\ell ,\ell' \in
\Lambda} \epsilon_{\ell\ell'}^\Lambda x_\ell x_{\ell'},
\end{equation}
in the same sense as $I_\Lambda$ given by (\ref{19}) corresponds to
$H_\Lambda$ given by (\ref{sch}). Now we introduce the periodic
kernels (c.f., (\ref{34}))
\begin{equation} \label{mm9}
{\pi}_\Lambda^{\rm per} (B) = \frac{1}{Z^{\rm
per}_\Lambda}\int_{\mathit{\Omega}}\exp\left[- {I}^{\rm
per}_\Lambda (\omega_\Lambda) \right]\mathbb{I}_B (\omega_\Lambda
\times 0_{\Lambda^c} ) \chi_\Lambda({\rm d}\omega_\Lambda), \ \
B\in \mathcal{B}(\mathit{\Omega}),
\end{equation}
\[
Z^{\rm per}_\Lambda = \int_{\mathit{\Omega}}\exp\left[- {I}^{\rm
per}_\Lambda (\omega_\Lambda) \right] \chi_\Lambda({\rm
d}\omega_\Lambda).
\]
Thereby, for every box $\Lambda$, the above ${\pi}_\Lambda^{\rm
per}$ is a probability measure on $\mathit{\Omega}^{\rm t}$. By
$\mathcal{L}_{\rm box}$ we denote the sequence of boxes (\ref{mm8})
indexed by $L\in \mathbb{N}$. For a given $\alpha \in \mathcal{I}$,
let us choose $\vartheta, \varkappa
>0$ such that the estimate (\ref{67d}) holds.
\begin{lemma} \label{mmlm}
For every box $\Lambda$, $\alpha \in \mathcal{I}$, and $\sigma\in
(0, 1/2)$, the measure ${\pi}_\Lambda^{\rm per}$ obeys the estimate
\begin{equation} \label{ees}
\int_{\mathit{\Omega}} \|\omega\|_{\alpha, \sigma}^2
\pi_\Lambda^{\rm per} ({\rm d}\omega) \leq C_{\ref{ees}}.
\end{equation}
Thereby, the sequence $\{{\pi}_\Lambda^{\rm per}\}_{\Lambda \in
\mathcal{L}_{\rm box}}$ is $\mathcal{W}^{\rm t}$-relatively compact.
\end{lemma}
\begin{proof}
For $\ell\in \Lambda$ such that $\{\ell' \in \mathbb{L}  \ | \ |\ell
- \ell'|=1\} \subset \Lambda$, we set $\Delta_\ell =
\mathbb{L}\setminus \{\ell \}$. Then let $\nu_\ell^\Lambda$ be the
projection of ${\pi}_\Lambda^{\rm per}$ onto
$\mathcal{B}(\mathit{\Omega}_{\Delta_\ell})$. Let also $\nu_\ell
(\cdot|\xi)$, $\xi\in \mathit{\Omega}$ be the following probability
measure on the single-spin space $\mathit{\Omega}_{\{\ell\}} =
C_\beta$
\begin{equation} \label{sig}
\nu_\ell ({\rm d}\omega_\ell |\xi )  = \frac{1}{N_\ell (\xi)}
\exp\left\{J \sum_{\ell'}\epsilon_{\ell \ell'}(\omega_\ell,
\xi_{\ell'})_{L^2_\beta} -  \int_0^\beta V( \omega_\ell (\tau)){\rm
d}\tau \right\} \chi({\rm d}\omega_\ell).
\end{equation}
Then (c.f., (\ref{35})) desintegrating $\pi_\Lambda^{\rm per}$ we
get
\begin{equation} \label{mm11}
{\pi}_\Lambda^{\rm per} ({\rm d}\omega) =\nu_\ell ({\rm
d}\omega_\ell |\omega_{\Delta_\ell} )\nu_\ell^\Lambda ({\rm
d}\omega_{\Delta_\ell}).
\end{equation}
As in Lemma \ref{4lm} and Corollary \ref{1co} one proves that the
measure $\nu_\ell(\cdot |\xi)$ obeys
\[
\int_{C_\beta}\exp\left\{\lambda_\sigma |\omega_\ell
|^2_{C_\beta^\sigma} + \varkappa |\omega_\ell |^2_{L_\beta^2}
\right\}\nu_\ell ({\rm d}\omega_\ell|\omega_{\Delta_\ell} ) \leq
\exp\left\{C_{\ref{53}} + \vartheta J \sum_{\ell' } \epsilon_{\ell
\ell'}|\omega_{\ell'}|^2_{L^2_\beta} \right\},
\]
where $\lambda_\sigma$, $\varkappa$, and $\vartheta$ are as in
(\ref{53}), (\ref{55}). Now we integrate both sides of this
inequality with respect to $\nu_\ell^\Lambda$ and get, c.f.,
(\ref{66}), (\ref{67d})
\[
n_\ell^{\rm per}(\Lambda) \ \stackrel{\rm def}{=} \  \log\left\{
\int_{\mathit{\Omega}}\exp [\lambda_\sigma |\omega_\ell
|^2_{C_\beta^\sigma} + \varkappa |\omega_\ell
|^2_{L_\beta^2}]\pi_\Lambda^{\rm per}({\rm d}\omega) \right\} \leq
C_{\ref{66a}}.
\]
Then the estimate (\ref{ees}) is obtained in the same way as
(\ref{61a}) was proven. The relative
$\mathcal{W}_\alpha$-compactness of $\{\pi_\Lambda^{\rm
per}\}_{\Lambda \in \mathcal{L}_{\rm per}}$ follows from (\ref{ees})
and the compactness of the embeddings $\mathit{\Omega}_{\alpha ,
\sigma} \hookrightarrow \mathit{\Omega}_{\alpha'}$, $\alpha <
\alpha'$. The $\mathcal{W}^{\rm t}$-compactness is a consequence of
by Lemma \ref{tanlm}.
\end{proof}
\begin{lemma} \label{nmlm} Every $\mathcal{W}^{\rm t}$-accumulation
point $\mu^{\rm per}$ of the sequence $\{{\pi}_\Lambda^{\rm
per}\}_{\Lambda \in \mathcal{L}_{\rm per}}$ is a Euclidean Gibbs
measure of the ${\rm low}$-reference model.
\end{lemma}
\begin{proof}
Let $\mathcal{L}\subset \mathcal{L}_{\rm per}$ be the subsequence
along which $\{{\pi}_\Lambda^{\rm per}\}_{\Lambda \in \mathcal{L}}$
converges to $\mu^{\rm per} \in \mathcal{P}(\mathit{\Omega}^{\rm
t})$. Then $\{\nu_\ell^{\Lambda}\}_{\Lambda \in \mathcal{L}}$
converges to the projection of $\mu^{\rm per}$ on
$\mathcal{B}(\mathit{\Omega}_{\Delta_\ell})$. Employing the Feller
property (Lemma \ref{2lm}) we pass in (\ref{mm11}) to the limit
along this $\mathcal{L}$ and apply both its sides to a function $f
\in C_{\rm b}(\mathit{\Omega}^{\rm t})$. This yields that $\mu^{\rm
per}$ has the same one-point conditional distributions as the
Euclidean Gibbs measures of the reference model. But according to
Theorem 1.33 of \cite{Ge}, page 23, every Gibbs measure is uniquely
defined by its conditional distributions corresponding to one-point
sets $\Lambda = \{\ell \}$ only.
\end{proof}
Now we are at a position to prove that (\ref{rf10}) holds if $\beta>
\beta_*$. As the $low$-reference model is translation invariant, in
the case of uniqueness the unique element of the set of its
Euclidean Gibbs states should be ergodic. Periodic states, which one
obtains as accumulation points of the sequence $\{{\pi}_\Lambda^{\rm
per}\}_{\Lambda \in \mathcal{L}_{\rm box}}$, are automatically
translation invariant but they can be nonergodic.  By the von
Neumann ergodic theorem (page 244 of \cite{Simonl}), periodic
Euclidean Gibbs states are nonergodic if
\begin{equation} \label{sig7}
\lim\inf_{\mathcal{L}_{\rm box}} P_\Lambda (\beta) \ \stackrel{\rm
def}{=} \ P(\beta) >0,
\end{equation}
where
\begin{equation} \label{op}
P_\Lambda (\beta) = \int_{\mathit{\Omega}}
\left(\frac{1}{|\Lambda|}\sum_{\ell \in \Lambda}\omega_\ell (0)
\right)^2 {\pi}^{\rm per}_\Lambda ({\rm d}\omega).
\end{equation}
\begin{lemma} \label{phlem}
Let the assumptions of Theorem \ref{phtm} be satisfied and
$\beta_*$ be the solution of (\ref{cja}). Then for $\beta >
\beta_*$ the order parameter $P(\beta)$ is positive.
\end{lemma}
\begin{proof}
Here we mainly follow \cite{Kond}, see also Theorem 8.1 in
\cite{AKKR}.  By means of the infrared estimates one obtains (see
equation (8.13) in \cite{AKKR}) that for every box $\Lambda$
\begin{eqnarray} \label{sig9}
& & P(\beta) \geq \int_{\mathit{\Omega}} [\omega_\ell
(0)]^2{\pi}^{\rm per}_\Lambda ({\rm d}\omega) \\ & & \qquad \ \ -
\frac{1}{2 (2\pi)^d} \int_{(- \pi, \pi]^d} \left[2 m J
E(p)\right]^{-1/2} \coth\left(\beta \sqrt{J E(p)/2m} \right){\rm
d}p, \nonumber
\end{eqnarray}
where the function $E(p)$ is given by (\ref{cj}) whereas $m$ and $J$
are as in (\ref{cja}). Therefore, we have to find an appropriate
bound for the first term in (\ref{sig9}). For any $\ell$, one can
take the box $\Lambda$ such that the Euclidean distance from this
$\ell$ to $\Lambda^c$ be greater than $1$. Then by Corollary
\ref{wzco} and Lemma \ref{lblm} one gets
\[
\int_{\mathit{\Omega}} [\omega_\ell (0)]^2{\pi}^{\rm per}_\Lambda
({\rm d}\omega) \geq {K}^{\rm low}_\ell (0,0)\geq t_*.
\]
 Hence, for $\beta
>\beta_*$, the right-hand side of (\ref{sig9}) is positive.
\end{proof}
\vskip.1cm \noindent \textbf{Proof of Theorem \ref{phtm}:} By the
latter lemma the periodic state of the $low$-reference model is
nonergodic at $\beta>\beta_*$ yielding (\ref{rf10}).$\square$

\subsection{Proof of Theorem \ref{5tm}}

\label{ss5.2}
 By Corollary \ref{compco} it is enough to prove the uniqueness for
 the $upp$-reference model, which by Lemma \ref{bhpn} is equivalent
 to
\begin{equation} \label{rf20}
\mu^{\rm upp}_{+} (\omega_\ell (0)) = 0, \quad {\rm for} \ {\rm all}
\ \ \beta>0 \ \ {\rm and} \ \ \ell.
\end{equation}
Given $\Lambda \Subset \mathbb{L}$, we introduce the matrix
$(T^\Lambda_{\ell \ell'})_{\ell , \ell'\in \mathbb{L}}$ as follows.
We set $T^\Lambda_{\ell \ell'} = 0$ if either of $\ell , \ell'$
belongs to $\Lambda^c$. For $\ell , \ell' \in \Lambda$,
\begin{equation} \label{w10}
T^\Lambda_{\ell \ell'} = \sum_{\ell_1 \in \Lambda} J_{\ell \ell_1}
\int_0^\beta \pi_\Lambda^{\rm upp}\left[\omega_{\ell_1} (\tau)
\omega_{\ell'} (\tau') \left\vert 0 \right. \right]{\rm d}\tau'.
\end{equation}
By (\ref{mats1}) the above integral is independent of $\tau$.

\begin{lemma} \label{newlm}
If (\ref{52}) is satisfied, there exists $\alpha \in \mathcal{I}$,
such that for every $\Lambda \Subset \mathbb{L}$, the matrix
$(T^\Lambda_{\ell \ell'})_{\ell , \ell'\in \mathbb{L}}$ defines a
bounded operator in the Banach space $l^\infty (w_\alpha)$.
\end{lemma}
\begin{proof}
The proof will be based on a generalization of the method used in
\cite{AKKR02b} for proving Lemma 4.7. For $t\in [0, 1]$, let
$\varpi^{(t)}_{\Lambda}\in \mathcal{P}(\mathit{\Omega}_\Lambda))$ be
defined by (\ref{w4z}) with $\Delta = \Lambda$ and each $V_\ell
(\omega_\ell (\tau))$ replaced by $v ([\omega_\ell(\tau)]^2)$, where
the latter function is the same as in (\ref{2}). Then by (\ref{rf2})
\begin{equation} \label{rf30}
\varpi_\Lambda^{(0)} = \prod_{\ell \in \Lambda}\pi_{\ell}^{\rm upp}
(\cdot|0), \quad
 \varpi_\Lambda^{(1)} = \pi_\Lambda^{\rm
upp}(\cdot |0), \quad {\rm for} \ {\rm any} \ \ \Lambda \Subset
\mathbb{L}.
\end{equation}
Thereby, we set
\begin{equation} \label{w14a}
T^\Lambda_{\ell \ell'} (t) = \sum_{\ell_1} J_{\ell \ell_1}
\int_0^\beta \varpi_\Lambda^{(t)}\left[\omega_{\ell_1} (\tau)
\omega_{\ell'} (\tau')  \right]{\rm d}\tau' \quad t\in [0, 1].
\end{equation}
One can show that for every fixed $\ell, \ell'$, the above
$T^\Lambda_{\ell \ell'} (t)$ are differentiable on the interval
$t\in (0, 1)$ and continuous at its endpoints where (see
(\ref{w70}))
\begin{equation} \label{w15a}
 T^\Lambda_{\ell \ell'} (0) = J_{\ell \ell'}
K^{\rm upp}_{\ell'} \leq J_{\ell \ell'}/ m \mathit{\Delta}^2, \quad
\ T^\Lambda_{\ell \ell'}(1) = T^\Lambda_{\ell \ell'}.
\end{equation}
 Computing the derivative we get
\begin{eqnarray} \label{w16}
\frac{\partial }{\partial t} T_{\ell \ell'}^\Lambda (t) & = &
\frac{1}{2} \sum_{\ell_1, \ell_2, \ell_3 } J_{\ell \ell_1} J_{\ell_2
\ell_3} \int_0^\beta \int_0^\beta U^\Lambda_{\ell \ell' \ell_2
\ell_3} (t ,
\tau, \tau' , \tau_1 , \tau_1) {\rm d}\tau' {\rm d}\tau_1 \\
& + & \sum _{\ell_1 } T^\Lambda_{\ell \ell_1} (t) T^\Lambda_{\ell_1
\ell'} (t), \nonumber
\end{eqnarray}
where $U^\Lambda_{\ell \ell' \ell_1 \ell_2} (t , \tau, \tau' ,
\tau_1 , \tau_1)$ is  the Ursell function which obeys the estimate
(\ref{v2}) since the function $v$ is convex. Hence, except for the
trivial case $J_{\ell \ell'} \equiv 0$, the first term in
(\ref{w16}) is strictly negative. Bearing in mind (\ref{52}) let us
show that there exists $\alpha \in \mathcal{I}$ such that
\begin{equation} \label{w16a}
\hat{J}_0 < \hat{J}_\alpha < m \mathit{\Delta}^2.
\end{equation}
Here the cases of $J_{\ell \ell'}$ obeying (\ref{24a}) and the
power-like decaying $J_{\ell \ell'}$ should be considered
separately. In the first case we have $\mathcal{I} = (0,
\overline{\alpha})$ and the above $\alpha$ exist in view of
\begin{equation} \label{26m}
\lim_{\alpha \rightarrow 0+} \hat{J}_\alpha = \hat{J}_0,
\end{equation}
which readily follows from (\ref{24a}), (\ref{24b}).
 In the second case the weights are defined by (\ref{24d}) with a
 positive $\varepsilon$, which we are going to use to ensure
 (\ref{w16a}). To indicate the dependence of $\hat{J}_\alpha$ on
 $\varepsilon$ we write
 $\hat{J}_\alpha^{(\varepsilon)}$. Simple calculations yield
 \[
0 < \hat{J}^{(\varepsilon)}_\alpha - \hat{J}_0 \leq \varepsilon
\alpha d \hat{J}_\alpha^{(1)}.
\]
Thereby, we fix $\alpha \in \mathcal{I}$ and choose $\varepsilon$ to
obey $\varepsilon < m \mathit{\Delta}^2 / \alpha d
\hat{J}_\alpha^{(1)}$. This yields (\ref{w16a}).

Let us consider the following Cauchy problem
\begin{equation} \label{w17}
\frac{\partial }{\partial t} L_{\ell \ell'} (t) = \sum _{\ell_1}
L_{\ell \ell_1} (t)  L_{\ell_1 \ell'} (t), \quad L_{\ell \ell'} (0)
= \lambda J_{\ell \ell'}, \ \ \ell, \ell' \in \mathbb{L},
\end{equation}
where $\lambda \in (1/ m \mathit{\Delta}^2 , 1/\hat{J}_\alpha) $,
with $\alpha \in \mathcal{I}$ chosen to obey (\ref{w16a}). For such
$\alpha$, one can solve the problem (\ref{w17}) in the space
$l^\infty(w_\alpha)$ (see Remark \ref{1rm}) and obtain
\begin{equation} \label{w17a}
L(t) = \lambda J \left[I - \lambda t J \right]^{-1}, \quad
\|L(t)\|_{l^\infty (w_\alpha)} \leq \frac{\lambda \hat{J}_\alpha}{1
- \lambda t \hat{J}_\alpha}.
\end{equation}
where $I$ is the identity operator. Now let us compare (\ref{w16})
and (\ref{w17}) considering the former expression as a differential
equation subject to the initial condition (\ref{w15a}). Since the
first term in (\ref{w16}) is negative, one can apply Theorem V, page
65 of \cite{Walter} and obtain $T_{\ell \ell'}^\Lambda < L_{\ell
\ell'} (1)$, which in view of (\ref{w17a}) yields the proof.
\end{proof} \vskip.1cm \noindent
\textbf{Proof of Theorem \ref{5tm}:} \ For $\ell, \ell_0$, $\Lambda
\Subset \mathbb{L}$, such that $\ell \in \Lambda$, and $t\in [0,
1]$, we set
\begin{equation}\label{w19}
\psi_\Lambda (t) = \int_{\mathit{\Omega}} \omega_{\ell} (0)\pi^{\rm
upp}_\Lambda ({\rm d}\omega | t \xi^{\ell_0}) ,
\end{equation}
where $\xi^{\ell_0}$ is the same as in (\ref{v3}). The function
$\psi_\Lambda$ is obviously differentiable on the interval $t \in
(-1, 1)$ and continuous at its endpoints.
 Then
\begin{equation} \label{w20}
0 \leq \psi_\Lambda (1) \leq  \sup_{t \in [0, 1]} \psi_\Lambda' (t).
\end{equation}
The derivative is
\begin{equation} \label{w21} \psi_\Lambda' (t) =
\sum_{\ell_1 \in \Lambda , \  \ell_2 \in \Lambda^c} J_{\ell \ell_1}
\int_0^\beta \pi_\Lambda^{\rm upp}\left[\omega_{\ell_1}(0) \omega_{
\ell_2} ( \tau) \left\vert t \xi^{\ell_0} \right. \right]
\eta_{\ell_2} {\rm d}\tau,
\end{equation}
where the `external field' $\eta_{\ell'} = \left[b \log(1 + |\ell' -
\ell_0|)\right]^{1/2}$ is positive at each site. Thus, we may use
(\ref{w5}) and obtain
\begin{equation} \label{w22}
\psi_\Lambda' (t) \leq \sum_{l' \in \Lambda^c} T^\Lambda_{\ell
\ell'} \eta_{\ell'}.
\end{equation}
By Assumption \ref{a2} (b), $\eta \in l^\infty (w_\alpha)$ with any
$\alpha >0$, then employing Lemma \ref{newlm}, the estimate
(\ref{w17a}) in particular, we conclude that the right-hand side of
(\ref{w22}) tends to zero as $\Lambda \nearrow \mathbb{L}$, which by
(\ref{soc1}) and (\ref{w19}), (\ref{w20}) yields (\ref{rf20}).
$\square$

\section{Uniqueness at Nonzero External Field}
\label{7s}

In  statistical mechanics phase transitions may be associated with
nonanalyticity of thermodynamic characteristics considered as
functions of the external field $h$. In special cases one can
oversee at which values of $h$ this nonanaliticity can occur. The
Lee-Yang theorem states that the only such value  is $h=0$; hence,
no phase transitions can occur at nonzero $h$. In the theory of
classical lattice models these arguments were applied in
\cite{LML,LPe,LP}. We refer also to sections 4.5, 4.6 in \cite{GJ}
and sections IX.3 -- IX.5 in \cite{Si74} where applications of such
arguments in quantum field theory and classical statistical
mechanics are discussed.

In the case of lattice models with the single-spin space
$\mathbb{R}$ the validity of the Lee-Yang theorem depends on the
properties of the self-interaction potentials. For the polynomials
$V(x) = x^4 + a x^2$, $a\in \mathbb{R}$, the Lee-Yang theorem holds,
see e.g., Theorem IX.15 on page 342 in \cite{Si74}. But no other
examples of this kind were known, see the discussion on page 71 in
\cite{GJ}. Below we give a sufficient condition for self-interaction
potentials to have the corresponding property and discuss some
examples. Here we use the family $\mathcal{F}_{\rm Laguerre}$
defined by (\ref{m1}). We also prove a number of lemmas, which allow
us to employ the arguments based on the Lee-Yang theorem to our
quantum model and hence to prove Theorem \ref{6tm}.

\subsection{The Lee-Yang property}
\label{ss5.3}

Recall that the elements of $\mathcal{F}_{\rm Laguerre}$ can be
continued to entire functions $\varphi:\mathbb{C}\rightarrow
\mathbb{C}$, which have no zeros outside of $(-\infty, 0]$.
\begin{definition} \label{lydf}
A probability measure $\nu$ on the real line is said to have the
Lee-Yang property if there exists $\varphi\in \mathcal{F}_{\rm
Laguerre}$ such that
\[
\int_\mathbb{R} \exp(xy)\nu({\rm d}y) = \varphi (x^2).
\]
\end{definition}
\noindent In \cite{Koz1}, see also Theorem 2.3 in \cite{KOU}, the
following fact was proven.
\begin{proposition} \label{lypn}
Let the function $u:\mathbb{R} \rightarrow \mathbb{R}$ be such that
for a certain $b\geq 0$, its derivative obeys the condition $b + u'
\in \mathcal{F}_{\rm Laguerre}$. Then the probability measure
\begin{equation} \label{w23}
\nu({\rm d}y) = C \exp[- u(y^2)]{\rm d}y,
\end{equation}
has the Lee-Yang property.
\end{proposition}
\vskip.1cm \noindent This statement gives a sufficient condition,
the lack of which was mentioned on page 71 of the book \cite{GJ}.
The example of a polynomial given there for which the corresponding
classical models undergo phase transitions at nonzero $h$,
 in our
notations is $u (t) = t^3 - 2t^2 + (\alpha + 1)t$, $\alpha >0$. It
certainly does not meet the condition of Proposition \ref{lypn}.
Turning to the model described by Theorem \ref{6tm} we note that,
for $ v(t)
 = t^3 + b^{(2)} t^2 + b^{(1)} t$, the function $u (t) = v (t) + a
 t/2$ obeys the conditions of Proposition \ref{lypn} if and only if
$b^{(2)} \geq 0$ and $b^{(1)} + a/2 \leq [b^{(2)}]^2/3$. Therefore,
 according to Theorem \ref{6tm}
 we have $|\mathcal{G}^{\rm t}|=1$ at $h \neq 0$ and
$2b^{(1)} + a < 0$, $b^{(2)} \geq 0$. On the other hand, for this
model, by Theorem \ref{phtm} one has a phase transition at $h=0$ and
the same coefficients of $v$.

Set
\begin{equation} \label{w24}
f(h^2) = \int_{\mathbb{R}^n} \exp\left[ h \sum_{i=1}^n x_i +
\sum_{i,j=1}^n M_{ij}x_i x_j \right]\prod_{i=1}^n \nu({\rm d}x_i),
\quad h \in \mathbb{R}.
\end{equation}
By Theorem 3.2 of \cite{LS}, we have the following
\begin{proposition} \label{ly1pn}
If in (\ref{w24}) $M_{ij} \geq 0$ for all $i,j = 1, \dots , n$, and
the measure $\nu$ is as in Proposition \ref{lypn}, then the function
$f$, if exists, belongs to $\mathcal{F}_{\rm Laguerre}$. It
certainly exists if $u'$ is not constant.
\end{proposition}
\noindent Now let the potential $V$  obey the conditions of Theorem
\ref{6tm}. Then the partition function $Z_\Lambda (0)$ (here $0$
means $\xi = 0$) given by (\ref{33}) is an even function of $h$.
Define
\begin{equation} \label{w25}
p_\Lambda (h) = \frac{1}{|\Lambda|}\log Z_\Lambda (0), \quad
 \varphi_\Lambda (h^2) = p_\Lambda (h) .
\end{equation}
\begin{lemma} \label{lylm}
If $V$  obeys the conditions of Theorem \ref{6tm}, the function
$\exp\left(|\Lambda|\varphi_\Lambda\right)$ belongs to
$\mathcal{F}_{\rm Laguerre}$.
\end{lemma}
\begin{proof}
With the help of the lattice approximation technique the function
$\exp\left(|\Lambda|\varphi_\Lambda\right)$ may be approximated by
$f_N$, $N\in\mathbb{N}$, having the form (\ref{w24}) with the
measures $\nu$ having of the form (\ref{w23}) with $u(t) = v(t) + a
t/2$, $v$ is as in (\ref{m2}), and non-negative $M_{ij}$ (see
Theorem 5.2 in \cite{AKKR}). For every $h\in \mathbb{R}$, $f_N (h^2)
\rightarrow \exp\left(|\Lambda|\varphi_\Lambda (h^2)\right)$ as $N
\rightarrow +\infty$. The entire functions $f_N$ are ridge, with the
ridge $[0, +\infty)$. For sequences of such functions, their
point-wise convergence on the ridge implies via the Vitali theorem
(see e.g., \cite{Si74}) the uniform convergence on compact subsets
of $\mathbb{C}$, which yields the property stated (for more details,
see \cite{KozHol,KW}).
\end{proof}

\subsection{Existence of pressure}
\label{ss5.4} The results obtained in this subsection are valid for
any translation invariant model with $\nu=1$, obeying Assumption
\ref{vas} and hence possessing the properties described by Theorems
\ref{1tm} -- \ref{httm}.

Along with (\ref{w25}) we introduce
\begin{equation} \label{c1}
p_\Lambda (h, \xi) = \frac{1}{|\Lambda|} \log Z_\Lambda (\xi), \quad
\xi \in \mathit{\Omega}^{\rm t}.
\end{equation}
Then $p_\Lambda (h) = p_\Lambda (h, 0)$. For $\mu \in
\mathcal{G}^{\rm t}$, we set
\begin{equation} \label{c2}
p^\mu_\Lambda (h) = \int_{\mathit{\Omega}}p_\Lambda (h , \xi)
\mu({\rm d}\xi).
\end{equation}
If for a cofinal sequence $\mathcal{L}$, the limit
\begin{equation} \label{c3}
p^\mu (h) \ \stackrel{\rm def}{=} \ \lim_{\mathcal{L}}p^\mu_\Lambda
(h),
\end{equation}
exists, we shall call it pressure in the Gibbs state $\mu$. We shall
also consider $\lim_{\mathcal{L}} p_\Lambda (h)$ with $p_\Lambda
(h)$ given by (\ref{w25}). To obtain such limits we impose certain
conditions on the sequences $\mathcal{L}$. Given $l = (l_1 , \dots
l_d)$, $l' = (l'_1 , \dots l'_d)\in \mathbb{L}= \mathbb{Z}^d$, such
that $l_j < l'_j$ for all $j=1, \dots , d$, we set
\begin{equation} \label{de1}
\Gamma = \{ \ell \in \mathbb{L} \ | \ l_j \leq  \ell_j \leq l'_j, \
\ {\rm for} \ {\rm all}\  j = 1 , \dots , d\}.
\end{equation}
For this parallelepiped, let $\mathfrak{G}(\Gamma)$ be the family of
all pair-wise disjoint translates of $\Gamma$ which covers
$\mathbb{L}$. Then for $\Lambda \Subset \mathbb{L}$, we set
$N_{-}(\Lambda|\Gamma)$ (respectively, $N_{+}(\Lambda|\Gamma)$) to
be the number of the elements of $\mathfrak{G}(\Gamma)$ which are
contained in $\Lambda$ (respectively, which have non-void
intersections with $\Lambda$). Then we introduce the following (see
\cite{Ru})
\begin{definition} \label{rdf}
Given $\mathcal{L}$ is a van Hove sequence if for every $\Gamma$,
\begin{equation} \label{de2}
(a) \ \  \lim_{\mathcal{L}} N_{-}(\Lambda |\Gamma) = +\infty; \quad
\quad (b) \ \ \lim_{\mathcal{L}}\left( N_{-}(\Lambda |\Gamma)/
N_{+}(\Lambda |\Gamma)\right)= 1.
\end{equation}
\end{definition}
Given $R>0$ and $\Lambda \Subset \mathbb{L}$, let $\partial^{+}_R
\Lambda$ be the set of all $\ell \in \Lambda^c$, such that ${\rm
dist}(\ell, \Lambda) \leq R$. Then for a van Hove sequence
$\mathcal{L}$ and any $R>0$, one has $\lim_{\mathcal{L}}
|\partial^{+}_R \Lambda|/|\Lambda| = 0$, yielding
\begin{equation} \label{c5}
\lim_{\mathcal{L}} \frac{1}{|\Lambda|}\sum_{\ell \in \Lambda, \ell'
\in \Lambda^c} J_{\ell \ell'} = 0.
\end{equation}
 The existence of van Hove sequences means
amenability of the graph $(\mathbb{L}, E)$, $E$ being the set of all
pairs $\ell, \ell'$, such that $|\ell - \ell'|=1$. For nonamenable
graphs, phase transitions with $h\neq 0$ are possible; hence,
statements like Theorem \ref{6tm} do not hold, see \cite{JS,Ly}. Now
we are at a position to prove the existence of the pressure for all
$\mu\in \mathcal{G}^{\rm t}$. It will be done in two subsequent
lemmas.
\begin{lemma} \label{conlm}
The limiting pressure $p(h) \ \stackrel{\rm def}{=} \
\lim_{\mathcal{L}}p_\Lambda (h)$ exists for every van Hove sequence
$\mathcal{L}$. It is independent of the particular choice of
$\mathcal{L}$.
\end{lemma}
\begin{proof}
For $t \geq 0$, $\xi \in \mathit{\Omega}^{\rm t}$, and $\Delta
\subset \Lambda$, let $ \varpi_{\Lambda, \Delta}^{(t)}$,
$Y_{\Lambda, \Delta}(t)$ be defined by (\ref{w4z}) with the
potentials $V_\ell = V$ having the form (\ref{m2}). Then we define
\begin{equation} \label{c6}
f_{\Lambda, \Delta} (t) = \frac{1}{|\Lambda|} \log Y_{\Lambda,
\Delta} (t), \quad t\geq 0.
\end{equation}
This function is differentiable and
\begin{eqnarray} \label{c7}
g_{\Lambda, \Delta} (t) \ \stackrel{\rm def}{=} \ f'_{\Lambda,
\Delta} (t) & = & \frac{1}{2|\Lambda|}\sum_{\ell ,\ell' \in \Delta}
J_{\ell \ell'} \varpi^{(t)}_{\Lambda , \Delta}[(\omega_\ell ,
\omega_{\ell'})_{L^2_\beta} ] \\ & + & \frac{1}{|\Lambda|}\sum_{\ell
\in \Delta ,\ell' \in \Lambda \setminus  \Delta} J_{\ell \ell'}
\varpi^{(t)}_{\Lambda, \Delta} [(\omega_\ell ,
\omega_{\ell'})_{L^2_\beta}]\geq 0. \nonumber
\end{eqnarray}
Here we used that $ \varpi^{(t)}_{\Lambda, \Delta} [(\omega_\ell ,
\omega_{\ell'})_{L^2_\beta}] \geq 0,$ which follows from the GKS
inequality (\ref{w3b}). The function $g_{\Lambda, \Delta}$ is also
differentiable and
\begin{equation} \label{c8}
  g'_{\Lambda, \Delta} (t) \geq 0,
\end{equation}
which may be proven similarly by means of the GKS inequality
(\ref{w3c}). Therefore,
\begin{equation} \label{c8a}
f_{\Lambda, \Delta} (0) \leq f_{\Lambda, \Delta} (1) \leq
g_{\Lambda, \Delta} (1).
\end{equation}
Now we take here $\Delta = \Lambda$ and obtain by (\ref{61q}) that
for any $\alpha \in \mathcal{I}$,
\begin{equation} \label{w32}
\log Y_{\{\ell\}, \{\ell\}} (0) \leq p_\Lambda (h) \leq
 \hat{J}_0  C_{\ref{61q}} (0)/2.
\end{equation}
By the translation invariance the lower bound in (\ref{w32}) is
independent of $\ell$. Therefore, the set $\{p_\Lambda
(h)\}_{\Lambda \Subset \mathbb{L}}$ has accumulation points. For one
of them, $p (h)$, let $\{\Gamma_n\}_{n \in \mathbb{N}}$ be the
sequence of parallelepipeds such that $p_{\Gamma_n} (h) \rightarrow
p(h)$ as $n \rightarrow +\infty$. Let also $\mathcal{L}$ be a van
Hove sequence. Given $n \in \mathbb{N}$ and $\Lambda \in
\mathcal{L}$, let $\mathfrak{L}_n^{-} (\Lambda)\subset
\mathfrak{G}(\Gamma_n)$ (respectively, $\mathfrak{L}_n^{+}
(\Lambda)\subset \mathfrak{G}(\Gamma_n)$) consist of the translates
of $\Gamma_n$ which are contained in $\Lambda$ (respectively, which
have non-void intersections with $\Lambda$). Let also
\begin{equation} \label{c8z}
\Lambda_n^{\pm} = \bigcup_{\Gamma \in \mathfrak{L}_n^{\pm}}
\Gamma.
\end{equation}
Now we take in (\ref{c6}) first $\Delta = \Lambda_n^{-}$, then
$\Delta = \Lambda$, $\Lambda = \Lambda_n^{+}$ and obtain by
(\ref{c8a})
\begin{equation} \label{c8w}
\frac{|\Lambda_n^{-}|}{|\Lambda|} p_{\Lambda_n^{-}} (h) \leq
p_\Lambda (h) \leq \frac{|\Lambda_n^{+}|}{|\Lambda|}
p_{\Lambda_n^{+}} (h).
\end{equation}
Let us estimate $p_{\Lambda_n^{\pm}}(h) - p_{\Gamma_n}(h)$. To this
end we introduce for $t\geq 0$, c.f., (\ref{w4z}),
\begin{eqnarray} \label{c8v}
 X_{\Lambda_n^{-}} (t) & = & \int_{\mathit{\Omega}} \exp \left\{
\frac{1}{2}\sum_{\Gamma \in \mathfrak{L}_n^{-}} \sum_{\ell, \ell'
\in \Gamma} J_{\ell \ell'}
(\omega_{\ell},\omega_{\ell'})_{L_\beta^2} \right.\\  \quad & + &
\left. t \sum_{\Gamma , \Gamma' \in \mathfrak{L}_n^{-}, \ \Gamma
\neq \Gamma'} \sum_{\ell \in \Gamma} \sum_{\ell' \in \Gamma'}
J_{\ell \ell'} (\omega_{\ell},\omega_{\ell'})_{L_\beta^2}\right.
\nonumber \\
& + & \left. \sum_{\ell \in \Lambda_n^{-}}\int_0^\beta\left[ h
\omega_\ell (\tau) - v([\omega_\ell(\tau)]^2) \right]{\rm d}\tau
\right\}\chi_\Lambda ({\rm d}\omega), \nonumber
\end{eqnarray}
and
\begin{equation} \label{c8t}
f_{\Lambda_n^-}(t) = \frac{1}{|\Lambda_n^{-}|}\log
X_{\Lambda_n^{-}} (t).
\end{equation}
Then
\begin{equation} \label{c8q}
f_{\Lambda_n^-}(1) = p_{\Lambda_n^{-}} (h), \quad
f_{\Lambda_n^-}(0) = \frac{|\Gamma_n|}{|\Lambda_n^{-}|}
\sum_{\Gamma \in \mathfrak{L}_n^{-}} p_{\Gamma}(h) = p_{\Gamma_n}
(h).
\end{equation}
Observe that $p_{\Gamma}(h) = p_{\Gamma_n} (h)$ for all $\Gamma \in
\mathfrak{G}(\Gamma_n)$ follows from the translation invariance of
the model. Thereby,
\begin{eqnarray} \label{c8s}
0 & \leq & p_{\Lambda_n^{-}} (h) - p_{\Gamma_n} (h) \leq
f'_{\Lambda_n^{-}} (1) \\ & = & \frac{1}{|\Lambda_n^{-}|}
\sum_{\Gamma , \Gamma' \in \mathfrak{L}_n^{-}, \ \Gamma \neq
\Gamma'} \sum_{\ell \in \Gamma} \sum_{\ell' \in \Gamma'} J_{\ell
\ell'} \pi_{\Lambda_n^{-}} \left(
(\omega_{\ell},\omega_{\ell'})_{L_\beta^2}|0\right) \nonumber \\ &
\leq & \frac{1}{|\Lambda_n^{-}|} \sum_{\Gamma  \in
\mathfrak{L}_n^{-}} \sum_{\ell \in \Gamma} \sum_{\ell' \in \Gamma^c}
J_{\ell \ell'} \pi_{\Lambda_n^{-}} \left(
(\omega_{\ell},\omega_{\ell'})_{L_\beta^2}|0\right)\nonumber
\\
& & \leq \hat{J}(\Gamma_n)C_{\ref{61q}} (0) , \nonumber
\end{eqnarray}
where we used the estimate (\ref{61q}) and set
\begin{equation} \label{c8o}
\hat{J}(\Gamma_n) = \frac{1}{|\Gamma_n|} \sum_{\ell \in \Gamma_n }
\sum_{\ell' \in \Gamma_n^c} J_{\ell \ell'} =
\frac{1}{|\Gamma|}\sum_{\ell \in \Gamma } \sum_{\ell' \in
\Gamma^c} J_{\ell \ell'}, \quad {\rm for} \ {\rm every} \ \Gamma
\in \mathfrak{G}(\Gamma_n).
\end{equation}
In deriving (\ref{c8s}) we took into account that the function
(\ref{c8t}) has positive first and second derivatives, c.f.,
(\ref{c7}) and (\ref{c8}). By literal repetition one proves that
(\ref{c8s}) holds also for $p_{\Lambda_n^{+}} (h) - p_{\Gamma_n}
(h)$. In view of (\ref{c5}) the above $\hat{J}(\Gamma_n)$ may be
made arbitrarily small by taking big enough $\Gamma_n$. Thereby, for
any $\varepsilon>0$, one can choose $n\in \mathbb{N}$ such that the
following estimates hold (recall that $p_{\Gamma_n} \rightarrow p$
as $n \rightarrow +\infty$)
\begin{equation} \label{c80}
|p_{\Gamma_n} (h) - p(h)| < \varepsilon/3, \quad 0 \leq
p_{\Lambda_n^-} (h) - p_{\Gamma_n} (h) \leq p_{\Lambda_n^+} (h) -
p_{\Gamma_n} (h) < \varepsilon/3.
\end{equation}
As $\mathcal{L}$ is a van Hove sequence, one can pick up $\Lambda
\in \mathcal{L}$ such that
\[
\max\left\{\left(\frac{|\Lambda_n^{+}|}{|\Lambda|}- 1 \right)
p_{\Lambda_n^{+}} (h); \left(1 - \frac{
|\Lambda_n^{-}|}{|\Lambda|} \right) p_{\Lambda_n^{+}} (h) \right\}
< \varepsilon/3,
\]
which is possible in view of (\ref{w32}). Then for the chosen $n$
and $\Lambda \in \mathcal{L}$, one has
\begin{eqnarray*}
& & |p_{\Lambda} (h) - p(h) | \leq |p_{\Gamma_n} (h) - p(h) | +
p_{\Lambda_n^+} (h)- p_{\Gamma_n} (h) \\ & & \quad  +
\max\left\{\left(\frac{|\Lambda_n^{+}|}{|\Lambda|}- 1 \right)
p_{\Lambda_n^{+}} (h); \left(1 - \frac{
|\Lambda_n^{-}|}{|\Lambda|} \right) p_{\Lambda_n^{+}} (h) \right\}
< \varepsilon,
\end{eqnarray*}
which obviously holds also for all $\Lambda'\in \mathcal{L}$ such
that $\Lambda \subset \Lambda'$.\end{proof} \vskip.1cm
\begin{lemma} \label{con1lm}
For every $\mu \in \mathcal{G}^{\rm t}$ and any van Hove sequence
$\mathcal{L}$,
\[
\lim_{\mathcal{L}}p^{\mu}_\Lambda (h) = p (h).
\]
\end{lemma}
\begin{proof}
By the Jensen inequality one obtains for $t_1 , t_2 \in \mathbb{R}$,
$\xi \in \mathit{\Omega}^{\rm t}$,
\begin{eqnarray*}
Z_{\Lambda}((t_1 + t_2)\xi) \geq Z_{\Lambda}(t_1 \xi) \exp\left\{
t_2 \sum_{\ell \in \Lambda , \ell' \in \Lambda^c} J_{\ell \ell'}
\pi_\Lambda \left[(\omega_\ell , \xi_{\ell'})_{L^2_\beta} \left\vert
t_1 \xi \right. \right]\right\}.
\end{eqnarray*}
We set here first $t_1 = 0, \ t_2 = 1$, then $t_1 = - t_2 = 1$, and
obtain after taking logarithm and dividing by $|\Lambda|$
\begin{eqnarray} \label{c9}
 p_\Lambda (h) & + & \frac{1}{|\Lambda|} \sum_{\ell \in \Lambda,
\ell' \in \Lambda^c} J_{\ell \ell'} \pi_\Lambda \left[(\omega_\ell ,
\xi_{\ell'})_{L^2_\beta}| 0 \right] \leq p_\Lambda (h , \xi) \\
& & \quad \qquad \quad \leq p_\Lambda (h) + \frac{1}{|\Lambda|}
\sum_{\ell \in \Lambda, \ell' \in \Lambda^c} J_{\ell \ell'}
\pi_\Lambda \left[(\omega_\ell , \omega_{\ell'})_{L^2_\beta}| \xi
\right], \nonumber
\end{eqnarray}
where we used that $\pi_\Lambda \left[(\omega_\ell ,
\omega_{\ell'})_{L^2_\beta}| \xi \right] = \pi_\Lambda
\left[(\omega_\ell , \xi_{\ell'})_{L^2_\beta}| \xi \right]$, see
(\ref{34}). Thereby, we integrate (\ref{c9}) with respect to $\mu\in
\mathcal{G}^{\rm t}$, take into account (\ref{40}), and obtain after
some calculations the following
\begin{eqnarray} \label{c10}
 p_\Lambda (h) & - & \frac{1}{|\Lambda|} \sum_{\ell \in \Lambda,
\ell' \in \Lambda^c} J_{\ell \ell'}\pi_\Lambda
\left(|\omega_\ell|_{L^2_\beta} \left\vert 0 \right.
\right)\mu\left(
|\xi_{\ell'}|_{L^2_\beta}\right) \leq p_\Lambda^\mu \\
& & \quad \qquad \quad \leq p_\Lambda (h) + \frac{1}{|\Lambda|}
\sum_{\ell \in \Lambda, \ell' \in \Lambda^c} J_{\ell \ell'} \mu
\left((\omega_\ell, \omega_{\ell'})_{L^2_\beta} \right). \nonumber
\end{eqnarray}
By means of Theorem \ref{2tm} (respectively, Lemma \ref{klm}), one
estimates $\mu \left((\omega_\ell, \omega_{\ell'})_{L^2_\beta}
\right)$, $\mu\left( |\xi_{\ell'}|_{L^2_\beta}\right)$
(respectively, $\pi_\Lambda (|\omega_\ell|_{L^2_\beta}|0)$) by
positive constants independent of $\ell, \ell'$. Thereby, the
property stated follows from (\ref{c5}) and Lemma \ref{conlm}.
\end{proof}

\subsection{Proof of Theorem \ref{6tm}} \label{sqs5.5}

By Lemma \ref{lylm}, for every $\Lambda \Subset \mathbb{L}$,
$p_\Lambda(h)$ can be extended to a function of  $h \in \mathbb{C}$,
holomorphic in the right and left open half-planes. By standard
arguments, see e.g., Lemma 39, page 34
 of \cite{KozHol}, and Lemma \ref{conlm}
it follows that the limit of such extensions $p(h)$ is holomorphic
in certain subsets of those half-planes containing the real line,
except possibly for the point $h =0$. Therefore, $p(h)$ is
differentiable at each $h \neq 0$. This yields by Lemma \ref{con1lm}
that for every $\mu\in \mathcal{G}^{\rm t}$, the corresponding
$p^\mu(h)$ enjoys this property and the derivatives of all these
functions coincide at every $h \neq 0$. In particular,
\begin{equation} \label{X}
\frac{\partial }{\partial h} p^{\mu_+} (h) = \frac{\partial
}{\partial h} p^{\mu_-} (h).
\end{equation}
For every $\mu \in \mathcal{G}^{\rm t}$ and $\Lambda \Subset
\mathbb{L}$, one has
\begin{eqnarray} \label{X1}
\frac{\partial }{\partial h} p^\mu_\Lambda (h)& = &
\int_{\mathit{\Omega}} \frac{\partial }{\partial
h}\left(p^\mu_\Lambda (h, \xi) \right)\mu({\rm d}\xi) \\ & = &
\frac{1}{|\Lambda|} \sum_{\ell \in \Lambda}\int_0^\beta
\int_{\mathit{\Omega}} \pi_\Lambda \left[ \omega_\ell
(\tau)|\xi\right]\mu({\rm d}\xi) { \rm d}\tau  \nonumber \\ & = &
\frac{1}{|\Lambda|} \sum_{\ell \in \Lambda}\int_0^\beta \mu\left[
\omega_\ell (\tau)\right]{\rm d}\tau \nonumber
\end{eqnarray}
Both extreme measures $\mu_{\pm}$ are translation and shift
invariant. Then combining (\ref{X}) and (\ref{X1}) one obtains
$\mu_{+} (\omega_\ell (0)) = \mu_{-} (\omega_\ell (0))$ for any $h
\neq 0$. By Lemma \ref{bhpn} this gives the proof. $\square$
\vskip.5cm \noindent \textit{\textbf{Acknowledgement:} The authors
are grateful to Sergio Albeverio, Tomasz Komorowski, Yuri
Kondratiev, Michael R\"ockner, and Zdzis{\l}aw Rychlik for many
valuable and stimulating discussions on the matter of the article.}

\end{document}